# The *R*-process Alliance: Actinide Abundances, Variation, and Evolution in Metal-poor Stars

Shivani P. Shah[1,2], Rana Ezzeddine[2], Erika M. Holmbeck[3], Alexander P. Ji[4,5,6], Vinicius M. Placco[7], Ian U. Roederer[1,8], Mohammad K. Mardini[9], Sam A. Usman[4,5,10], Avrajit Bandyopadhyay[2], Timothy C. Beers[8,11], Anna Frebel[8,12,13], Terese T. Hansen[14], Charli M. Sakari[15], and Chris Sneden[16]

[1] 1 North Carolina State University, Department of Physics and Astronomy, Raleigh, NC 27695, USA; spshah7@ncsu.edu
[2] University of Florida, Department of Astronomy, 211 Bryant Space Science Center, Gainesville, FL 32611, USA
[3] Lawrence Livermore National Laboratory, 7000 East Avenue, Livermore, CA 94550, USA
[4] Department of Astronomy & Astrophysics, University of Chicago, 5640 S. Ellis Avenue, Chicago, IL 60637, USA
[5] Kavli Institute for Cosmological Physics, University of Chicago, 5640 S. Ellis Avenue, Chicago, IL 60637, USA
[6] NSF-Simons AI Institute for the Sky (SkAI), 172 E. Chestnut Street, Chicago, IL 60611, USA
[7] NSF NOIRLab, Tucson, AZ 85719, USA
[8] Joint Institute for Nuclear Astrophysics—Center for the Evolution of the Elements (JINA-CEE), USA
[9] Department of Physics and Kavli Institute for Astrophysics and Space Research, Massachusetts Institute of Technology, Cambridge, MA 02139, USA
[10] Department of Physics and Astronomy, Colgate University, Hamilton, NY 13346, USA
[11] Department of Physics and Astronomy, University of Notre Dame, Notre Dame, IN 46556, USA
[12] Department of Physics, Massachusetts Institute of Technology, 77 Massachusetts Avenue, Cambridge, MA 02139, USA
[13] Kavli Institute for Astrophysics and Space Research, Massachusetts Institute of Technology, 77 Massachusetts Avenue, Cambridge, MA 02139, USA
[14] Astronomy Department, Stockholm University, Roslagstullsbacken 21, 114 21 Stockholm, Sweden
[15] Department of Physics and Astronomy, San Francisco State University, San Francisco, CA 94132, USA
[16] Department of Astronomy and McDonald Observatory, The University of Texas, Austin, TX 78712, USA
*Received 2026 April 11; revised 2026 June 10; accepted 2026 June 10; published 2026 September 28*

## Abstract

The actinides, including thorium (Th), are the heaviest observable elements synthesized in the Universe, holding clues to the extremes of the astrophysical and nuclear conditions of *r*-process sites. We present Th abundances based on high-resolution spectroscopy for 47 metal-poor stars, the largest homogeneously analyzed sample to date. The chemical evolution of Th exhibits a decrease in dispersion in [Th/H] and [Th/Fe] from ∼0.6 dex at the lowest metallicities to ∼0.2 dex at higher metallicities. We also find that Th and the lanthanides Eu and Dy are coproduced remarkably well, with average [Th/Eu] ∼ 0.0 across $-3.0 \lesssim$ [Fe/H] $\lesssim -1.5$, as well as across stars with $0.0 \lesssim$ [Eu/Fe] $\lesssim 2.5$. Even so, the absolute range of log $\epsilon$(Th/Eu) is 1.02 dex, with an observed standard deviation of ±0.20 dex and an intrinsic standard deviation of ±0.11 dex at the lowest metallicities. We infer that 68% of *r*-process events have log $\epsilon$(Th/Eu) yields that only vary within a factor of ±1.3 or ±30%, while 5% of *r*-process events have log $\epsilon$(Th/Eu) yields that vary by factors >3.3 approaching ∼10. This serves as a strong constraint for the nuclear and astrophysical models of *r*-process sites and suggests that achieving an *r*-process site that both is prompt and produces a robust log $\epsilon$(Th/Eu) ratio is a challenge for current models.

*Materials only available in the online version of record: machine-readable tables*

## 1. Introduction

The rapid neutron capture process (*r*-process) is responsible for creating roughly half the abundances of elements heavier than iron in the solar system (E. M. Burbidge et al. 1957; A. G. W. Cameron 1957). At early cosmic times ($\gtrsim$12 Gyr ago), it is responsible for creating most of the abundances of elements heavier than iron, before the slow neutron capture process (*s*-process) starts to contribute substantially (P. A. Seeger et al. 1965; R. G. Gratton & C. Sneden 1994; A. McWilliam 1998; S. Honda et al. 2004; J. Simmerer et al. 2004; Á. Skúladóttir et al. 2019; L. Lombardo et al. 2025). In fact, the *r*-process is believed to be solely responsible for the creation of the heaviest group of elements, the actinides, which include the elements thorium, uranium, and plutonium, throughout cosmic time (E. M. Burbidge et al. 1957; D. D. Clayton & M. E. Rassbach 1967; C. Freiburghaus et al. 1999; M. Busso et al. 2001; O. Korobkin et al. 2012, although see A. Choplin et al. 2022, 2025). Moreover, some *r*-process elements are even known to be critical for life and habitability on Earth and exoplanets (e.g., thorium, uranium, iodine; C. Jaupart et al. 2007; C. T. Unterborn et al. 2015; F. Nimmo et al. 2020; H. Luo et al. 2024). However, identifying the primary astrophysical site(s) of the *r*-process and the associated nuclear and astrophysical properties has remained a long-standing challenge, with implications for our understanding of the origin of elements, Galactic formation and evolution, and our own cosmic origin.

A fundamental means of answering this question has been chemical-abundance information derived from the spectra of metal-poor (MP; [Fe/H][17] < −1.0) and very metal poor (VMP; [Fe/H] < −2.0) stars (T. C. Beers & N. Christlieb 2005).

[17] $[A/B] = \log(N_A/N_B)_{\rm Star} - \log(N_A/N_B)_{\odot}$, where $N$ is the number density of the elements of interest, $A$ and $B$.

Currently, 40%–50% of these stars have been found to be $r$-process-enhanced (RPE) stars (e.g., P. S. Barklem et al. 2005; T. T. Hansen et al. 2018; C. M. Sakari et al. 2018; R. Ezzeddine et al. 2020; E. M. Holmbeck et al. 2020; A. Bandyopadhyay et al. 2024b), defined to have $r$-process elemental abundances in excess of twice the Fe abundance compared to the Sun, i.e., [Eu/Fe] $> +0.3$, as well as [Eu/Ba] $> 0.0$ to exclude stars with significant contributions from the $s$-process (T. C. Beers & N. Christlieb 2005; E. M. Holmbeck et al. 2020). In particular, VMP RPE stars are believed to be preceded by only an individual or a few $r$-process-enrichment events (A. Frebel 2018). Thus, the chemical signatures of RPE stars provide a detailed view of the elements created in $r$-process-enrichment events and their corresponding yields. Such studies have already provided a range of clues. For example, the diversity in the abundance patterns of light $r$-process elements ($38 \leqslant$ atomic number, $Z \leqslant 51$; e.g., C. Travaglio et al. 2004; S. Honda et al. 2006, 2007; P. François et al. 2007; C. J. Hansen et al. 2012; M. Aoki et al. 2017; I. U. Roederer et al. 2022a) has indicated the prevalence of a separate type of $r$-process, termed as limited-$r$. On the other hand, the universality in the abundance pattern of the lanthanide elements ($57 \leqslant Z \leqslant 71$; e.g., C. Sneden et al. 2000a, 2009; J. Westin et al. 2000; I. U. Roederer et al. 2022b; M. Racca et al. 2025), along with the correlation of these elements with $r$-process enrichment (i.e., [Eu/Fe]), has indicated the deposition of transuranic isotope fission fragments in the lanthanides (N. Vassh et al. 2020; I. U. Roederer et al. 2023). Signatures of $r$-process elements across a range of metallicities have also been used to assemble the picture of their buildup and evolution in the Universe, shedding light on the occurrence rates, timescales, and prevalence of different $r$-process sites (e.g., C. Battistini & T. Bensby 2016; B. Côté et al. 2019; Á. Skúladóttir et al. 2019; T. Mishenina et al. 2022; X. Ou et al. 2024).

In that respect, the abundances of actinide elements ($89 \leqslant Z \leqslant 103$) have held a special interest. Although lanthanides (e.g., Eu, Dy) and possibly even third-peak $r$-process elements ($76 \leqslant Z \leqslant 83$) exhibit a universal pattern across RPE stars (e.g., I. U. Roederer et al. 2022b; S. P. Shah et al. 2024; A. Alencastro Puls et al. 2025; T. T. Hansen et al. 2025; M. Racca et al. 2025), the same has not been observed for actinides, which require especially neutron-rich conditions to be synthesized (e.g., E. M. Holmbeck et al. 2019b). The actinide element most easily detected and therefore most widely observed is thorium (Th; $Z = 90$). The detection of Th in a VMP RPE star, CS 31082–001, showed that the $\log \epsilon$(Th/Eu) ratio in this star is 0.28 dex or 1.9 times greater than that observed in other VMP RPE stars and the Sun, which follow the universal $r$-process pattern (R. Cayrel et al. 2001; V. Hill et al. 2002). The Th level in this star is so enhanced that it has an implied stellar age of <0 Gyr, using the radioactive property of $^{232}$Th with a half-life of 14.5 Gyr. Not only is this age unphysical, but the star should have an age >9 Gyr given its VMP status. Thus, the idea was introduced that actinides do not always conform to the universal $r$-process pattern, and the actinide-to-lanthanide yields must vary between $r$-process events.

Since the discovery of CS 31082–001, ∼30% of MP stars with actinide detections have shown similarly high levels of Th, now termed as "actinide-boost" stars (L. Mashonkina et al. 2014; E. M. Holmbeck et al. 2019b; V. M. Placco et al. 2023). On the other hand, the discovery of DES J033523−540407 in an ultrafaint dwarf galaxy, for which the $\log \epsilon$(Th/Eu) ratio is 0.34, or 2.2 times lower than typically observed in RPE stars and the Sun, indicated the potential existence of "actinide-deficient" stars, which have actinide-derived radioactive ages greater than the age of the Universe, i.e., >13.8 Gyr (A. P. Ji & A. Frebel 2018). In fact, the absolute observed range of the $\log \epsilon$(Th/Eu) ratio is currently 0.8 dex. The exact delineations of these actinide-boost/actinide-deficient classes of stars depend on the assumed $\log \epsilon$(Th/Eu) zero-age ratio or production ratio (PR), which could be from the solar $r$-process pattern, corrected for the Sun's age (e.g., L. Mashonkina et al. 2014), or from an $r$-process nucleosynthesis model (e.g., H. Schatz et al. 2002; K. Farouqi et al. 2010). In particular, the presence of actinide-boost and actinide-deficient stars indicated PRs outside of the solar-calibrated values. All in all, the observed variations in actinide abundances, along with the requirement of very neutron-rich conditions to synthesize actinides, made it clear that Th abundance determinations in MP stars serve as unique constraints to understanding the extremes of $r$-process conditions, such as the mass fraction of neutron-rich ejecta, magnetic field strengths, and fission cycling (S. Wanajo et al. 2002, 2024; S. Wanajo 2007; M. Eichler et al. 2019; E. M. Holmbeck et al. 2019b; K. A. Lund et al. 2024).

Thus far, Th abundances have been derived for only ∼40 MP stars. These abundances have been obtained over 30 yr by different groups using different abundance-determination techniques and atomic data. Therefore, Th abundances have been limited and inhomogeneous. In this paper, we determine reliable Th abundances for 47 RPE stars—the largest homogeneous sample to date. We use this sample, along with the existing literature Th abundances, to explore the yields of one of the heaviest naturally occurring elements in the Universe and its chemical evolution.

The remainder of this paper is organized as follows: We describe the data and methods for the determination of stellar parameters and abundance estimates in Section 2, the results in Section 3, the discussion in Section 4, and the conclusions in Section 5.

## 2. Data and Methods

### 2.1. Stellar Sample and Spectral Quality

Our sample consists of all stars identified as RPE in the first, third, and fourth $R$-Process Alliance data releases (T. T. Hansen et al. 2018; R. Ezzeddine et al. 2020; E. M. Holmbeck et al. 2020), along with a few stars internally identified as RPE but not published in the data releases. The spectra for the data releases were collected between 2016 and 2020. Additionally, several stars were followed-up for higher-quality data between 2016 and 2024. If available, we used the higher-quality data. Various instruments were used for data collection: the MIKE spectrograph (R. Bernstein et al. 2003) on the 6.5 m Magellan II Clay telescope at Las Campanas Observatory (LCO), the echelle spectrograph on the du Pont 2.5 m telescope at LCO, and the TS23 echelle spectrograph (R. G. Tull et al. 1995) on the Harlan J. Smith 2.7 m telescope at McDonald Observatory. The spectra have variable data quality, with combinations of 0.″35, 0.″5, 0.″7, and 1.″0 slit widths used with 1 × 1, 2 × 1, and 2 × 2 binning for Magellan/MIKE; the widths of the slits 1.″2 and 1.″8 used with the 1 × 1 binning for the TS23; and 1.″0 used with the 2 × 1 binning for the du Pont spectrograph. We limit our sample to stars with signal-to-noise ratio (SNR) > 30 at ∼4000 Å. We also only report Th abundances with $\geqslant 3\sigma$ detections and those for which other blend abundances are reliably determined and the spectral region fits well. As a result, for the stars with Th abundances reported here, the SNR ranges from 33 to 312, with the mean and standard deviation of 133 ± 75 and the resolving power, *R*, ranging between 40,000 and 80,000.

Reliable Th abundances are challenging and observationally expensive to obtain. Therefore, while it is desirable to have a purely homogeneous sample for analysis, we find it more beneficial to combine our sample with the literature sample for a larger sample size. We justify this with the results from our analysis of four benchmark stars, for which we obtained Th abundances consistent with that reported in the literature (Section 2.4.2). We acknowledge biases in the results of the combined samples and discuss them briefly in Section 4.5.

The literature sample was compiled with the help of JINABase (A. Abohalima & A. Frebel 2018), in addition to a general literature search, resulting in 47 stars that we included in this work. The literature stars, abundances, and sources are listed in Table B1. They include stars in the Milky Way (MW) halo, dwarf galaxies, and globular clusters. We chose not to include Th abundances of the following stars: 17 stars from J. Ren et al. (2012), due to the low resolving power and SNR of the spectra, resulting in low detection significance of the Th absorption signature; 2MASS J20093393−3410273 and 2MASS J20492765−5124440 from M. Racca et al. (2025), due to caution about their low effective temperatures; M4 and M5 globular cluster stars, which are *s*-process-enhanced (D. Yong et al. 2008a, 2008b); M15 K83 of the M15 globular cluster, which does not have a reliable Th abundance (C. Sneden et al. 2000b); and HD 74462, HD 108317, HD 122956, and HD 204543 from I. U. Roederer et al. (2009), since they appear to have contributions from the *s*-process.

### 2.2. Radial Velocity Correction and Normalization

The Magellan/MIKE and du Pont spectra were reduced from 2D to 1D using the Carnegie Python Distribution (D. D. Kelson 1998, 2003; D. D. Kelson et al. 2000), while the McDonald spectra were reduced using standard `IRAF`[18] packages (D. Tody 1986, 1993; M. Fitzpatrick et al. 2025). We then used `LESSPayne`[19] (A. P. Ji et al. 2020a; A. P. Ji et al. 2025) for further processing and analysis of the spectra. `LESSPayne` performs semiautomatic spectral analysis by combining the capabilities of `Payne4MIKE` (Y.-S. Ting et al. 2019) and `Spectroscopy Made Harder (SMHr)`[20] (A. R. Casey 2014). We specifically used the capability of `Payne4Mike` to provide theoretical masks for normalization with estimated stellar parameters from a full spectrum fit. These masks are then used within the `SMHr` architecture for normalization with a cubic spline function (we do not use the stellar parameters estimated by `LESSPayne` for abundance determination). We corrected for the radial velocity of each reduced exposure by cross-correlating the 5150–5200 Å region with a rest-frame spectrum of the red giant HD 122563. In the cases where the red-chip spectrum could not be used, we cross-correlated using the 4000–4100 Å region.

### 2.3. Stellar Parameters

Following the standard *R*-Process Allicance (RPA) procedure (e.g., I. U. Roederer et al. 2018, 2024; V. M. Placco et al. 2023), we used photometric effective temperatures and geometric surface gravities for the stars. To determine $T_{\rm eff}$, we used the six color−[Fe/H]−$T_{\rm eff}$ relations provided by A. Mucciarelli et al. (2021). We obtained the *G*, BP, and RP magnitudes from Gaia Data Release 3 (DR3; Gaia Collaboration et al. 2023) and the $K_s$ magnitude from the Two Micron All Sky Survey (2MASS; R. M. Cutri et al. 2003). We dereddened these magnitudes using 3D reddening estimates, $E(B-V)$, from the `bayestar2017` version of the dustmaps[21] application (G. M. Green 2018). In cases where stars were outside the footprint of the 3D dust maps, we used $E(B-V)$ estimates from E. F. Schlafly & D. P. Finkbeiner (2011). For [Fe/H], we used spectroscopic estimates of [Fe II/H] from the RPA data releases. We resampled the input parameters, including magnitude, metallicity, and reddening, from their corresponding error distributions, which we assumed to be Gaussian, and recalculated $T_{\rm eff}$ $10^4$ times. We used the resulting median value of $T_{\rm eff}$ for each color relation and the weighted mean of $T_{\rm eff}$ from all six color relations as the final $T_{\rm eff}$. The statistical uncertainty on $T_{\rm eff}$ is the standard deviation of this distribution. For the systematic uncertainty on $T_{\rm eff}$ we repeated this procedure with the A. Alonso et al. (1999), I. Ramírez & J. Meléndez (2005), and L. Casagrande et al. (2010) calibrations and take the standard deviation of the median $T_{\rm eff}$ values from each relation.

We calculated $\log g$ estimates using the fundamental relation given in I. U. Roederer et al. (2018), with distances from Gaia DR3. More details about this method are described in I. U. Roederer et al. (2018). For the statistical uncertainty on $\log g$, we take the standard deviation of $10^4$ samples of the input parameters. Additionally, we assume a 150 K uncertainty on the $T_{\rm eff}$ as an input parameter to the $\log g$ fundamental relation to account for systematics in $\log g$. After the first estimates of $T_{\rm eff}$ and $\log g$, we estimated [Fe/H] with equivalent width (EW) measurements of Fe II lines from our spectra and removing lines with abundances >±0.27 dex from the mean [Fe II/H] abundance (this number is arbitrary; however, we found that it took care of most outliers). We repeated the above steps to redetermine $T_{\rm eff}$ and $\log g$, which we adopted for the rest of the analysis. Using the adopted $T_{\rm eff}$ and $\log g$, we fit for model metallicity ([M/H]) and microturbulent ($\xi$) velocity by removing any trends of Fe I line abundances with respect to reduced EWs. Most of our stars are red giants, except four, which appear to be horizontal branch stars. The typical total uncertainty on $T_{\rm eff}$ is ±60 K, and the typical uncertainty on $\log g$ is ±0.08 dex. We assumed a fiducial systematic uncertainty of 0.2 km s$^{-1}$ for $\xi$ and ±0.2 dex for [M/H] for all stars (the statistical uncertainties for $\xi$ and [M/H] are negligible). The final stellar parameters and their uncertainties are listed in Table B2.

[18] NOIRLab IRAF is distributed by the Community Science and Data Center at NSF NOIRLab, which is managed by the Association of Universities for Research in Astronomy (AURA) under a cooperative agreement with the US National Science Foundation.

[19] https://github.com/alexji/LESSPayne

[20] https://github.com/andycasey/smhr

[21] https://dustmaps.readthedocs.io/en/latest/

**Table 1**
Atomic Parameters of Transitions Used for Key Elements

| Wavelength (Å) | Species | $\chi$ (eV) | log *gf* |
|---|---|---|---|
| 4118.77 | Co I | 1.05 | −0.48 |
| 4121.32 | Co I | 0.92 | −0.33 |
| 4137.64 | Ce II | 0.52 | 0.40 |
| 4562.36 | Ce II | 0.48 | 0.21 |
| 4012.24 | Nd II | 0.63 | 0.81 |
| 4109.45 | Nd II | 0.32 | 0.35 |
| 4303.57 | Nd II | 0.00 | 0.08 |
| 4156.08 | Nd II | 0.18 | 0.16 |
| 4177.32 | Nd II | 0.06 | −0.10 |
| 4073.12 | Dy II | 0.54 | −0.32 |
| 4449.70 | Dy II | 0.00 | −1.03 |
| 4129.72 | Eu II | 0.00 | 0.22 |
| 4205.04 | Eu II | 0.00 | 0.21 |
| 4019.13 | Th II | 0.00 | −0.23 |

### *2.4. Abundance Analysis*

We determined abundances with EW and spectral synthesis fitting techniques using the 1D LTE radiative transfer code MOOG[22] (C. A. Sneden 1973; C. Sneden et al. 2012), with scattering included[23] (J. S. Sobeck et al. 2011) and using the ATLAS9 plane-parallel atmospheres (F. Castelli & R. L. Kurucz 2003). LESSPayne was used to semiautomatically fit EWs, interpolate model atmospheres, run MOOG, and fit synthesis models. In particular, we customized LESSPayne to specifically fit Th abundances and estimate Th uncertainties semiautomatically for a large sample of stars.

We curated our atomic line list based on I. U. Roederer et al. (2018) and A. P. Ji et al. (2020a) line lists, with atomic data obtained from linemake[24] (V. M. Placco et al. 2021a, 2021b). Our line list is highly tailored so that a homogeneous set of lines can be used in all stars. The final set of lines used and their corresponding atomic data are tabulated in Table 1 for select elements.

For determining abundances with spectral synthesis, LESSPayne uses SMHr's least-squares minimization routine, which simultaneously fits for the local continuum, smoothing, and minute radial velocity shifts over a 10 Å region. We repeated the fitting procedure for all lines six times, updating the elemental abundances after every iteration. LESSPayne stores the derived spectral properties in an SMHr-like session file, which can be opened as an interactive GUI to manually fine-tune the spectral synthesis fits. We individually inspected all lines and adjusted the fits when needed. We used isotopic ratios of *r*-process elements from C. Sneden et al. (2008) and took absolute solar abundances from M. Asplund et al. (2009).

#### *2.4.1. Fitting the Th II λ4019 Line*

We fit for the abundance of Th after the abundances of all other key elements were finalized. We used the strongest Th II transition at 4019.13 Å, since it is the most likely line to be detected and to be reliably fit in a large sample of MP stars. This line is part of an absorption feature at ∼4019.1 Å, primarily formed from blended lines of Fe I, Co I, $^{13}$CH, Ce II, and Ni I in MP stars. We list the transitions and atomic data used to model this absorption feature in Table B3 and depict the blends individually in Figure 1.

To constrain the Fe I blends, we used the mean abundance of Fe I and Fe II lines. Since Fe I generally has an order of magnitude more lines, this mean abundance is typically close to the abundance of Fe I. We find that the strongest Fe I line in the region, at 4019.04 Å, primarily impacts the blue wing of the absorption feature. The uncertainty on the oscillator strength of this line is substantial, with the National Institute of Standards and Technology's Atomic Spectra Database (NIST ASD; A. Kramida et al. 2022) assigning a grade of "E," i.e., >50%. Therefore, in cases where the mean Fe abundance derived from the star does not fit the blue wing of the absorption feature at 4019.1 Å we manually adjust the Fe abundance to fit those pixels.

Another species blended in the feature is the $^{13}$CH molecule with absorption lines in this region at 4018.98 and 4019.15 Å. We determined the abundance of CH with the 4300 Å G band, which consists mainly of $^{12}$CH absorption lines. To constrain the isotopic ratio $^{12}$C/$^{13}$C, we used the $^{13}$CH absorption lines at 4217.70 and 4217.73 Å. We fixed the abundance of CH to the abundance derived from the G band and fit the isotopic ratio using least squares applied to models with $^{12}$C/$^{13}$C isotopic ratios varying between 0.0 and 1.0 in steps of 0.01. We then refit the 4300 Å band with the new isotopic ratio to obtain the final abundance of CH. We find that this technique works well and that the final isotopic ratios follow the expected trends, based on the evolutionary stage of each star (R. G. Gratton et al. 2000).

We also found that the feature at 4019.3 Å is poorly fitted for most stars with the default atomic data of the lines in the region. Although this absorption feature does not significantly impact the estimated Th abundance for stars observed with high spectral resolving power, its substantial strength could impact stars with lower resolving power. The main known contributor to the absorption feature is a Co I line with hyperfine structure (HFS). However, for most stars, the HFS underestimates the line depth of the absorption feature. It is possible that the oscillator strength of the HFS line—which is currently taken from R. L. Kurucz (2011), who also calculated the HFS—is inaccurate and needs to be reexamined. One option is to change the abundance of Co I to fit this feature. A second option is to add a fabricated absorption line for a chosen element that does not have any other absorption lines in this region and then adjust the strength of this line to fit the feature. We tested both options and find that they yield similar Th abundances for most stars within ±0.05 dex. The absorption profile of the synthetic spectra with the two options is also very similar. We decided to use a fabricated line to avoid even the smallest impact on other Co I HFS features in this region. The spectral sysnthesis fits to a subset of stars are shown in Figure 2.

#### *2.4.2. Testing with Benchmarks*

We tested our methods by determining the Th abundance for four benchmark stars and comparing them with the Th abundance obtained in the literature. Here we used CS 31082–001 (V. Hill et al. 2002), HE 1523–0901 (A. Frebel et al. 2007), RAVE J203843.2–002333 (V. M. Placco et al. 2017), and 2MASS J09544277+5246414 (E. M. Holmbeck et al. 2018; S. P. Shah et al. 2023). For this comparison, we

[22] https://www.as.utexas.edu/~chris/moog.html
[23] https://github.com/alexji/moog17scat
[24] https://github.com/vmplacco/linemake

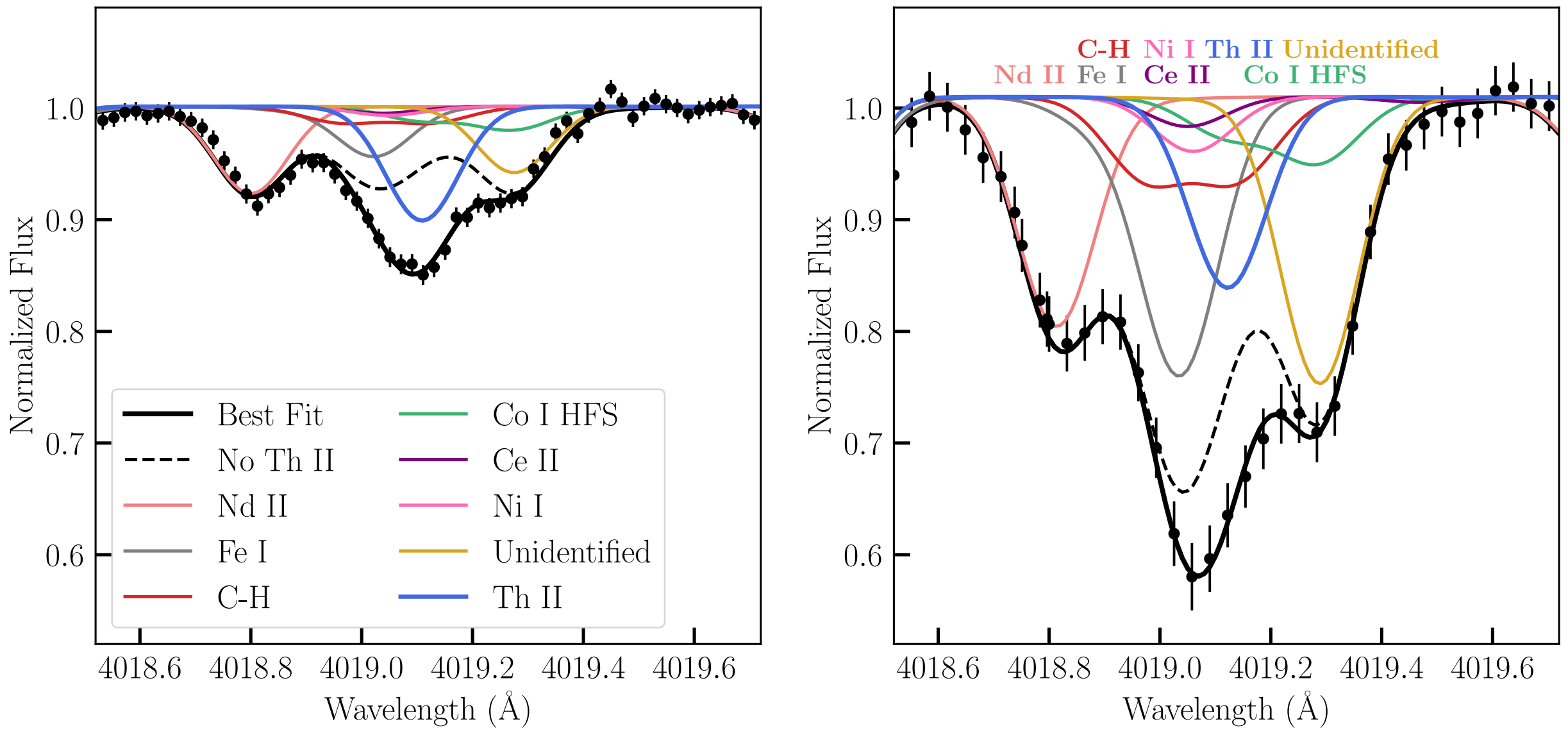

**Figure 1.** Spectra of a VMP star, 2MASS J19215077−4452545 ($T_{\rm eff}$ = 4430 K, log $g$ = 0.39, [M/H] = −2.79), and an MP star, 2MASS J22041814−0232101 ($T_{\rm eff}$ = 4506 K, log $g$ = 1.07, [M/H] = −1.73), shown with black points for the $\lambda$4019 absorption feature. The best-fit model for Th is shown with a solid black line. Colored lines show synthetic spectra with the abundance of only one blend element included, while the abundances of all other elements are set to $-\infty$. This plot depicts the contribution of various species, including Th II, to the $\lambda$4019 absorption feature.

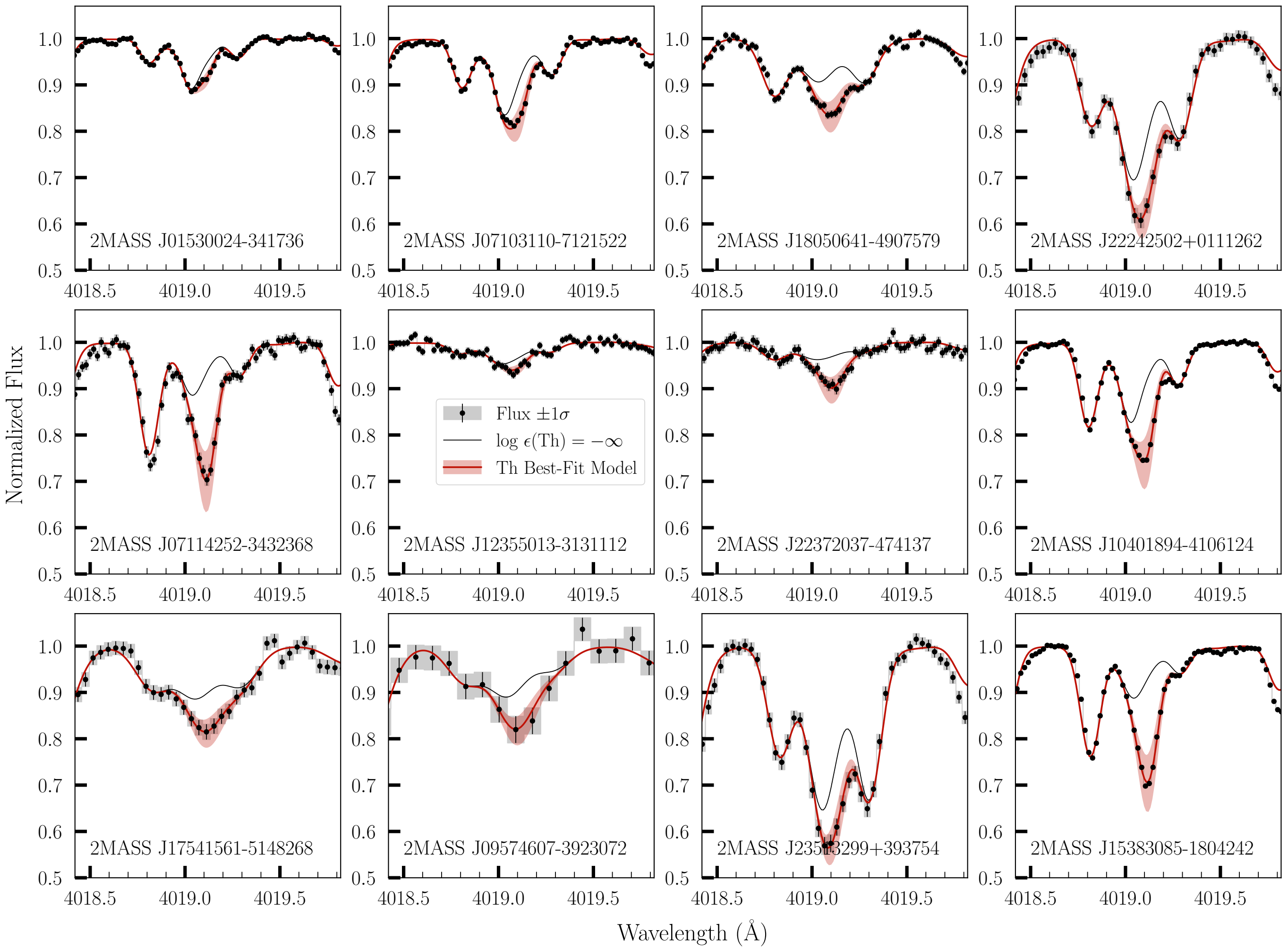

**Figure 2.** Spectral synthesis fits to the Th II $\lambda$4019.13 absorption line for a subset of the stars analyzed in this work. The spectral data are shown with black points, with black error bars indicating $\pm 1\sigma$ of photon noise (note that error bars may not be visible for higher-quality data). The width of the filled gray region indicates the width of the resolution elements, while the height is set by the photon noise. The solid red line traces the best-fit model, with the shaded red region representing the $\pm 0.2$ dex change in abundance.

**Table 2**
Abundance and Isotopic Ratio Comparison between This Work and the Literature of Various Elements for Benchmark Stars

| | Source | J0954+5246[a] | J2038–0023[b] | HE 1523–0901[c] | CS 31082–001[d] |
|---|---|---|---|---|---|
| log $\epsilon$(Fe I) | This Work | 4.46 ± 0.13 | 4.45 ± 0.13 | 4.62 ± 0.13 | 4.55 ± 0.10 |
| | Other | 4.55 ± 0.14 | 4.59 ± 0.12 | 4.50 ± 0.20 | 4.60 ± 0.13 |
| log $\epsilon$(CH) | This Work | 4.95 ± 0.20 | 5.21 ± 0.20 | 5.19 ± 0.20 | 5.78 ± 0.20 |
| | Other | 4.97 ± 0.20 | 5.08 ± 0.20 | 5.14 | 5.82 ± 0.05 |
| $^{12}C/^{13}C$ | This Work | 3.55 | 4.56 | 4.00 | 32.00 |
| | Other | 4.00 | ⋯ | ∼3.00–4.00 | >20.00 |
| log $\epsilon$(Co) | This Work | 1.74 ± 0.0 | 1.85 ± 0.01 | 1.96 ± 0.00 | 2.17 ± 0.00 |
| | Other | 1.94 ± 0.41 | 2.25 ± 0.04 | ⋯ | 2.28 ± 0.11 |
| log $\epsilon$(Eu) | This Work | −1.20 ± 0.01 | −0.93 ± 0.04 | −0.63 ± 0.03 | −0.80 ± 0.01 |
| | Other | −1.16 ± 0.02 | −0.75 ± 0.04 | −0.62 | −0.76 ± 0.11 |
| log $\epsilon$(Dy) | This Work | −0.48 ± 0.04 | −0.36 ± 0.02 | 0.07 ± 0.0 | −0.03 ± 0.01 |
| | Other | −0.47 ± 0.11 | −0.33 ± 0.03 | 0.02 | −0.21 ± 0.13 |
| log $\epsilon$(Th) | This Work | −1.96 | −1.41 | −1.11 | −1.08 |
| | Other | −1.92 | −1.27 | −1.20 | −0.98 |

**Notes.** Source of other work:
[a] S. P. Shah et al. (2023),
[b] V. M. Placco et al. (2017),
[c] A. Frebel et al. (2007),
[d] V. Hill et al. (2002).

adopted the stellar parameters reported in the literature (specific references are listed in Table 2). We list the resulting derived abundances and isotopic ratios in Table 2, along with the corresponding values from the literature. We find good agreement between the abundances of all the elements for all four stars, validating our methodology.

### 2.4.3. Uncertainty Analysis

We took into account different sources of uncertainty for each element's abundances, including standard deviation, photon noise and model fitting, stellar parameters, and—specifically for Th—blend uncertainties. Our methods are inspired by A. P. Ji et al. (2020b), with modifications.

We determined the total uncertainty for all abundances except Th using the following equation:

$$\sigma_{\text{tot}}^2 = \sigma_{\text{std}}^2 + \sigma_{\text{SP}}^2 + \sigma_{\text{stat}}^2, \tag{1}$$

where $\sigma_{\text{std}}$ is the standard deviation of the abundances derived from multiple lines, $\sigma_{\text{stat}}$ is the average statistical abundance uncertainty of the lines used, and $\sigma_{\text{SP}}$ is the average uncertainty of the lines from the adopted stellar parameters. For each line $i$, $\sigma_{\text{stat},i}$ is obtained by changing the abundance so that $\Delta\chi^2 = 1\sigma$, where $\sigma$ is the uncertainty in the data due to photon noise. For each line $\sigma_{\text{SP},\ i}$, is obtained using the following equation:

$$\sigma_{\text{SP},\,i}^2 = \delta_{T_{\text{eff}}}^2 + \delta_{\log g}^2 + \delta_{\xi}^2 + \delta_{[\text{M/H}]}^2, \tag{2}$$

where $\delta$ components were determined by separately increasing the stellar parameters $T_{\text{eff}}$, $\log\ g$, $\xi$, and [M/H] by their respective uncertainties and reanalyzing the absorption line.

For the total uncertainty on Th abundances, which is determined with only a single line, we used the following equation:

$$\sigma_{\text{Th}}^2 = \sigma_{\text{blend}}^2 + \sigma_{\text{SP}}^2 + \sigma_{\text{stat}}^2. \tag{3}$$

For Th, $\sigma_{\text{SP}}$ was determined by separately increasing each stellar parameter by their respective uncertainties, refitting lines of all elements, and then refitting the Th II $\lambda$4019 absorption line. Using a few stars, we found that the $^{12}C/^{13}C$ isotopic ratios are minimally changed when stellar parameters are changed, so we did not rederive them in this process. $\sigma_{\text{stat}}$ was determined similarly to other elements.

We determined $\sigma_{\text{blend}}$ for Th by changing the abundances of CH, Fe, and Co separately and then rederiving the Th abundance. Specifically, we increase CH abundance by 0.1 dex (in absence of a straightforward and accurate uncertainty estimate for molecules), Fe abundance by $\sigma_{\text{std}}$ of Fe I, and Co abundance by $\sigma_{\text{std}}$ of Co I. The resulting uncertainty components CH, Fe, and Co were added in quadrature to obtain $\sigma_{\text{blend}}$. We only take into account Fe, Co, and CH since they are the dominant and most impactful blends in the $\lambda$4019 feature; +0.1 dex changes to Ni and Ce abundances alter Th abundances by −0.02 dex at most, possibly due to the low metallicity of the stars. We do not account for uncertainty from the fabricated line, since its abundance is derived to precisely fit the data pixels in this region.

Finally, we determined the uncertainties in the [X/Y] and log $\epsilon$(X/Y) ratios using Equation (4) when neither element is Th and Equation (5) when one of the elements is Th:

$$\sigma_{[\text{X/Y}]} = \sigma_{\text{stat, X}}^2 + \sigma_{\text{stat, Y}}^2 + \sigma_{\text{std, X}}^2 + \sigma_{\text{std, Y}}^2 + \sum_{\text{SP}} \delta_{\text{X},SP} - \delta_{\text{Y, SP}} \tag{4}$$

$$\sigma_{[\text{Th/Y}]} = \sigma_{\text{stat, Th}}^2 + \sigma_{\text{stat, Y}}^2 + \sigma_{\text{blend, Th}}^2 + \sigma_{\text{std, Y}}^2 + \sum_{\text{SP}} \delta_{\text{Th,SP}} - \delta_{\text{Y, SP}}. \tag{5}$$

In addition to these uncertainties, systematic uncertainties originating from atomic parameters are present. We estimate an uncertainty in the log *gf* value of the Th II line at 4019 Å of 0.04 dex (H. Nilsson et al. 2002). However, we do not add this to the total uncertainty, since it will shift all abundances equally and systematically by a minimal amount.

### 2.5. Detection Thresholds of Thorium

A large sample of Th abundances has made it possible to infer its yields and chemical evolution in the early Universe. However, given the relatively weak and blended nature of the Th II $\lambda4019$ line, it is critical to understand the detection thresholds (or limits) and subsequent biases to the results.

To estimate the detection thresholds of Th, we followed the procedure from I. U. Roederer (2013). We determined the detection limits for a cool red giant star with $T_{\rm eff} = 4500$ K, $\log g = 1.0$, and $\xi = 2.0$ km s$^{-1}$. In reality, our sample extends in $T_{\rm eff}$ from ∼5600 to ∼4300 K and in log $g$ from 3.0 to 0.5. We used the Cayrel formula (R. Cayrel 1988; R. Cayrel et al. 2004) as in S. P. Shah et al. (2024) to estimate the minimum EW detectable with $3\sigma$ significance in two scenarios: (1) a spectrum with resolving power $R \sim 30{,}000$ and SNR = 60, and (2) a spectrum with $R \sim 60{,}000$ and SNR = 80. These two scenarios represent the typical data qualities that observers in the neutron-capture community aim to obtain for studying heavy-element abundances, although there are several stars in the literature and our sample with better and worse data qualities. For these two scenarios, we obtained $\mathrm{EW}_{\rm limit} \sim 6.0$ and 3.0 mÅ, respectively. Using these EWs and the curve-of-growth method with MOOG, we obtained the corresponding "minimum detectable" abundances with the Th II $\lambda4019$ line in the range of [Fe/H] = −4.0 to −1.0. In the cases where an abundance ratio $\log\epsilon$(Th/X) is of interest, we determined the detection threshold of the ratio using the median abundance of the element X from our sample. These thresholds are indicated in Sections 3.2, 3.3, and the Appendix A.

We then used these detection thresholds as guidelines to determine whether there are apparent trends in the data that may have been biased by the limits of Th detection. Specifically, we consider the data to be minimally impacted if the mean abundance values in the relevant bins are higher than the detection thresholds by $3\sigma$, where $\sigma$ is the standard deviation of the abundances in the bins. This is a simple test, especially given the limited scenarios considered for the detection thresholds, including only one set of stellar parameters, two sets of data qualities, and, for Th/X ratios, a constant abundance value for the element X. The Cayrel formula is also idealistic, as errors from continuum placement and blends are not considered. However, for the present data and the scope of this paper, we find this test to be sufficiently effective.

## 3. Results

### 3.1. A New Sample of Thorium Abundances

We obtained Th abundances for 47 RPE stars, including 43 new stars and 4 stars from the literature (I. U. Roederer et al. 2018, 2024; Z. Hackshaw et al. 2026, in preparation; M. K. Mardini et al. 2026, in preparation). This is the largest sample of homogeneously determined Th abundances in MP stars, almost doubling the current literature sample. Abundances and uncertainties for Fe, Eu, Dy, and Th are listed in Tables B3, B4, B5, and B6, respectively. We adopt the classifications from E. M. Holmbeck et al. (2020), so that non-RPE stars have [Eu/Fe] $\leqslant 0.3$, $r$-I stars have $+0.3 <$ [Eu/Fe] $\leqslant +0.7$, $r$-II stars have $+0.7 <$ [Eu/Fe] $\leqslant +2.0$, and $r$-III stars have [Eu/Fe] $> +2.0$. Additionally, $r$-I, $r$-II, and $r$-III stars are defined to have [Eu/Ba] $> 0.0$. While not formally defined in the literature, we also required non-RPE stars presented in this work and selected from the literature to have [Eu/Ba] $> 0.0$ to exclude significant $s$-process contributions. With this, our Th sample consists of 27 VMP stars and 20 MP stars, including 1 non-RPE star, 24 $r$-I stars, 21 $r$-II stars, and 1 $r$-III star. For comparison, the Th abundances in the literature span 37 VMP and 10 MP stars, including 8 non-RPE stars, 6 $r$-I stars, 32 $r$-II stars, and 1 $r$-III star. The [Th/H] and [Th/Fe] distributions of our sample and the literature sample are shown in Figure 3. We find that the distributions of our sample differ from those of the literature sample. However, these differences can be explained by the underlying selection effects: our sample has more MP and $r$-I stars (see the bottom panels of Figure 3), which have lower [Th/Fe] and higher [Th/H] values than $r$-II and MP stars.

Note that Th detections in the literature and this work are biased toward stars with high $r$-process enrichment, given by [Eu/Fe], i.e., most stars with Th detections are RPE. This is seen in Figure 4, where we show the [Eu/H] and [Eu/Fe] values of all stars in the first, second, third, and fourth data releases with white markers, and stars with Th detected in this work are highlighted with red markers. Stars in the literature with Th detected are also shown. Most stars with Th detected have [Eu/Fe] $> 0.3$; this is due to an intentional selection bias in the literature and this work to increase the success of detecting Th. Therefore, along with the bias from the detection thresholds (Section 2.5), we must also take into account the bias in the sample selection when interpreting results in the following sections.

### 3.2. Relationship with [Fe/H]

The relationships of [Th/H] and [Th/Fe] with [Fe/H] are shown in Figure 5. As expected, we observe [Th/H] to increase as a function of metallicity, representing the increasing absolute abundance of cosmic Th as a function of [Fe/H] or, equivalently, the age of the Universe. Some stars with [Fe/H] $\sim -1.5$ approach solar [Th/H] ratio, and one star, 2MASS J22132050–5137385, surpasses this value at a metallicity of [Fe/H] $\sim -2.2$ and is discussed in more detail by I. U. Roederer et al. (2024). The running mean and standard deviation, shown in black, are determined using sliding bins of 16 stars, with 8 stars overlapping between the bins. We suspect that the sample is highly biased at [Fe/H] $\lesssim -2.2$, since the Th detection thresholds pass within two standard deviations of the mean [Th/H] of these bins (Section 2.5). Therefore, stars with [Th/H] $< -3.0$ could exist but would require dedicated efforts to be discovered and have their Th abundance reliably determined.

In contrast, we find that [Th/Fe] exhibits a slight decreasing trend with increasing metallicity, before flattening out. We suspect that this trend is also biased by the detection thresholds at [Fe/H] $\lesssim -2.2$, as well as by selection biases toward stars with high [Eu/Fe], as seen in Figure 4. In fact, given the one-to-one correlation between Th and Eu (Section 3.3), we default to expecting that the chemical-evolution picture of [Th/Fe] would be similar to the classic trumpet-shaped evolution seen for [Eu/Fe] in Figure 4, with a flat trend from [Fe/H] $\sim -3.0$ to $\sim -1.5$ and reducing dispersion at higher metallicities, as homogenized levels of [Th/Fe] are reached in the interstellar medium (ISM) (see also A. McWilliam et al. 1995; C. Travaglio et al. 1999; K. A. Venn et al. 2004; B. Côté et al. 2019; E. M. Holmbeck et al. 2020; C. Kobayashi et al. 2020; E. L. Lombardo et al. 2025, and references therein). However, it is also possible that the true [Th/Fe] trend takes a different form. We use simple inverse modeling to find that, given the detection limits, both a "true" decreasing trend and a "true" flat trend in [Th/Fe] can reproduce the observed decreasing trend (see the Appendix A and Figure A1). On the other hand, a "true" increasing trend at low metallicities, like that observed for Ba and Sr (e.g., A. McWilliam 1998; P. S. Barklem et al. 2005; I. U. Roederer 2013; C. Kobayashi et al. 2020; L. Lombardo et al. 2025, and references therein), results in a flatter trend than what we observe.

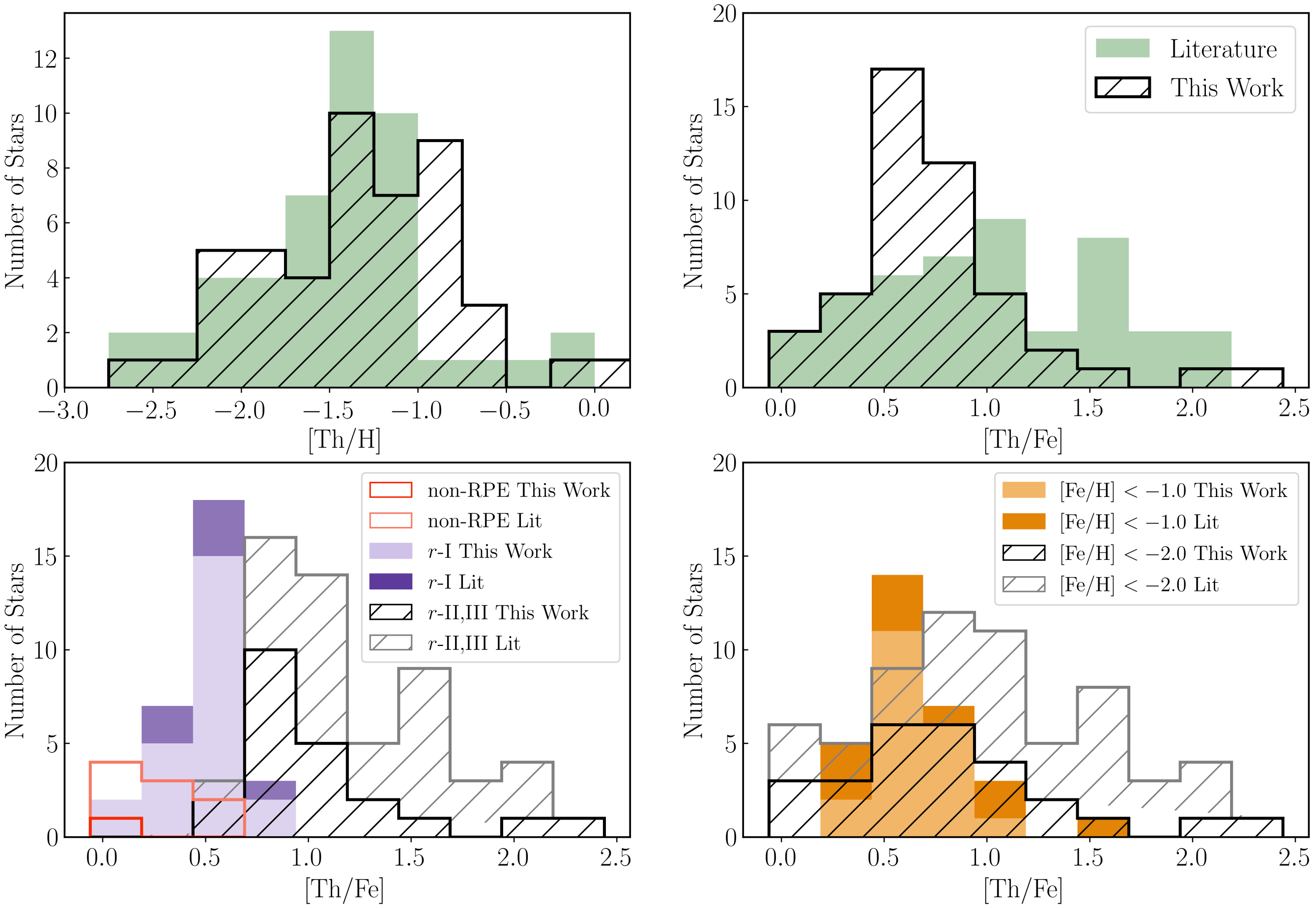

**Figure 3.** The top panels show [Th/H] and [Th/Fe] distributions from this work and the literature. Bottom left panel: [Th/Fe] divided by $r$-process enrichment classes: non-RPE ([Eu/Fe] $\leqslant +0.3$) in red shades, $r$-I ($+0.3 <$ [Eu/Fe] $\leqslant +0.7$) in purple shades, $r$-II ($+0.7 <$ [Eu/Fe] $\leqslant +2.0$) and $r$-III stars ([Eu/Fe] $> +2.0$) in gray shades. All stars in the literature sample and from this work also have [Eu/Ba] $> 0.0$. Bottom right panel: [Th/Fe] divided by metallicity classes: MP ([Fe/H] $< -1.0$) in yellow shades and VMP ([Fe/H] $< -2.0$) in gray shades. For both bottom panels, the distributions from our work and the literature are stacked for each class.

Although the true trend of [Th/Fe] may still be elusive, the decrease in the dispersion of [Th/Fe] is clear, with the standard deviation evolving from ∼0.6 dex at the lowest metallicities (a similar value was found for [Eu/Fe] by K. Brauer et al. 2021) to ∼0.2 dex at the highest metallicities. This follows the expectation from chemical evolution, where lower-metallicity stars are tracing early discrete and stochastic enrichment events, while higher-metallicity stars begin tracing homogenized [Th/Fe] levels in the ISM from the continuous evolution of >30 enrichment events (e.g., K. A. Venn et al. 2004; K. Brauer et al. 2021; A. Frebel & A. P. Ji 2023; X. Ou et al. 2024). Determining the exact metallicity at which the chemical composition of the stars transitions from being enrichment dominated to evolution dominated is beyond the scope of this paper. Here we simply consider this transition to occur at [Fe/H] $= -2.2$, given the sharp drop in the standard deviation of [Th/Fe] abundances at this metallicity. This value is also consistent with the use of the eXtreme Deconvolution Gaussian Mixture Modeling algorithm, `XDGMM`[25] (J. Bovy et al. 2011; J. T. Vanderplas et al. 2012) in two dimensions of [Th/Fe] versus [Fe/H] and with two assumed Gaussian components. For the remainder of the paper, we refer to the [Fe/H] $< -2.2$ region as the enrichment regime and the [Fe/H] $\geqslant -2.2$ region as the evolution regime. The mean and standard deviation of [Th/Fe] in the enrichment and evolution regimes are $+1.09 \pm 0.55$ and $+0.69 \pm 0.33$, respectively.

### *3.3. Relationship with Eu and Dy*

We find that [Th/Fe] versus [Eu/Fe] and [Th/Fe] versus [Dy/Fe] follow a remarkably tight correlation. This is shown in Figure 6, where data from this sample and the literature are plotted, along with binned means and standard deviation. In fact, the average trend of [Th/Fe] versus [Eu/Fe] consistently follows the solar $r$-process model and absolute ratios (M. Asplund et al. 2009; N. Prantzos et al. 2020). This was also observed in I. U. Roederer et al. (2009) with 20 RPE stars. Here we show, for the first time, that the correlation is present for most classes of *r*-process enrichment, including non-RPE, *r*-I, *r*-II, and *r*-III (although there are limited data for non-RPE and *r*-III stars).

[25] 11.2.2.1. astroML.density_estimation.XDGMM — astroML 1.0 documentation.

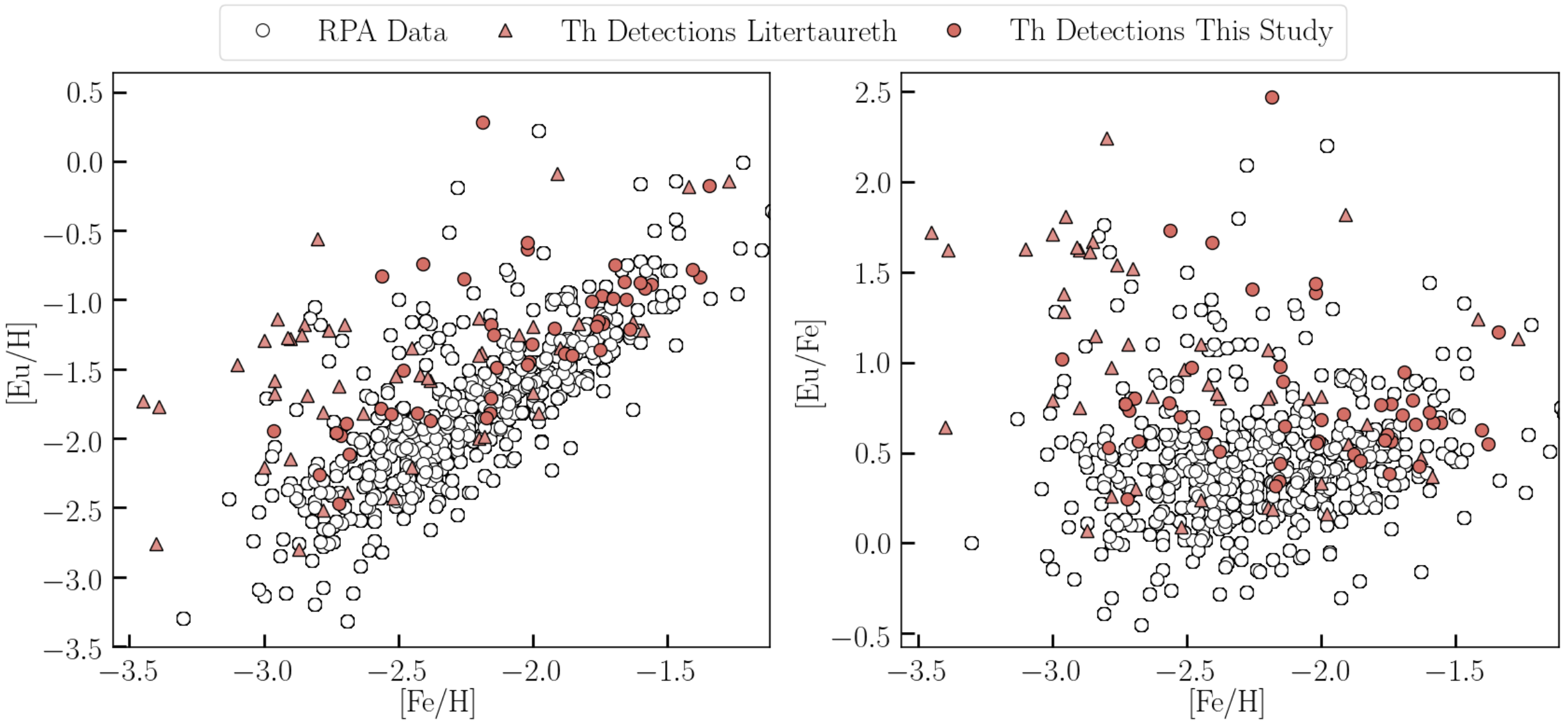

**Figure 4.** [Eu/H] and [Eu/Fe] as a function of [Fe/H] are shown for stars from the first, second, third, and fourth RPA data releases with white circles. Stars with Th abundances derived in this work are highlighted with red circles (note that the [Eu/H], [Eu/Fe], and [Fe/H] values of these stars are as rederived in this work and not as reported in the RPA data releases). [Eu/H] and [Eu/Fe] for stars with Th abundances derived in the literature are also shown with red triangles. Both the literature sample and our sample of Th abundances are biased toward stars with higher *r*-process enrichment, given by [Eu/Fe].

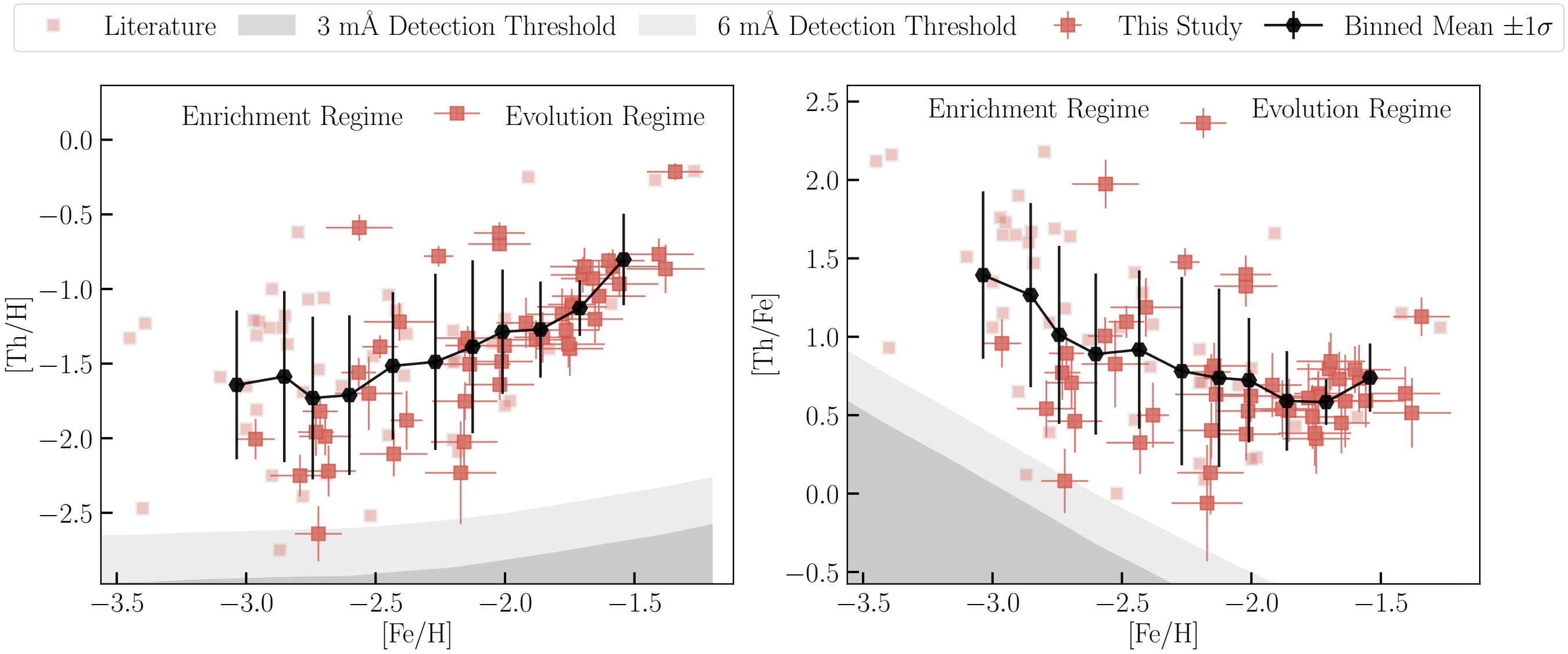

**Figure 5.** Evolution of [Th/H] and [Th/Fe], as a function of [Fe/H], for stars from this work (bright red) and the literature (light red). Regions below the detection thresholds are shown with shaded gray areas. Mean and standard deviation values in sliding bins of 16 stars with an overlap of 8 stars are shown with a solid black line and black error bars, respectively. We define the enrichment regime as [Fe/H] $< -2.2$ and the evolution regime as [Fe/H] $\geqslant -2.2$.

Additionally, we find that the $\log\epsilon$(Th/Eu) ratios are approximately constant as a function of metallicity and also follow the solar *r*-process model and absolute ratios (Figure 7). This was also indicated by I. U. Roederer et al. (2009). We argue that any dips and increases in the trend, as well as the slightly elevated values of the VMP stars, are within the standard deviation of other bins and/or can be explained by the detection limits (e.g., Section 2.5 and the Appendix A). We determine the mean ($\pm$ standard error) $\log\epsilon$(Th/Eu) of all bins to be $-0.50 \pm 0.02$, of [Fe/H] $< -2.2$ bins (enrichment regime) to be $-0.45 \pm 0.2$, and of [Fe/H] $\geqslant -2.2$ bins (evolution regime) to be $-0.55 \pm 0.01$. All of these values are within $3\sigma$ of each other, as well as the solar $\log\epsilon$(Th/Eu) *r*-process model ratio of $-0.48$ (N. Prantzos et al. 2020) and solar absolute ratio of $-0.50$ (M. Asplund et al. 2009). We conclude that there is no statistically significant difference in the mean ratios of $\log\epsilon$(Th/Eu) in the range of metallicities found in our sample.

Similarly, $\log\epsilon$(Th/Dy) also exhibits an approximately constant trend as a function of metallicity, along with the remarkable correlation of [Th/Fe] versus [Dy/Fe]. However, slight departures are observed from the solar *r*-process model and absolute ratios for both trends. In Figure 7, while the [Fe/H] $< -2.5$ stars are close to the solar ratios, more metal-rich stars exhibit lower $\log\epsilon$(Th/Dy) ratios by $\sim$0.2 dex. This is possibly due to the contribution of other neutron-capture processes to the production of Dy at higher metallicities (85% of solar Dy is attributed to the *r*-process, whereas 95% of solar Eu and 100% of solar Th is attributed to the *r*-process; N. Prantzos et al. 2020). We also consider that a positive offset in the Dy abundances of our sample relative to the literature sample may exist, contributing to the deviations. The exact cause is unclear and could be differences in the Dy II transitions used and/or differences in the log *gf* values.

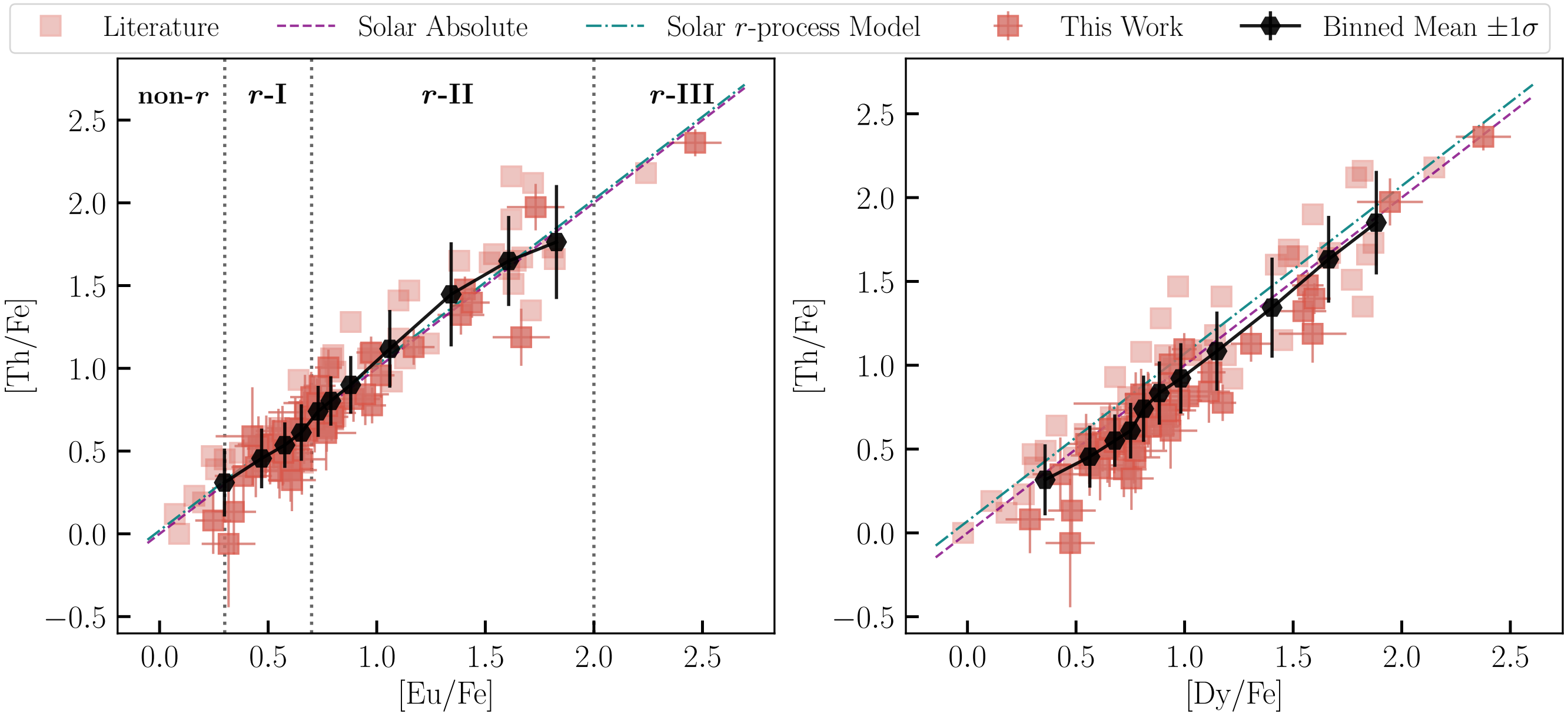

**Figure 6.** [Th/Fe] vs. [Eu/Fe] and [Th/Fe] vs. [Dy/Fe]. Dashed purple lines represent the absolute solar ratio of $\log\epsilon$(Th/Eu) and $\log\epsilon$(Th/Dy), while the dashed blue line represents the *r*-process model ratios of the same. Non-RPE, *r*-I, *r*-II, and *r*-III classes are labeled in the left panel. Mean and standard deviation values in sliding bins of 16 stars with 8 stars overlapping are shown by a solid black line and black error bars, respectively.

### 3.4. Dispersion in $\log\epsilon$(Th/Eu) Ratios

Although the mean trend of the ratios $\log\epsilon$(Th/Eu) is constant as a function of metallicity, there is a distinct variation around this mean (Figure 7). The absolute range of $\log\epsilon$(Th/Eu) is 1.02 dex, although driven primarily by a few VMP stars. This range is slightly higher than previously reported in the literature (e.g., L. Mashonkina et al. 2014; A. P. Ji & A. Frebel 2018; E. M. Holmbeck et al. 2019a; V. M. Placco et al. 2023), due to the star 2MASS J14534137 +0040467 in our sample (Mardini et al. 2026, in preparation). A similarly high absolute range is observed for $\log\epsilon$(Th/Dy) of 1.03 dex, while the absolute range of $\log\epsilon$(Dy/Eu) is lower at 0.60 dex.

Moreover, for the first time, we note an evolution in the standard deviation of $\log\epsilon$(Th/Eu) from 0.18 dex in the enrichment regime to 0.10 dex in the evolution regime. A less pronounced evolution may also be present for the $\log\epsilon$(Th/Dy) ratio. However, a similar evolution is not observed for the $\log\epsilon$(Dy/Eu) ratio, which exhibits a consistently low standard deviation of $\sim$0.10 dex throughout the metallicity range. Since Dy and Eu are known to be produced in a universal ratio in RPE stars (see, e.g., C. Sneden et al. 2000a; A. Frebel 2018; J. J. Cowan et al. 2021, and references therein), we suspect that the dispersion in $\log\epsilon$(Dy/Eu) is primarily driven by observational uncertainties. On the other hand, we consider the evolution observed for $\log\epsilon$(Th/Eu) and $\log\epsilon$(Th/Dy) a possible signature of varying actinide-to-lanthanide yields of *r*-process events, such that the dispersion signature is most pronounced in the enrichment regime and diminishes in the evolution regime.

Given the implications, we seek to further characterize the $\log\epsilon$(Th/Eu) variations statistically. We fit the $\log\epsilon$(Th/Eu) distributions of the full sample and of the enrichment-regime sample with Gaussian mixture models (GMMs). For both cases, we tried one-, two-, and three-component GMMs; however, we found that the differences in the Akaike information criterion (AIC) values—which estimate the relative amount of information loss while encouraging goodness of fit—are small. Given the still small sample, we chose to fit the distributions with one Gaussian component and obtained a mean ($\mu$) and standard deviation ($\sigma$) of $-0.49 \pm 0.17$ dex for the full sample and $-0.43 \pm 0.20$ dex for the enrichment-regime sample. The corresponding probability distribution functions (pdf's) are shown in Figure 8 with solid lines. Using the statistics for the full sample, the model suggests that for [Fe/H] $< -1.4$ ($\sim$MP) stars 32% of the observations have $\log\epsilon$(Th/Eu) values $>\mu + \sigma = -0.32$ or $<\mu - \sigma = -0.66$ and 5% of the observations have $\log\epsilon$(Th/Eu) $>\mu + 2\sigma = -0.15$ or $<\mu - 2\sigma = -0.83$. Interestingly, the $\pm1\sigma$ limits of the full sample coincide with the current actinide-deficient ($-0.33$) and actinide-boost ($-0.66$) limits, based on the $\log\epsilon$(Th/Eu) PR of $-0.33$ from the *r*-process waiting-point calculations of H. Schatz et al. (2002). The statistics of the enrichment-regime sample tell a similar story: 5% of $\log\epsilon$(Th/Eu) have $>\mu + 2\sigma = -0.03$ or $<\mu - 2\sigma = -0.83$ for [Fe/H] $< -2.2$ ($\sim$VMP) stars.

While the above model provides a means to characterize the *observed* $\log\epsilon$(Th/Eu) variation, we ask how much of the variance is *intrinsic*, i.e., reflecting the variation in the yields of *r*-process events, as opposed to observational uncertainties. This question especially gains relevance, as the average uncertainty of the values of $\log\epsilon$(Th/Eu) is $\pm$ 0.16 dex, which is similar to the standard deviation obtained from the fit GMMs. We used Bayesian analysis to fit for the mean ($\mu$) and intrinsic standard deviation ($\sigma_{\rm int}$) of the log $\epsilon$(Th/Eu) sample. The likelihood function for each data point is given in Equation (6).

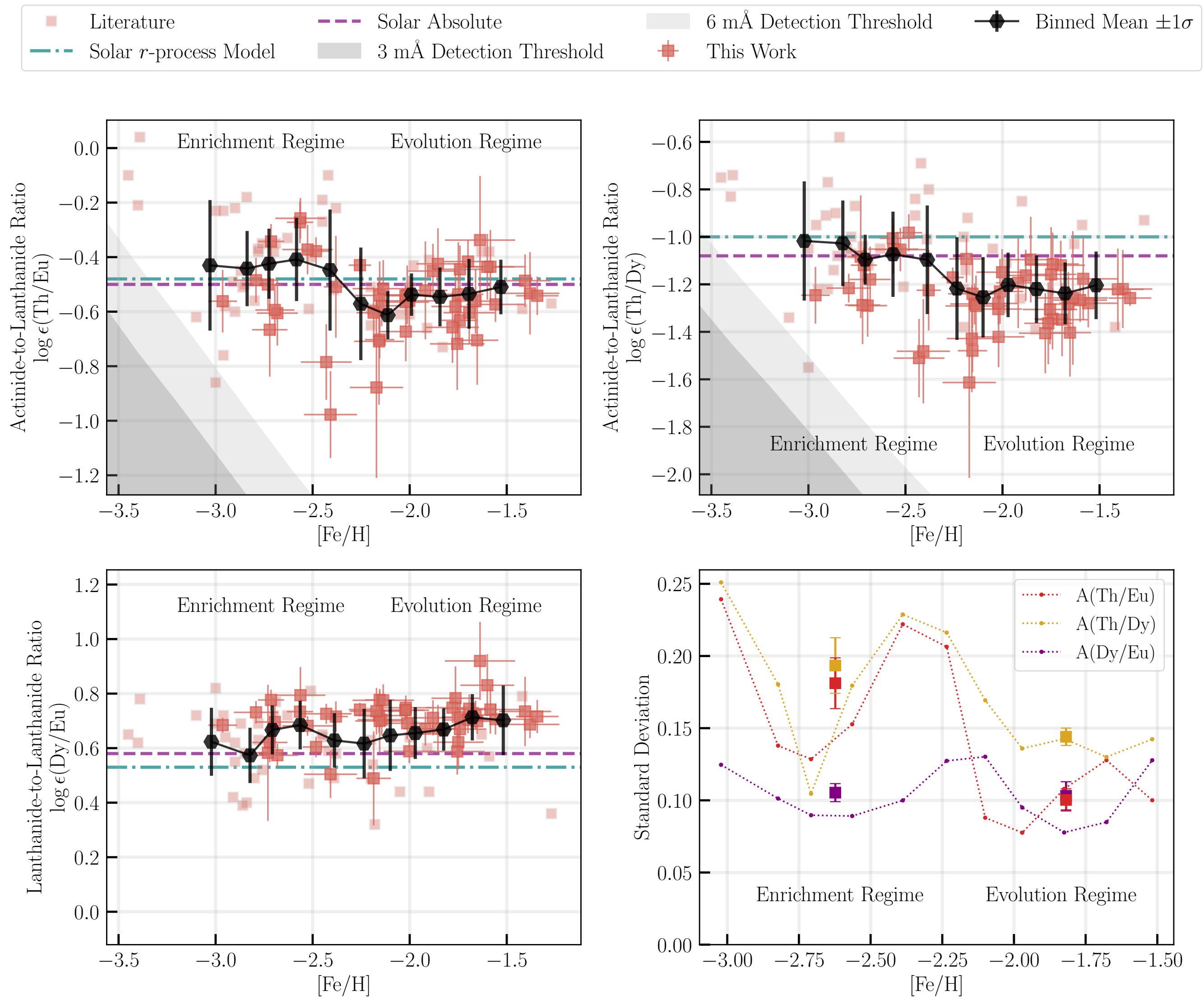

**Figure 7.** log $\epsilon$(Th/Eu), log $\epsilon$(Th/Dy), and log $\epsilon$(Dy/Eu), as a function of [Fe/H]. Mean and standard deviation values in sliding bins of 16 stars with 8 stars overlapping are shown with a solid black line and black error bars, respectively. The bottom right panel explicitly shows the standard deviation of each bin with dotted lines for the three ratios. The squares with error bars indicate the mean of the standard deviation in the enrichment regime and evolution regime separately, along with the standard errors of the means.

$$L_i = P(x| \quad \log \epsilon\,(\mathrm{Th/Eu})_i,\, \sigma_i) \cdot P(x| \quad \mu,\, \sigma_{\rm int}) \equiv \frac{1}{\sqrt{2\pi(\sigma_i + \sigma_{\rm int})^2}} \exp \frac{(x_i - \mu)^2}{2(\sigma_i + \sigma_{\rm int})^2}. \qquad (6)$$

It considers both a Gaussian probability distribution for the data point with observed uncertainty, $\sigma_i$, and a Gaussian probability distribution for all data points with standard deviation, $\sigma_{\rm int}$ (e.g., E. N. Kirby et al. 2011; A. P. Ji et al. 2023; A. M. Luna et al. 2025). We optimize the sum of the log-likelihood of all data points ($\mathcal{L} = \sum_i^n \ln L_i$) with Markov Chain Monte Carlo (MCMC), using `emcee` (D. Foreman-Mackey et al. 2013). We used 32 walkers with 5000 steps for each walker and burned the first 100 steps. For our prior, we assumed that $-3.0 < \mu < 0.2$ and $0.0 \leqslant \sigma_{\rm int} < 2.0$.

The resulting posterior distributions of $\mu$ and $\sigma_{\rm int}$ are given in Figure 9. They look relatively Gaussian, so we use the mean of these distributions as the best-fit values. For the full sample (Figure 9(a)), the best-fit $\mu$ with the $\pm 1\sigma$ confidence level is $-0.50 \pm 0.02$ and the best-fit $\sigma_{\rm int}$ is $0.08 \pm 0.02$ dex. For the enrichment regime (Figure 9(b)), we obtained $\mu = -0.43 \pm 0.03$ and $\sigma_{\rm int} = 0.11 \pm 0.04$ dex. The best-fit means are similar to that obtained by averaging the bin means (Section 3.4) and from the GMM fits, whereas the best-fit standard deviations are lower by ∼0.10 dex. The corresponding pdf's of log $\epsilon$ (Th/Eu) characterized by these best-fit values are shown in Figure 8 with dashed lines. Note that a similar analysis applied to the evolution regime resulted in best-fit $\mu$ of $-0.55 \pm 0.02$ and best-fit $\sigma_{\rm int}$ of $0.00 \pm 0.03$, providing further evidence that the abundances in this regime are homogenized owing to chemical evolution (here we changed our prior to $-2.0 \leqslant \sigma_{\rm int} \leqslant 2.0$ to ensure that the posterior distribution is Gaussian).

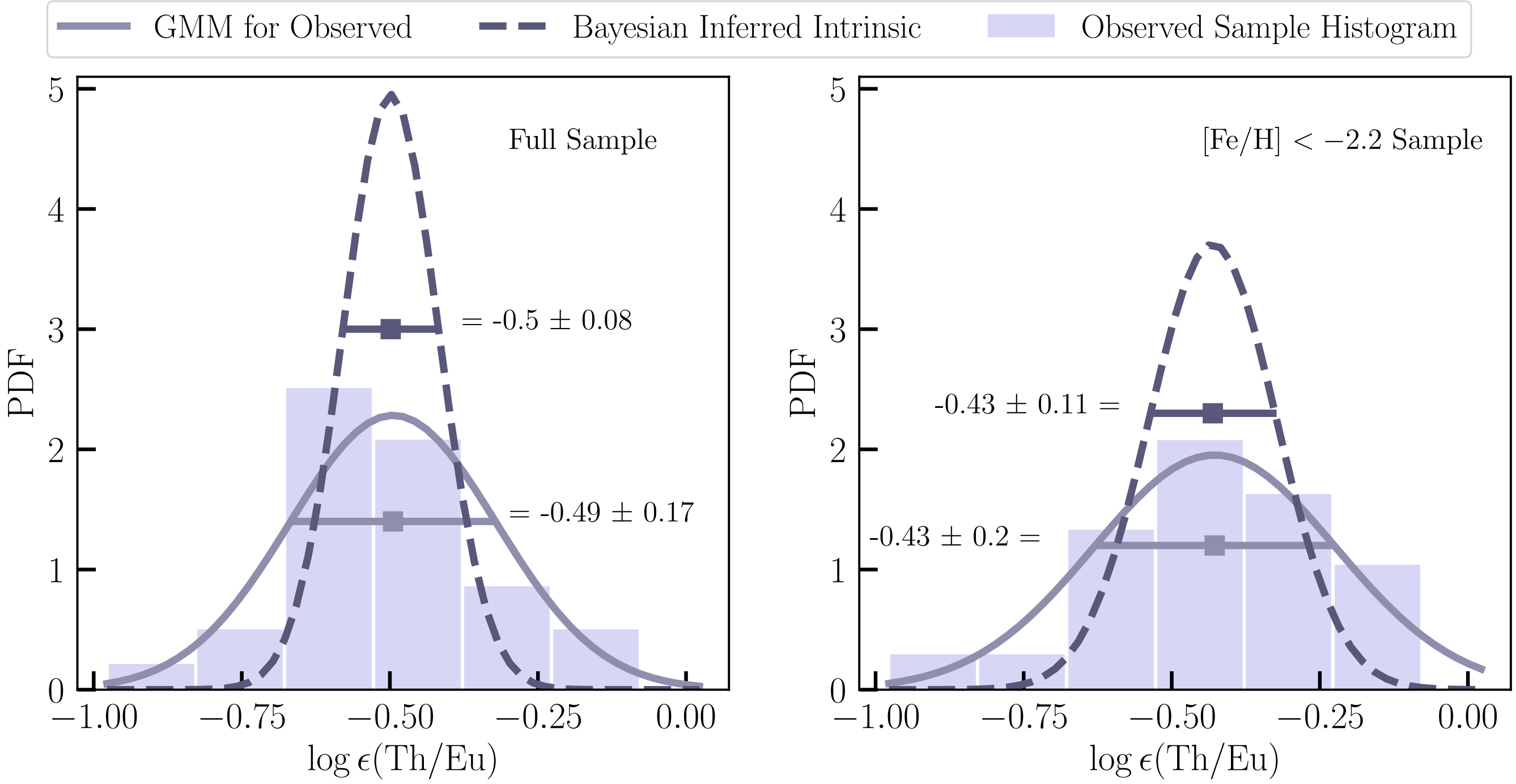

**Figure 8.** Left panel: the normalized distribution of log $\epsilon$(Th/Eu) for the full sample is shown with the binned histogram. The pdf of the observed log $\epsilon$(Th/Eu) values, estimated with a one-component GMM model, is shown using a solid purple line, with mean and $\pm 1\sigma$ indicated. The pdf of the intrinsic log $\epsilon$(Th/Eu) values is shown using a dashed purple line, with mean and intrinsic standard deviation, $\pm 1\sigma_{\rm int}$, indicated. Right panel: same as the left panel, but for [Fe/H] $< -2.2$ enrichment-regime stars.

Thus, using the results from the enrichment regime, we find that a standard deviation of $\pm$ 0.11 dex (or a factor of $\pm$1.3 or $\pm$30%) in the log $\epsilon$(Th/Eu) ratio is intrinsic, with a possible astrophysical origin. Conservatively, *no* intrinsic variation can not be ruled out with a $3\sigma$ confidence (see the marginalized distribution in Figure 9(b)); however, the enrichment-regime sample is small. On the other hand, the analysis of the full sample suggests a $3\sigma$ significance for an intrinsic variation of $\geqslant$0.02 dex. In general, the intrinsic variation of the log $\epsilon$(Th/Eu) ratio is not substantial, and we discuss it further in Section 4.2, with possible implications discussed in Section 4.4.

Finally, we revisit the numbers of actinide-boost and actinide-deficient stars with the new sample of Th abundances. With our sample and the literature sample combined, 15% stars are actinide-boost stars, 14% are actinide-deficient stars, and 71% are actinide-normal stars. The share of actinide-boost and actinide-deficient stars is quite symmetric, representing $> \pm 1\sigma$ tails of the log $\epsilon$(Th/Eu) distribution, as discussed above. On the other hand, considering only the enrichment-regime stars, 30% are actinide-boost stars, 11% are actinide-deficient stars, and 59% are actinide-normal stars. This high share of actinide-boost stars is in keeping with literature estimates. However, we consider two effects here: (1) the detection thresholds in the enrichment regime have likely shifted the log $\epsilon$(Th/Eu) distribution to higher values, and (2) many of the actinide-boost stars are from the literature, with possible biases toward higher values. In fact, when considering only our sample in the enrichment regime, we find that 13% are actinide-boost stars, 20% are actinide-deficient stars, and 67% are actinide-normal stars, indicating a more symmetric distribution.

## 4. Discussion

### 4.1. Thorium Evolution and Average Abundance

Figures 5 and 7 show the largest high-quality homogeneous measurements of Th abundances in MP stars, the first since I. U. Roederer et al. (2009) and J. Ren et al. (2012). Combined with the literature sample, the overall picture is that Th likely mimics the classic Eu "trumpet" plot (e.g., E. M. Holmbeck et al. 2020; C. Kobayashi et al. 2020, and shown in Figure 4): high [Th/Fe] scatter at low [Fe/H] $< -2.2$ that decreases to unresolved [Th/Fe] scatter at higher [Fe/H]. This is the first time the decrease in Th scatter has been measured.

In the context of Eu chemical evolution, the trumpet shape has been explained in two ways. The first is that the most MP stars are stochastically sampling variable Eu yields, and the change in the scatter can be explained by averaging out to a mean Eu yield as the number of *r*-process events increases, i.e., a transition from stochastic chemical enrichment to continuous chemical evolution (K. A. Venn et al. 2004; K. Brauer et al. 2021; A. Frebel & A. P. Ji 2023; X. Ou et al. 2024). The second is that the stellar halo is composed of a mix of dwarf galaxies that have different chemical-evolution trends and relatively low intrinsic scatter but superimpose to form the trumpet shape (e.g., T. Tsujimoto & T. Shigeyama 2014; Y. Ishimaru et al. 2015; T. Ojima et al. 2018; L. Cavallo et al. 2023; Y. Hirai et al. 2025). It is currently unclear which mechanism dominates, and both may be at play (i.e., the early chemical evolution of dwarf galaxies has larger scatter than at later times). The same two mechanisms can likely be used to understand Th chemical evolution, because Th closely tracks Eu, with log $\epsilon$(Th/Eu) $\sim -0.50$ or [Th/Eu]$\sim$0.00 (Figure 7).

Although one may be tempted to interpret the average [Th/Fe] ratio in Figure 5, the values themselves are almost certainly influenced by selection. We find that in the enrichment regime $\langle[\mathrm{Th/Fe}]\rangle_{[\mathrm{Fe/H}]<-2.2} = 1.1$, while in the evolution regime $\langle[\mathrm{Th/Fe}]\rangle_{-2.2\leqslant[\mathrm{Fe/H}]\lesssim-1.4} = 0.7$. However, it is clear that Th nondetections cause [Th/Fe] to be biased high in the enrichment regime (see the Appendix A and Figure 5). In the evolution regime, our cut to RPE stars likely also biases the [Th/Fe] abundances to higher values (Section 3.1 and Figure 4); for instance, adding in the 18 *s*-process-enhanced stars with measured Th abundances from D. Yong et al. (2008a, 2008b) and I. U. Roederer et al. (2009) lowers the mean value of [Th/Fe] to 0.53. A lower mean value of [Th/Fe] would naturally connect to studies of Th chemical evolution at higher metallicities of $[\mathrm{Fe/H}] \geqslant -1.0$, which have found that [Th/Fe] decreases from an MP plateau value around 0.6 to solar ratios by $[\mathrm{Fe/H}] = 0$ (R. B. Botelho et al. 2019; T. Mishenina et al. 2022; A. Azhari et al. 2025).

Thus, we overall find that Th chemical evolution is almost identical to Eu chemical evolution. However, we note that the $\log\epsilon(\mathrm{Th/Eu})$ ratio is still poorly constrained in stars with low *r*-process enrichment and that this correlation could break, as discussed in Section 4.2.

### 4.2. Coproduction of the Actinides and the Lanthanides

The observed trends between Th and lanthanides, Eu and Dy (Section 3.3), strongly suggest the coproduction of actinides and lanthanides in main *r*-process events. For simplicity, we will use Eu as a characteristic tracer of lanthanides in the following discussion.

We find that the actinides and lanthanides are consistently coproduced on average across (a) varying *r*-process enrichment levels or strengths (e.g., I. U. Roederer et al. 2023) of $0.0 \lesssim [\mathrm{Eu/Fe}] \lesssim 2.5$ (Figure 6) and (b) metallicities of $-3.0 \lesssim [\mathrm{Fe/H}] \lesssim -1.4$ (Figure 7). Moreover, the $\log\epsilon(\mathrm{Th/Eu})$ values are distributed normally with an intrinsic standard deviation of $\pm0.11$ dex. Assuming that the *r*-process elemental yields of each star with $[\mathrm{Fe/H}] < -2.2$ originate primarily from a single *r*-process event, we expect 68% of main *r*-process events to have $\log\epsilon(\mathrm{Th/Eu})$ yields varying by factors $\leqslant 2.6$ (i.e., $\leqslant \pm1.3$ or $\leqslant \pm 30\%$) at early cosmic times; this variation is not substantial. These lines of evidence suggest a very robust and nearly universal actinide-to-lanthanide ratio in the main *r*-process events of $\log\epsilon(\mathrm{Th/Eu}) \sim -0.50$ or $[\mathrm{Th/Eu}] \sim 0.0$. At the same time, exceptions exist, with 5% of main *r*-process events expected to have actinide-to-lanthanide yields varying by factors $>3.3$, approaching factors of 10. The origin of these variations or lack thereof is a significant question, and we discuss possibilities and implications for *r*-process sites in Section 4.4.

Interestingly, the continuation of a consistent average [Th/Eu] ratio at higher metallicities is contested. For instance, studies by E. F. del Peloso et al. (2005) and A. Azhari et al. (2025) have indicated a constant flat trend for stars in the range $-1.0 < [\mathrm{Fe/H}] < +0.4$. However, the values of [Th/Eu] in A. Azhari et al. (2025) are systematically offset by $\sim+0.24$ dex from $[\mathrm{Th/Eu}] = 0.0$. On the other hand, T. Mishenina et al. (2022) report a decreasing trend in [Th/Eu] values with increasing metallicity. These differences could be due to the use of different absorption lines, or different effects of radioactivity and non−local thermodynamic equilibrium (NLTE) between the samples. Homogenizing future studies will play an important role in uncovering the true trend of [Th/Eu] at higher metallicities. Finally, although [Fe/H] is often used as a proxy for time, studies with accurate stellar ages of MP stars will be needed to determine whether there is in reality no dependence of actinide-to-lanthanide yields on cosmic time (e.g., A. Azhari et al. 2025).

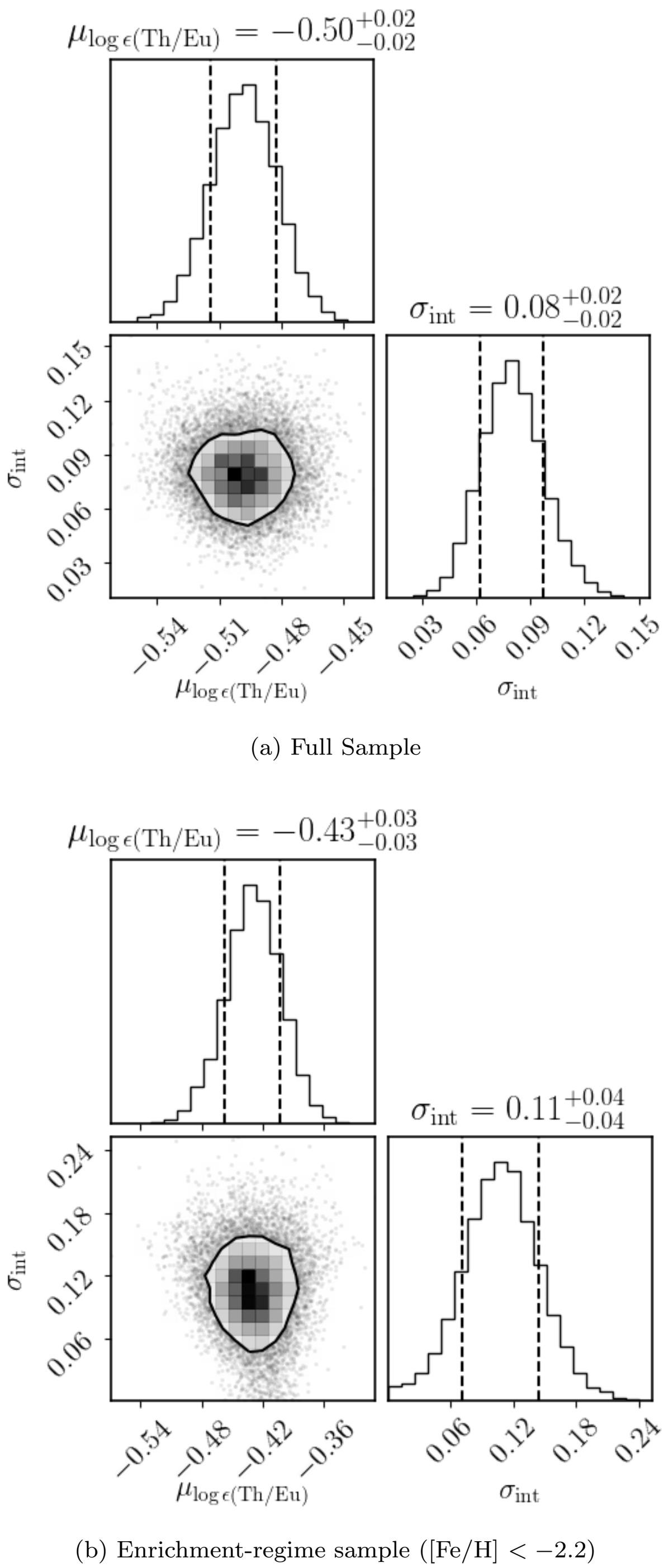

**Figure 9.** Posterior probability distributions of the mean ($\mu$) and intrinsic standard deviation ($\sigma_{\mathrm{int}}$) of the $\log\epsilon(\mathrm{Th/Eu})$ distribution with Bayesian analysis for the full sample (panel (a)) and the enrichment-regime sample (panel (b)). The contours mark the $1\sigma$ confidence regions in the 2D distributions, and the dashed lines represent the corresponding $\pm1\sigma$ confidence regions for the 1D marginalized distributions.

Observationally, the best-fit GMM model suggests a standard deviation of $\pm0.17$ dex in $\log\epsilon(\mathrm{Th/Eu})$ values for MP stars. Moreover, we have not found any significant evidence (such as multiple Gaussian components) in the current data to suggest that the actinide-boost and actinide-deficient stars are separate classes of stars with distinct origins, instead of $\geqslant\pm1\sigma$ tails of a continuous distribution of $\log\epsilon$ (Th/Eu) values. This was also suggested in E. M. Holmbeck et al. (2019a), who showed that a ∼0.8 dex range of $\log\epsilon$(Th/Eu) values can be reproduced by a continuous variation in the ejecta properties of *r*-process events (see also S. Wanajo et al. 2024). In that respect, future observations should expect 32% of MP RPE stars to be either actinide boost or actinide deficient, which will continue to be valuable probes of the dispersion in the $\log\epsilon$(Th/Eu) values. Higher-precision observations will further help to contribute to this question of whether actinide-boost and actinide-deficient stars are part of distinct distributions or origins.

However, the more interesting regime is now the 5% of MP RPE stars for which we expect the observed $\log\epsilon$(Th/Eu) ratio to be $>-0.15$ or $<-0.83$. Although the promise of future large spectroscopic surveys (e.g., 4MOST, R. S. de Jong et al. 2019; RPA, A. Bandyopadhyay et al. 2024b; WEAVE, G. Dalton et al. 2012) indicates that 5% of thousands of RPE stars is a substantial number, high-resolution data for these stars will not be available to determine Th reliably and precisely, requiring dedicated follow-up, which remains observationally costly. Unfortunately, there is currently no telltale sign for a high or low $\log\epsilon$(Th/Eu) value, but [Fe/H] $<-2.2$ and *r*-II stars tend to show higher dispersion than more metal-rich or *r*-I stars.

We also tested whether the observed variations $\log\epsilon$(Th/Eu) are correlated with the birth environment of the stars. We find that an equal portion of actinide-boost and actinide-deficient stars are in prograde orbits as retrograde orbits (∼40%). Other diagnostic plots such as $L_\perp$ versus $L_Z$ and $E_{\rm tot}$ versus $L_Z$, which have been used to tentatively classify stars as accreted versus in situ (e.g., P. Di Matteo et al. 2020; A. M. Matas Pinto et al. 2021; A. Bandyopadhyay et al. 2022, 2024a; V. Belokurov & A. Kravtsov 2024; S. Monty et al. 2024; M. Racca et al. 2025), also do not indicate a substantial difference in the birth environments of actinide-boost and actinide-deficient stars. However, a larger sample and a more detailed analysis are needed to confirm this, given the complicated overlap of several MW substructures in these phase-space plots (e.g., I. Jean-Baptiste et al. 2017; H. H. Koppelman et al. 2019; P. Di Matteo et al. 2020; R. P. Naidu et al. 2020).

Note that we specify consistent actinide-to-lanthanide coproduction in main *r*-process events (neutron-to-seed ratios $\geqslant$100 and neutron density $N_n \geqslant 10^{22}$ cm$^{-3}$). There has long been consideration of a second type of *r*-process, known as the limited *r*-process, weak *r*-process, or light-element primary process, often considered responsible for the chemical signatures of limited-*r* stars (e.g., C. Travaglio et al. 2004; T. C. Beers & N. Christlieb 2005; F. Montes et al. 2007; C. J. Hansen et al. 2012; A. Frebel 2018; J. J. Cowan et al. 2021, and references therein). However, this process is theorized to have rapid neutron captures, with low neutron-to-seed ratios ($\leqslant$100), so that elements heavier than the first *r*-process peak are not produced or are produced in limited amounts (e.g., S. Wanajo et al. 2001; N. Nishimura et al. 2017; A. Frebel 2018; S. Fujibayashi et al. 2023). Indeed, Th has not been detected in any limited-*r* star to date, while lanthanides such as Eu are commonly detected, although in low quantities (e.g., S. Honda et al. 2006, 2007; I. U. Roederer et al. 2010; E. M. Holmbeck et al. 2020; T. Xylakis-Dornbusch et al. 2024; H. Okada et al. 2026). On the other hand, A. Choplin et al. (2022, 2025) have shown that the intermediate neutron-capture process (*i*-process; J. J. Cowan & W. K. Rose 1977) can create actinides, reaching $N_n \geqslant 10^{15}$ cm$^{-3}$ in low-metallicity AGB stars. However, Th has not been detected in any star with a strong *i*-process signature (e.g., M. Hampel et al. 2016; I. U. Roederer et al. 2016, and references therein). Thus, the question of the exact nucleosynthesis conditions and the associated class of stars for which the actinide-to-lanthanide coproduction breaks or diverges is now of great interest. Additionally, note that there are only nine non-RPE and two *r*-III stars included here, so the evidence for coproduction is also not very strong for the extremes of *r*-process strengths in main *r*-process events.

### *4.3. Actinide-to-lanthanide Production Ratios*

When inferring the yields from Th abundances, it is important to consider that $^{232}$Th, the primary Th isotope observed in stellar spectra, is radioactive, with a half-life of 14.05 Gyr. Therefore, the initial ratio or the PR of $\log\epsilon$(Th/Eu) will be higher than what is observed at present in the stars. The distribution of PRs is of great interest, as it can provide strong constraints on the initial conditions of *r*-process-enrichment events, as well as possibly the nuclear properties of heavy isotopes (e.g., H. Schatz et al. 2002; S. Wanajo et al. 2002; S. Wanajo 2007; K. Farouqi et al. 2010; M. Eichler et al. 2019; E. M. Holmbeck et al. 2019a; K. A. Lund et al. 2023). Here we estimate actinide-to-lanthanide PRs ($\log\epsilon$(Th/Eu)$_{\rm PR}$), based on the observed $\log\epsilon$(Th/Eu) ratios and assumed stellar ages.

We assume three scenarios for the full sample: (1) all stars are 13 Gyr, (2) all stars are 9 Gyr, and (3) stars with [Fe/H] $<-2.2$ are 13 Gyr and stars with [Fe/H] $\geqslant-2.2$ are 9 Gyr. In a separate analysis, we consider only the enrichment-regime stars—all with stellar ages of 13 Gyr. These values are motivated by detailed studies of MP stars with asteroseismology (M. Valentini et al. 2019; D. Huber et al. 2024; J. R. Larsen et al. 2025) and isochrone fitting of subgiants (A. Bonaca et al. 2020; M. Xiang & H.-W. Rix 2022) and generally ensure a wide range of stellar ages. We then use Equation (7), which is a rearranged version of the radioactive-dating equation (R. Cayrel et al. 2001), to obtain the PR corresponding to each observed ratio.

$$\log\epsilon\,(\mathrm{Th/Eu})_{\rm PR} = \log\epsilon\,(\mathrm{Th/Eu})_{\rm observed} + \frac{t_{\star \rm age}\,\mathrm{Gyr}}{46.7}. \quad (7)$$

We assume that each star carries the signature of a single unique *r*-process event and the star formed shortly after the event. We plot the resulting PR distributions in Figure 10. The dashed black vertical line indicates the widely used theoretical $\log\epsilon$(Th/Eu)$_{\rm PR} = -0.33$ from the waiting-point calculations of H. Schatz et al. (2002). For the full sample, the distributions for all three cases are very similar, with only slight differences in shape. We employ a Bayesian analysis similar to that in Section 3.4, to determine the mean and intrinsic standard deviation, considering only observational uncertainties of the $\log\epsilon$(Th/Eu) ratios. In practice, two Gaussian components indicate a better fit to the Case 3 distribution; however, this is simply due to a bimodal distribution assumed for the stellar ages. We find that the mean $\log\epsilon$(Th/Eu)$_{\rm PR}$ and the intrinsic standard deviation of Case 3 are $-0.27 \pm 0.11$, and the means of Case 1 and Case 2 are similar, within 0.08 dex, to this mean. The means of all three distributions are close to the H. Schatz et al. (2002) PR. Therefore, compared to the observed log $\epsilon$(Th/Eu) ratios (Section 3.4), the PRs increase on average by only ~0.23 dex. This is due to the long half-life of the $^{232}$Th isotope, which makes the PR estimates relatively robust to the assumed stellar ages (a 4 Gyr difference in stellar age is equal to only 0.09 dex change in log $\epsilon$(Th/Eu)$_{\rm PR}$).

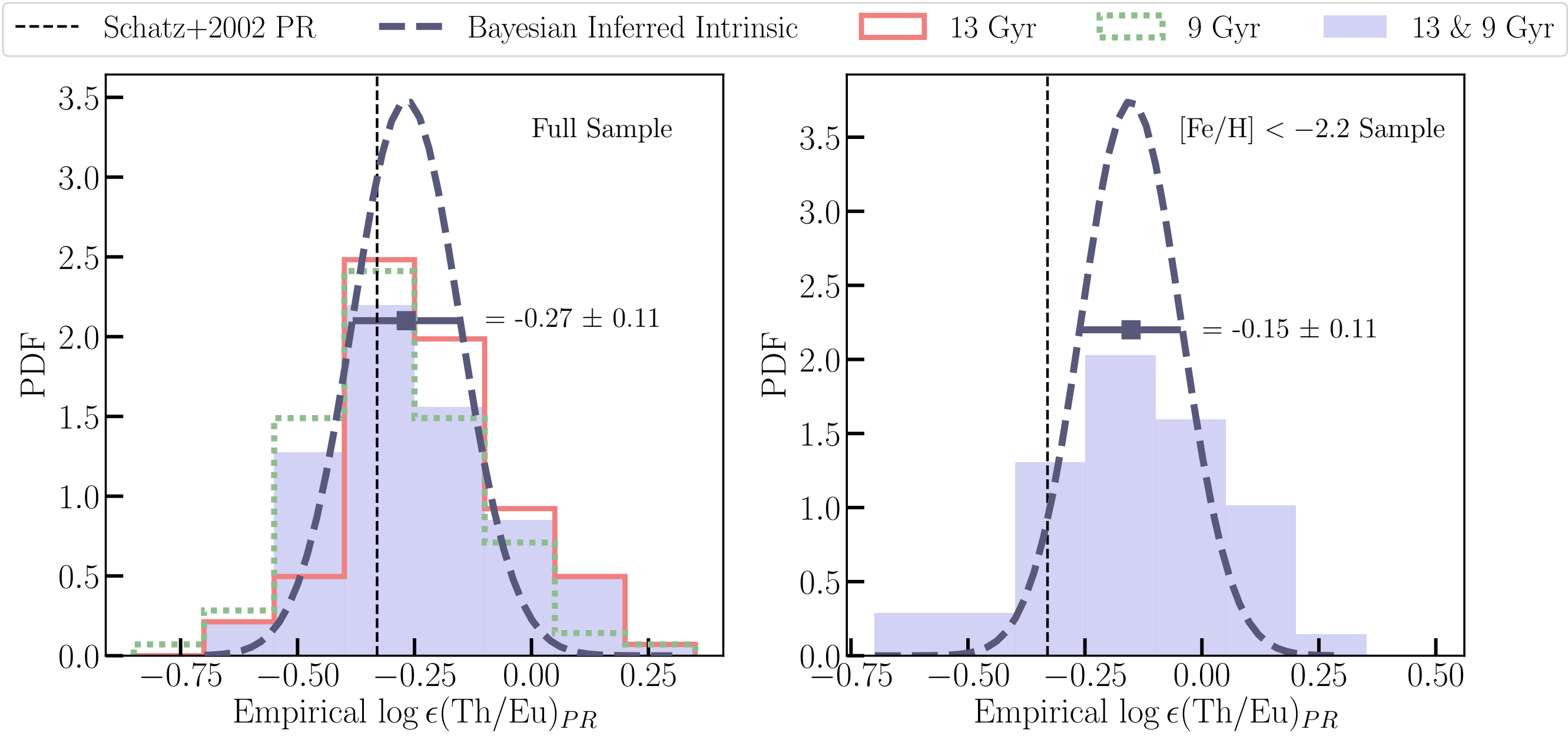

**Figure 10.** Left panel: distributions of the log $\epsilon$(Th/Eu) PR, estimated empirically for three cases: (1) assuming that all stars are 13 Gyr, shown in red; (2) assuming that all stars are 9 Gyr, shown in green; and (3) assuming that [Fe/H] $< -2.2$ stars are 13 Gyr and [Fe/H] $\geqslant -2.2$ stars are 9 Gyr, shown in purple. The pdf of the Case 3 distribution, characterized by a Bayesian estimate of the intrinsic standard deviation, is shown with the dashed purple line. Right panel: Same as the left panel, but only for the enrichment-regime stars ([Fe/H]$< -2.2$) and assumed stellar ages of 13 Gyr for all stars. Also shown in both panels is the PR from waiting-point calculations of H. Schatz et al. (2002) with a dashed black line.

The mean and intrinsic standard deviation in the enrichment regime are $-0.15 \pm 0.10$, slightly higher than those observed for the full sample and the H. Schatz et al. (2002) PR. This is likely due to sample bias from the detection thresholds. Therefore, we recommend that the mean and intrinsic standard deviation of Case 3 of the full sample, i.e., $-0.27 \pm 0.11$, are currently the most reliable empirical estimate of the log $\epsilon$(Th/Eu)$_{\rm PR}$ distribution of *r*-process-enrichment events. On the other hand, the extremes of the distribution, $-0.8 \lesssim$ log $\epsilon$(Th/Eu)$_{\rm PR} \lesssim +0.3$, are fairly robust between the samples and the distributions.

### 4.4. Implications for Astrophysical Conditions and r-process Sites

The presence of VMP and EMP RPE stars in the MW halo has required a prompt *r*-process site to operate in the early Universe with characteristic timescales $\lesssim$100 Myr after star formation (e.g., D. Argast et al. 2004; Y. Ishimaru et al. 2015; P. Beniamini et al. 2016; M. Safarzadeh et al. 2019b; P. Simonetti et al. 2019; C. Kobayashi et al. 2023). Moreover, the detection of Th in these stars requires that the prompt *r*-process site synthesize significant actinides. Our results add one more constraint: the log $\epsilon$(Th/Eu) yield variations must be $\leqslant \pm 0.11$ dex or $\leqslant \pm 30\%$ in 68% of the prompt *r*-process events. This constraint is in significant tension with the current models of candidate prompt *r*-process sites.

The electron fraction ($Y_e$) of *r*-process ejecta is a major factor in determining the *r*-process yields, with low $Y_e$ ($\lesssim$0.2) required to produce actinides (e.g., E. M. Holmbeck et al. 2019b). Interestingly, in very low $Y_e$ ($\lesssim$0.05) ejecta, the final log $\epsilon$(Th/Eu) ratio becomes constant, possibly due to the *r*-process flow near the nuclear drip line and multiple fission cycles through actinides, transuranic nuclei, and their daughter nuclei in the lanthanide region (e.g., J. Beun et al. 2008; O. Korobkin et al. 2012; A. Bauswein et al. 2013; S. Goriely 2015; M. Eichler et al. 2016; E. M. Holmbeck et al. 2019b; N. Vassh et al. 2020; S. Wanajo et al. 2024). However, simulations predict that *r*-process sites encompass a wide distribution of $Y_e$ ($0 \lesssim Y_e \lesssim 0.5$), with significant mass fractions at higher $Y_e$ that could synthesize Eu but not Th (e.g., N. Nishimura et al. 2017; P. Mösta et al. 2018; E. M. Holmbeck et al. 2019a). Furthermore, the $Y_e$ distribution is shaped by the astrophysical parameters of the system, which are realistically expected to vary between events. Therefore, a mass-averaged ratio of log $\epsilon$(Th/Eu) that is robust between *r*-process events does not follow in any obvious manner.

Among the leading candidates for a prompt *r*-process site are magnetorotationally driven jet supernovae (MRSNe) and collapsar disk winds,[26] which face similar challenges in producing robust log $\epsilon$(Th/Eu) ratios. First, while select models of these sites have successfully produced actinides (e.g., C. Winteler et al. 2012; N. Nishimura et al. 2015, 2017; D. M. Siegel et al. 2019; M. Reichert et al. 2021), more sophisticated models (e.g., in 3D, with detailed neutrino transport) generally do not produce sufficient low-$Y_e$ ejecta to even synthesize actinides (e.g., G. Halevi & P. Mösta 2018;

[26] See N. Soker (2026) for a review of other proposed *r*-process sites.

P. Mösta et al. 2018; J. M. Miller et al. 2020; O. Just et al. 2022; S. Fujibayashi et al. 2023; M. Reichert et al. 2023, 2024; S. Zha et al. 2024; D. Issa et al. 2025; M. Shibata et al. 2025). Second, even in cases where actinides are produced, the $\log\epsilon(\mathrm{Th/Eu})$ ratio depends on various parameters such as the magnetic field strength of the precollapse progenitors, explosion mechanism, and type of remnant, which shape the $Y_e$ distribution (e.g., N. Nishimura et al. 2015; M. Reichert et al. 2023; M. Shibata et al. 2025).

On the other hand, the dynamical ejecta from binary neutron star mergers (BNSMs) and neutron star–black hole mergers (NSBHMs) have significant mass fractions at low $Y_e \lesssim 0.20$ to synthesize actinides and possibly a robust $\log\epsilon(\mathrm{Th/Eu})$ ratio (e.g., O. Korobkin et al. 2012; A. Bauswein et al. 2013; J. d. J. Mendoza-Temis et al. 2015; E. M. Holmbeck et al. 2019b; S. Wanajo et al. 2024). However, their standard formation channels are delayed, and they have been repeatedly shown to not fit observations of *r*-process elements in MW stars at low metallicities (e.g., B. Wehmeyer et al. 2015; K. Hotokezaka et al. 2018; B. Côté et al. 2019; C. J. Haynes & C. Kobayashi 2019; C. Kobayashi et al. 2023; H.-Y. Chen et al. 2025; M. Saleem et al. 2025). Additionally, there is also the question of how the dynamical ejecta is combined with the wind/accretion-disk ejecta, which host systematically higher and wider $Y_e$ distributions, allowing order-of-magnitude variations in the $\log\epsilon(\mathrm{Th/Eu})$ ratio with relatively small changes in the distribution (e.g., E. M. Holmbeck et al. 2019a; E. M. Holmbeck & J. J. Andrews 2024; K. A. Lund et al. 2024; S. Wanajo et al. 2024; although see T. M. Sprouse et al. 2024; Y. Qiu et al. 2025). This is especially relevant for BNSMs because the wind/accretion-disk ejecta is predicted to have ∼10 times more mass than the dynamical ejecta (e.g., C. J. Krüger & F. Foucart 2020; A. Henkel et al. 2023).

In general, it seems difficult to have both a prompt *r*-process site and a robust $\log\epsilon(\mathrm{Th/Eu})$ ratio. We discuss some possible resolutions below.

#### 4.4.1. Possible Resolutions

At the basic level, regardless of the *r*-process site, models need to achieve significant mass fractions of neutron-rich ejecta, with $Y_e \lesssim 0.2$ as well as $Y_e \lesssim 0.05$. Only under these conditions can the $\log\epsilon(\mathrm{Th/Eu})$ ratio be consistent, set by nuclear physics, and robust to changes in other astrophysical parameters. In other words, the dominant mass fraction of Eu (lanthanides) and Th (actinides) must be produced in the low-$Y_e$ ejecta, with high-$Y_e$ ejecta ($\gtrsim$0.2) allowed to contribute minimally so that the variations in the final mass-averaged $\log\epsilon(\mathrm{Th/Eu})$ ratio are $\lesssim \pm 30\%$ between events. Alternatively, a very consistent set of astrophysical parameters will be required across *r*-process events, producing consistent $Y_e$ distributions and $\log\epsilon(\mathrm{Th/Eu})$ ratios; given current simulations, this scenario seems unlikely in MRSNe.

To enable MRSNe to meet low-$Y_e$ conditions, perhaps extreme conditions such as high magnetic fields and rotation rates of the precollapse iron core are more realistic than currently predicted by 1D stellar evolution codes (P. Mösta et al. 2018; J. J. Cowan et al. 2021; M. Reichert et al. 2021; S. Zha et al. 2024, and references therein). Similarly, for collapsars, high magnetic fields and accretion rates required to produce neutron-rich ejecta (e.g., J. M. Miller et al. 2020; D. Issa et al. 2025; O. Gottlieb et al. 2025) might be realistic. Moreover, it is possible that there is missing physics (e.g., M.-R. Wu et al. 2017; Y. Qiu et al. 2025) and/or numerical uncertainties in the simulations that need to be addressed (e.g., length and resolution of the simulations; T. M. Sprouse et al. 2024; M. Shibata et al. 2025).

Alternatively, the simplified delay time distribution ($\propto t^{-1}$) predicted for BNSMs and NSBHMs by binary population synthesis models might be steeper in the early Universe, allowing these *r*-process sites to be more prompt (e.g., M. Safarzadeh et al. 2019a; C. Kobayashi et al. 2023; P. Beniamini & T. Piran 2024; D. Maoz & E. Nakar 2025). This is possible given the range of uncertainties in the current binary evolution models and the possible dependence of the delay time distributions on metallicity (J. J. Andrews & I. Mandel 2019; F. S. Broekgaarden et al. 2022; I. Mandel & F. S. Broekgaarden 2022; C. Kobayashi et al. 2023). There is also indication that advances in yield estimates and galactic chemical models to include hierarchical galaxy formation and inefficient star formation may help reconcile BNSMs and/or NSBHMs as the main *r*-process channel(s) and ease this tension (e.g., Y. Ishimaru et al. 2015; Y. Komiya & T. Shigeyama 2016; F. van de Voort et al. 2020; I. Dvorkin et al. 2021; S. Wanajo et al. 2021; Y. Hirai et al. 2025.

Finally, there are sample biases in our data. In particular, given the selection for RPE stars and the detection thresholds for Th, it is possible that the intrinsic dispersion in $\log\epsilon(\mathrm{Th/Eu})$ ratio is larger than determined in this work. A targeted sample with high-precision data will be needed to address this.

### 4.5. Systematics from Sample Bias, Radioactivity, and NLTE

We now briefly discuss how the above results may be impacted by sample biases, radioactivity of Th, and unaccounted-for NLTE effects. The literature sample and our sample combined suffer biases from two effects: (1) detection limits of Th, and (2) most stars are RPE stars. We expect both of these effects to have biased all mean values of, e.g., [Th/H], [Th/Fe], and $\log\epsilon(\mathrm{Th/Eu})$ to higher values and all standard deviation values to lower values (see also Section 4.1 for comparison with [Eu/Fe], which has been determined in a larger variety of stars). Systematic studies of non-RPE and limited-*r* studies will be needed to address this bias. Another effect is the nonuniform proportion of literature stars and our sample of stars as a function of metallicity. In particular, most of the stars with [Fe/H] $< -2.7$ are from the literature. We expect Th abundances in these stars to be more inhomogeneously determined than in our sample. It is possible that our fiducial uncertainty of $\pm 0.20$ assigned to the literature values does not accurately capture the dispersion in the enrichment regime, with the dispersion being higher or lower. More homogeneous studies of Th in stars with [Fe/H] $< -2.7$ will play an important role (e.g., M. Racca et al. 2025).

Our discussion on the PRs showed that correcting for radioactivity increases Th abundances by 0.2–0.3 dex, depending on the stellar age (8–13.8 Gyr). In particular, it is possible that the abundances of the VMP stars increase more than those of MP stars, exacerbating the "decreasing" trend observed for [Th/Fe] and $\log\epsilon(\mathrm{Th/Eu})$ as a function of metallicity. We tested this effect using the Case 3 scenario from Section 4.3—flat trends can still not be ruled out since the Th detection limits also get corrected for radioactivity to higher values. It is possible that the $\log \epsilon$(Th/Eu) distribution corrected for radioactivity will change the intrinsic variation slightly; however, precise stellar ages (e.g., with an uncertainty of $\lesssim$1 Gyr) will be needed to assess this effect.

Finally, there has been only one NLTE study for Th II, which suggested a positive correction between 0.07 and 0.2 dex for the $\lambda$4019 line in cool MP and VMP giants with [Th/Fe] = +0.4 (L. Mashonkina et al. 2012). A larger-scale study is necessary to understand how each star will be impacted, but we can expect a slight systematic increase of all Th abundances. However, a 3D NLTE study of Eu has also indicated a correction +0.2 dex (N. Storm et al. 2025), so that the values $\log \epsilon$(Th/Eu) might be relatively unaffected.

## 5. Conclusion

We have presented homogeneous Th (along with Fe, Eu, and Dy) abundances for 47 MP stars, of which 46 are RPE and 1 is non-RPE. This sample brings the current number of *r*-process-enriched stars with Th abundances to almost ∼100. Combined with the literature sample, we obtain the first chemical-evolution picture of Th at low metallicities, marked by decreasing dispersions of [Th/H], [Th/Fe], and $\log \epsilon$(Th/Eu) with increasing metallicities (Figures 5 and 7), although the mean trend of [Th/Fe] is especially impacted by the selection effects (Section 4.1 and Figure A1).

We find that, on average, the actinides and lanthanides are coproduced with $\log \epsilon$(Th/Eu) $\sim -0.5$ or [Th/Eu] ∼ 0.0 in main *r*-process events of varying strengths and across metallicities (Section 4.2). This is based on the remarkable correlation of [Th/Fe] with [Eu/Fe] and [Dy/Fe] across $0 \lesssim$ [Eu/Fe] $\lesssim 2.5$ (Figure 6) and the approximately constant trend of $\log \epsilon$(Th/Eu) across $-3.0 \lesssim$ [Fe/H] $\lesssim -1.4$ (Figure 7). Moreover, the $\log \epsilon$(Th/Eu) is distributed normally with an observed standard deviation of ±0.17 dex for the full sample and ±0.20 dex for the [Fe/H] < −2.2 sample (Figure 8; see Figure 10 for this distribution corrected for the radioactivity of Th). Based on our best-fit model, future observations should expect 32% of MP stars to have $\log \epsilon$(Th/Eu) > −0.33 or $\log \epsilon$(Th/Eu) < −0.66, i.e., to be actinide boost or actinide deficient, respectively. However, the absolute range of $\log \epsilon$(Th/Eu) is 1.02 dex, and therefore the more interesting regime is now the 5% of observations for which we expect $\log \epsilon$(Th/Eu) > −0.15 or $\log \epsilon$(Th/Eu) < −0.83 and that will help to probe the very extremes of actinide-to-lanthanide yield variations.

To more accurately estimate the variation in actinide-to-lanthanide yields of *r*-process events, we also determined the *intrinsic* standard deviation in the $\log \epsilon$(Th/Eu) ratio of ±0.11 dex (Figure 8). Based on the best-fit model, we infer that 5% of main *r*-process events have $\log \epsilon$(Th/Eu) yield ratios varying by factors >3.3, approaching factors of ∼10 (Section 4.2). However, 68% of main *r*-process events have $\log \epsilon$(Th/Eu) yields varying by factors $\leqslant$2.6 (or $\leqslant$±1.3 or $\leqslant$±30%). This variation is very small, and we discuss implications for astrophysical conditions and *r*-process sites in Section 4.4. We especially highlight that achieving both a robust $\log \epsilon$(Th/Eu) ratio and a prompt *r*-process site presents a challenge for current models, and a resolution will have to be found.

Data from future large-scale spectroscopic surveys (e.g., 4MOST, R. S. de Jong et al. 2019; WEAVE, G. Dalton et al. 2012; MINCE, G. Cescutti et al. 2022; CERES, L. Lombardo et al. 2025; RPA, A. Bandyopadhyay et al. 2024b) will continue shaping the picture of actinide production and evolution in the Universe. We particularly recommend targeted surveys at [Fe/H] < −2.7 of RPE, non-RPE, and limited-*r* stars to push the detection limits of Th detection, as well as expand the picture of actinide and lanthanide coproduction for a larger variety of *r*-process events.

## Acknowledgments

S.P.S. acknowledges helpful conversations with Atul Kedia and support of the Charles Vincent and Heidi Cole McLaughlin Fellowship from the University of Florida. S.P.S. and R.E. acknowledge support from NASA Astrophysics Theory Program grant 80NSSC24K0899. This work is performed in part under the auspices of the US Department of Energy (DOE) by Lawrence Livermore National Laboratory under contract DE-AC52-107NA27344 and has been approved under release No. LLNL-JRNL-2016030. A.P.J. acknowledges support from the National Science Foundation (NSF) under grants AST-2307599 and AST-2510795 and the Alfred P. Sloan Foundation. This material is based on work supported in part by the US Department of Energy, Office of Science, Office of Nuclear Physics, under award No. DE-SC0023128 (CeNAM). The work of V.M.P. is supported by NOIRLab, which is managed by the Association of Universities for Research in Astronomy (AURA) under a cooperative agreement with the US National Science Foundation. I.U.R. acknowledges support from the US NSF (grant AST 2205847). S.A.U. acknowledges the support of the American Association of University Women through their American Dissertation Fellowship. T.T.H. acknowledges support from the Swedish Research Council (VR 2021-05556).

*Facilities:* Magellan: Clay (LDSS2 imaging spectrograph), Du Pont (Echelle Spectrophraph), Smith.

*Software:* `astropy` (Astropy Collaboration et al. 2013), `sklearn` (F. Pedregosa et al. 2011), `matplotlib` (J. D. Hunter 2007), `h5py` (A. Collette 2013), `numpy` (C. R. Harris et al. 2020), `corner` (D. Foreman-Mackey 2016), `scipy` (P. Virtanen et al. 2020).

## Appendix A
## Inverse Modeling the Evolution of [Th/Fe]

We used simple inverse modeling to determine cases that can reproduce the slightly decreasing trend observed for [Th/Fe] as a function of [Fe/H] (Section 3.2). We considered three cases: (A) the true [Th/Fe] trend is decreasing as a function of [Fe/H]; (B) the true [Th/Fe] trend is flat as a function of [Fe/H], inspired by the [Eu/Fe] trend observed in the literature (Figure 4); (C) the true [Th/Fe] trend is increasing as a function of [Fe/H], inspired by the [Sr/Fe] and [Ba/Fe] trends observed in the literature (e.g., A. McWilliam 1998; P. S. Barklem et al. 2005; I. U. Roederer 2013; C. Kobayashi et al. 2020; L. Lombardo et al. 2025, and references therein).

For Case A, we binned the observed [Th/Fe] values from [Fe/H] −3.05 to −1.5 in 0.1 dex intervals. For each bin, we determined the mean and standard deviation of [Th/Fe] and the mean [Fe/H]. We used an arbitrary standard deviation of 0.07 dex for [Fe/H] in each bin. For each bin, we generated 50 mock [Th/Fe] values and 50 mock [Fe/H] values from normal distributions characterized by these mean and standard deviations. The mock data are shown with gray data points in the left panel of Figure A1. For each bin, we determined the mock data that are above the detection thresholds (Section 2.5) and therefore will be observed; these are shown in dark gray. We then determined means and standard deviations of all [Th/Fe] and [Fe/H] mock values in sliding bins of 120 stars with 60 stars overlapping between the bins to obtain the "true" trend of the mock sample. Similarly, we determined the means and standard deviations for only the mock values that are above the detection threshold to obtain the "observed" trend of [Th/Fe] versus [Fe/H], shown in red. We find that the "observed" trend faithfully tracks the decreasing "true" trend, since the sample is not severely impacted by the detection thresholds.

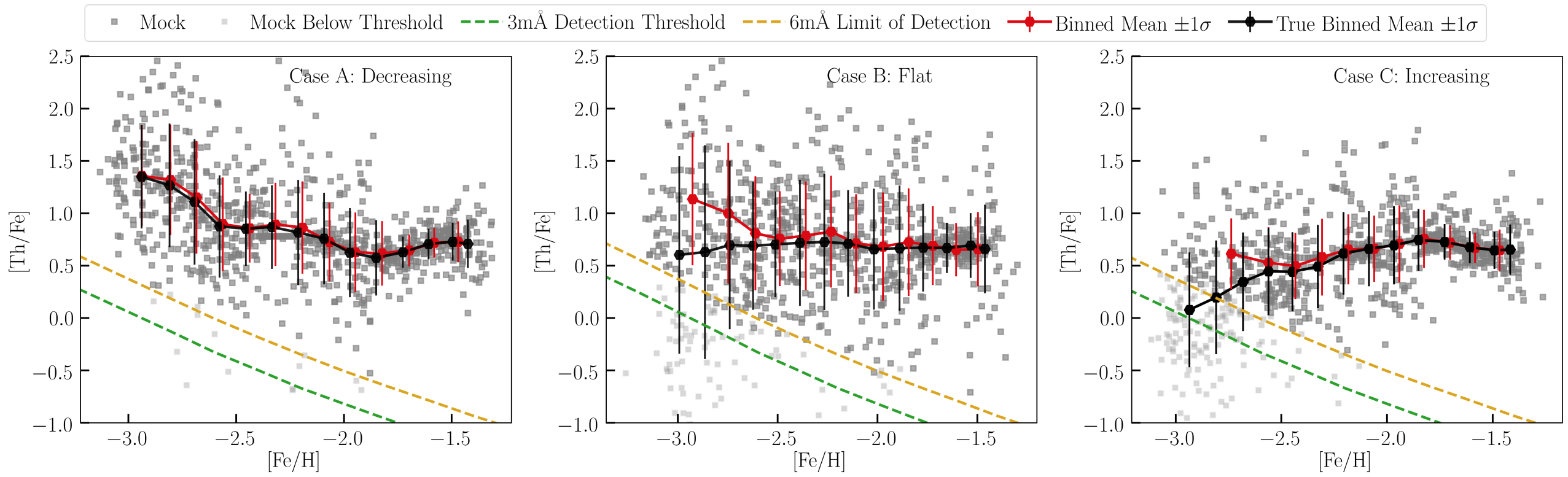

**Figure A1.** Gray (light and dark) markers are mock data points generated for three different inverse models for the "true" [Th/Fe] trend as a function of [Fe/H]. The solid black line and error bars trace the mean and standard deviation of the "true" trend, which include all data points. Dashed yellow and dashed green lines indicate the detection thresholds estimated in this work. The solid red line and error bars then trace the mean and standard deviation of the "observed" trend, which include only the data points that lie above the detection threshold (shown in dark gray).

We perform a similar exercise for Case C as a mirror image of Case B. In this exercise, the mean values of [Th/Fe] determined in the 0.1 dex bins of the observed data are subtracted from +0.69, which we consider to be the mean value of [Th/Fe] at all metallicities (Sections 3.2 and 4.1). This trick helps generate an increase in the "true" trend of the mock data (shown in black in the right panel of Figure A1), specifically at lower metallicities. There is also a slight decreasing trend at the higher metallicities because of the mirror imaging, but we ignore this in our interpretation. Applying the detection thresholds as above to classify the observed mock data, we show the resulting "observed" trend in red. We find that the "observed" trend traces the "true" trend only at higher metallicities, where the mock data are not impacted by the detection threshold. In fact, the "observed" trend at higher metallicities appears relatively flat.

For Case B, we assume that [Th/Fe] = +0.69 across all bins so that the "true" trend of the mock data is flat as a function of [Fe/H]. To create a classic trumpet shape for the mock data, we used the observed [Th/Fe] values, but with modifications. Specifically, we removed all data points below [Th/Fe] = +0.69, since these are the most affected by the detection thresholds and therefore could not provide a reliable measure of the dispersion for the mock data. Instead, we generate new values using a mirror image of the values above [Th/Fe] = +0.69, i.e., for each observed data point offset above [Th/Fe] = +0.69, we created a value similarly offset, but below [Th/Fe] = +0.69. The standard deviation for each bin is then computed from this modified dataset. We then generated the mock data by drawing 50 data points from normal distributions characterized by the above-obtained standard deviations in each bin and a mean [Th/Fe] of +0.69 for all bins. The corresponding "true" trend of the mock data and "observed" trend of the mock data are shown in the middle panel (B) of Figure A1 with black and red, respectively. We find that the "observed" trend does in fact show a slight decrease at the lowest metallicities, before it flattens out and begins tracing the "true" trend at higher metallicities.

In conclusion, we find that both the "true" decreasing (Case A) and "true" flat (Case B) trends in [Th/Fe] could reproduce the observed decreasing trend. On the other hand, the "true" increasing (Case C) trend results in a flatter trend than what was observed, although not significantly, and cannot yet be ruled out at present.

## Appendix B
## Data Tables

The literature stars, abundances, and sources are listed in Table B1. The stellar parameters and uncertainties derived for the stars in this sample are listed in Table B2. The transitions and atomic data used to model the spectral region of the $\lambda$4019 Å Th II line are listed in Table 4. The abundances and uncertainties of Fe, Eu, Dy, and Th derived for the stars in this sample are given in Tables B3, B4, B5, B6, and B7, respectively.

**Table B1**
Literature Abundances

| Name | [Fe/H] | log ϵ (Eu) | [Eu/Fe] | log ϵ (Dy) | [Dy/Fe] | log ϵ (Th) | [Th/Fe] | Source |
|---|---|---|---|---|---|---|---|---|
| 2MASS J00280692−2603042 | −2.85 | −0.66 | +1.67 | −0.08 | +1.67 | −1.16 | +1.67 | V. Hill et al. (2017) |
| 2MASS J00401252+2729247 | −2.72 | −1.10 | +1.10 | −0.48 | +1.14 | −1.52 | +1.18 | M. Racca et al. (2025) |
| 2MASS J00430527+1948592 | −1.90 | −0.83 | +0.55 | −0.39 | +0.41 | −1.24 | +0.64 | P. Saraf et al. (2023) |
| 2MASS J01021585−6143458 | −3.00 | −1.69 | +0.79 | −0.87 | +1.03 | −1.92 | +1.06 | I. U. Roederer et al. (2014) |
| 2MASS J01031817−2752499 | −2.69 | −1.87 | +0.30 | −1.10 | +0.49 | −2.22 | +0.45 | I. U. Roederer et al. (2014) |
| 2MASS J01293113−1600454 | −2.90 | −0.76 | +1.62 | −0.21 | +1.59 | −0.98 | +1.90 | V. Hill et al. (2002) |
| 2MASS J02172993−1903583 | −2.86 | −0.73 | +1.61 | −0.34 | +1.42 | −1.24 | +1.60 | M. Racca et al. (2025) |
| 2MASS J02462013−1518419 | −2.70 | −0.66 | +1.52 | 0.06 | +1.66 | −1.04 | +1.64 | M. Racca et al. (2025) |
| 2MASS J03010069+0616318 | −2.84 | −1.17 | +1.15 | −0.77 | +0.97 | −1.35 | +1.47 | D. K. Lai et al. (2008) |
| 2MASS J04090342−1553269 | −1.98 | −1.30 | +0.16 | −0.62 | +0.26 | −1.73 | +0.23 | I. U. Roederer et al. (2014) |
| 2MASS J06264076+0325298 | −1.59 | −0.70 | +0.37 | −0.13 | +0.36 | −1.08 | +0.49 | I. U. Roederer et al. (2014) |
| 2MASS J08045284+5740195 | −2.38 | −1.06 | +0.80 | −0.48 | +0.80 | −1.28 | +1.08 | Y. Lin et al. (2025) |
| 2MASS J09215727+5034047 | −2.05 | −0.73 | +0.80 | −0.29 | +0.66 | −1.34 | +0.69 | P. Saraf et al. (2023) |
| 2MASS J09544277+5246414 | −2.96 | −1.16 | +1.28 | ⋯ | ⋯ | −1.92 | +1.02 | S. P. Shah et al. (2023) |
| 2MASS J12213413−0328396 | −2.96 | −1.06 | +1.38 | −0.34 | +1.52 | −1.29 | +1.65 | W. Hayek et al. (2009) |
| 2MASS J12281686+1220411 | −2.20 | −1.48 | +0.20 | −0.99 | +0.11 | −1.99 | +0.19 | J. A. Johnson (2002) |
| 2MASS J12401407+0831380 | −2.52 | −1.91 | +0.09 | −1.44 | −0.02 | −2.50 | +0.00 | S. Honda et al. (2004) |
| 2MASS J12591993+0914356 | −2.45 | −1.69 | +0.24 | −1.05 | +0.30 | −1.96 | +0.47 | I. U. Roederer et al. (2010) |
| 2MASS J13164246+3622529 | −2.90 | −1.63 | +0.75 | −1.01 | +0.79 | −2.23 | +0.65 | J. Westin et al. (2000) |
| 2MASS J14235816+0801330 | −2.00 | −1.15 | +0.33 | −0.46 | +0.44 | −1.76 | +0.22 | J. A. Johnson & M. Bolte (2001) |
| 2MASS J14301385−2317388 | −1.83 | −0.65 | +0.66 | 0.01 | +0.74 | −1.38 | +0.43 | M. Racca et al. (2025) |
| 2MASS J14325334−4125494 | −2.76 | −0.70 | +1.54 | −0.18 | +1.48 | −1.05 | +1.69 | M. Racca et al. (2025) |
| 2MASS J15141890+0727028 | −2.42 | −1.02 | +0.88 | −0.43 | +0.89 | −1.12 | +1.28 | S. Honda et al. (2004) |
| 2MASS J15213995−3538094 | −2.80 | −0.04 | +2.24 | 0.45 | +2.15 | −0.60 | +2.18 | M. Cain et al. (2020) |
| 2MASS J15260106−0911388 | −2.95 | −0.62 | +1.81 | 0.02 | +1.87 | −1.20 | +1.73 | A. Frebel et al. (2007) |
| 2MASS J17281446+1730358 | −2.00 | −0.67 | +0.81 | −0.03 | +0.87 | −1.18 | +0.80 | J. J. Cowan et al. (2002) |
| 2MASS J18470646+7443316 | −1.63 | −0.64 | +0.47 | 0.01 | +0.54 | −1.02 | +0.59 | I. U. Roederer et al. (2014) |
| 2MASS J19161821−5544454 | −2.39 | −1.04 | +0.83 | −0.55 | +0.74 | −1.56 | +0.81 | M. Racca et al. (2025) |
| 2MASS J19451414−1729269 | −2.78 | −2.00 | +0.26 | −1.37 | +0.31 | −2.37 | +0.39 | I. U. Roederer et al. (2014) |
| 2MASS J20032253−1142028 | −3.45 | −1.21 | +1.72 | −0.56 | +1.79 | −1.31 | +2.12 | D. Yong et al. (2021) |
| 2MASS J20384318−0023327 | −2.91 | −0.75 | +1.64 | −0.33 | +1.48 | −1.24 | +1.65 | V. M. Placco et al. (2017) |
| 2MASS J20453454−1431151 | −2.87 | −2.28 | +0.07 | −1.59 | +0.18 | −2.73 | +0.12 | I. U. Roederer et al. (2014) |
| 2MASS J21091329−1310253 | −2.45 | −0.83 | +1.10 | −0.18 | +1.17 | −1.02 | +1.41 | M. Racca et al. (2025) |
| 2MASS J22170165−1639271 | −3.10 | −0.95 | +1.63 | −0.23 | +1.77 | −1.57 | +1.51 | C. Sneden et al. (2003) |
| 2MASS J22310218−3238365 | −2.51 | −1.03 | +0.96 | −0.22 | +1.19 | −1.43 | +1.06 | W. Hayek et al. (2009) |
| 2MASS J22545856−4209193 | −2.63 | −1.30 | +0.81 | −0.58 | +0.95 | −1.63 | +0.98 | L. Mashonkina et al. (2014) |
| 2MASS J23292881+3025578 | −2.19 | −0.86 | +0.81 | −0.32 | +0.77 | −1.46 | +0.71 | I. I. Ivans et al. (2006) |
| 2MASS J23303707−5626142 | −2.78 | −1.29 | +0.97 | −0.60 | +1.08 | −1.67 | +1.09 | L. Mashonkina et al. (2010) |
| 2MASS J23342669−2642140 | −3.40 | −2.24 | +0.64 | −1.62 | +0.68 | −2.45 | +0.93 | C. Siqueira Mello et al. (2014) |
| DES J033523–540407 | −3.00 | −0.77 | +1.71 | −0.08 | +1.82 | −1.63 | +1.35 | A. P. Ji & A. Frebel (2018) |
| Gaia DR3 6412626111278092544 | −1.91 | 0.43 | +1.82 | 1.03 | +1.84 | −0.23 | +1.66 | T. T. Hansen et al. (2021) |
| LAMOST J112456.61+453531.3 | −1.27 | 0.38 | +1.13 | 0.74 | +0.91 | −0.19 | +1.06 | Q. Xing et al. (2024) |
| M15 K341 | −2.20 | −0.88 | +0.80 | −0.25 | +0.85 | −1.47 | +0.71 | C. Sneden et al. (2000a) |
| M15 K462 | −2.20 | −0.61 | +1.07 | 0.12 | +1.22 | −1.26 | +0.92 | C. Sneden et al. (2000a) |
| M92 VII-18 | −2.18 | −1.47 | +0.19 | −1.15 | −0.07 | −2.07 | +0.09 | J. A. Johnson & M. Bolte (2001) |
| SPLUS J14244534−2542471 | −3.39 | −1.25 | +1.62 | −0.47 | +1.82 | −1.21 | +2.16 | V. M. Placco et al. (2023) |
| UMiCos82 | −1.42 | 0.34 | +1.24 | 1.13 | +1.45 | −0.25 | +1.15 | W. Aoki et al. (2007) |

**Note.** Table B1 is published in its entirety in the electronic edition of the *Astrophysical Journal*.

(This table is available in its entirety in machine-readable form in the online article.)

**Table B2**
Derived Stellar Parameters and Uncertainties

| Name | $T_{\rm eff}$ (K) | $\sigma_{T_{\rm eff}}$ | log *g* | $\sigma_{\log g}$ | [M/H] | $\sigma_{\rm [M/H]}$ | $\xi$ (km s$^{-1}$) | $\sigma_\xi$ |
|---|---|---|---|---|---|---|---|---|
| 2MASS J00101758−1735387 | 5426 | 93 | 3.07 | 0.07 | −2.29 | 0.20 | 1.58 | 0.20 |
| 2MASS J00193176+3141441 | 4654 | 67 | 1.24 | 0.09 | −1.88 | 0.20 | 2.15 | 0.20 |
| 2MASS J00524174−0902235 | 5309 | 52 | 2.10 | 0.08 | −1.41 | 0.20 | 1.99 | 0.20 |
| 2MASS J01320993+3300431 | 4956 | 66 | 1.75 | 0.08 | −1.74 | 0.20 | 2.05 | 0.20 |
| 2MASS J01425445−0904162 | 4525 | 49 | 1.17 | 0.10 | −2.00 | 0.20 | 1.96 | 0.20 |
| 2MASS J01530024−3417360 | 5289 | 100 | 3.06 | 0.07 | −1.65 | 0.20 | 1.41 | 0.20 |
| 2MASS J01542216+0341454 | 4309 | 47 | 0.61 | 0.12 | −2.16 | 0.20 | 2.14 | 0.20 |
| 2MASS J03422816−6500355 | 5049 | 57 | 2.21 | 0.08 | −2.15 | 0.20 | 1.80 | 0.20 |
| 2MASS J04141624−6009048 | 5001 | 58 | 2.16 | 0.09 | −2.56 | 0.20 | 1.72 | 0.20 |
| 2MASS J06320130−2026538 | 5197 | 82 | 2.53 | 0.08 | −1.67 | 0.20 | 1.62 | 0.20 |
| 2MASS J07103110−7121522 | 5041 | 62 | 2.52 | 0.08 | −1.54 | 0.20 | 1.44 | 0.20 |
| 2MASS J07114252−3432368 | 5012 | 54 | 2.09 | 0.08 | −2.06 | 0.20 | 1.80 | 0.20 |
| 2MASS J07352232−4425010 | 5324 | 49 | 2.89 | 0.08 | −1.74 | 0.20 | 1.51 | 0.20 |
| 2MASS J08474871−2204315 | 5004 | 57 | 2.50 | 0.08 | −1.69 | 0.20 | 1.51 | 0.20 |
| 2MASS J09185890−2311511 | 4721 | 71 | 1.53 | 0.09 | −1.98 | 0.20 | 1.87 | 0.20 |
| 2MASS J09574607−3923072 | 4922 | 62 | 1.73 | 0.08 | −1.64 | 0.20 | 2.17 | 0.20 |
| 2MASS J10251539−3554026 | 4955 | 85 | 1.94 | 0.10 | −1.72 | 0.20 | 1.72 | 0.20 |
| 2MASS J10401894−4106124 | 5004 | 51 | 2.17 | 0.08 | −1.60 | 0.20 | 1.63 | 0.20 |
| 2MASS J11093699−2005278 | 4775 | 55 | 1.67 | 0.09 | −1.74 | 0.20 | 1.80 | 0.20 |
| 2MASS J11404944−1615396 | 4508 | 71 | 1.04 | 0.11 | −1.92 | 0.20 | 2.00 | 0.20 |
| 2MASS J12044314−2911051 | 4408 | 44 | 0.67 | 0.11 | −2.35 | 0.20 | 2.25 | 0.20 |
| 2MASS J12091322−1415313 | 4500 | 55 | 1.11 | 0.11 | −2.02 | 0.20 | 1.96 | 0.20 |
| 2MASS J12170829+0415146 | 4488 | 40 | 1.07 | 0.23 | −2.69 | 0.20 | 2.41 | 0.20 |
| 2MASS J12355013−3131112 | 5019 | 65 | 2.02 | 0.08 | −2.38 | 0.20 | 1.90 | 0.20 |
| 2MASS J13494713−7423395 | 4850 | 60 | 1.53 | 0.09 | −2.61 | 0.20 | 2.06 | 0.20 |
| 2MASS J14100568−0701443 | 4609 | 45 | 1.42 | 0.26 | −2.16 | 0.20 | 2.19 | 0.20 |
| 2MASS J14534137+0040467 | 4424 | 56 | 1.02 | 0.14 | −2.41 | 0.20 | 2.34 | 0.20 |
| 2MASS J15383085−1804242 | 5028 | 47 | 2.21 | 0.08 | −2.02 | 0.20 | 1.69 | 0.20 |
| 2MASS J16110508−1107125 | 5165 | 71 | 2.37 | 0.09 | −1.76 | 0.20 | 1.70 | 0.20 |
| 2MASS J17405736−5339473 | 4530 | 86 | 1.11 | 0.11 | −2.11 | 0.20 | 1.96 | 0.20 |
| 2MASS J17541561−5148268 | 4740 | 73 | 1.51 | 0.10 | −1.85 | 0.20 | 2.11 | 0.20 |
| 2MASS J18050641−4907579 | 4337 | 44 | 0.53 | 0.13 | −2.68 | 0.20 | 2.37 | 0.20 |
| 2MASS J18294122−4504000 | 4888 | 38 | 1.71 | 0.09 | −2.53 | 0.20 | 2.03 | 0.20 |
| 2MASS J19192768−5959140 | 4264 | 36 | 0.42 | 0.14 | −2.72 | 0.20 | 2.58 | 0.20 |
| 2MASS J19215077−4452545 | 4430 | 50 | 0.39 | 0.20 | −2.79 | 0.20 | 2.49 | 0.20 |
| 2MASS J19291910−5528181 | 4309 | 57 | 0.97 | 0.13 | −2.17 | 0.20 | 2.30 | 0.20 |
| 2MASS J20050670−3057445 | 4529 | 47 | 0.91 | 0.12 | −2.99 | 0.20 | 2.35 | 0.20 |
| 2MASS J20554594−3155159 | 4627 | 52 | 1.06 | 0.10 | −2.73 | 0.20 | 2.39 | 0.20 |
| 2MASS J21051293−0439557 | 5103 | 61 | 2.11 | 0.08 | −1.58 | 0.20 | 1.64 | 0.20 |
| 2MASS J22041814−0232101 | 4506 | 62 | 1.07 | 0.10 | −1.73 | 0.20 | 2.27 | 0.20 |
| 2MASS J22132050–5137385 | 5508 | 44 | 2.31 | 0.08 | −2.18 | 0.20 | 2.16 | 0.20 |
| 2MASS J22190836−2333467 | 5077 | 62 | 1.56 | 0.09 | −2.48 | 0.20 | 2.07 | 0.20 |
| 2MASS J22242502+0111262 | 4853 | 108 | 1.68 | 0.09 | −1.38 | 0.20 | 1.93 | 0.20 |
| 2MASS J22372037−4741375 | 4709 | 63 | 1.42 | 0.10 | −2.72 | 0.20 | 2.26 | 0.20 |
| 2MASS J22562536−0719562 | 4547 | 62 | 1.31 | 0.12 | −2.14 | 0.20 | 2.07 | 0.20 |
| 2MASS J23451760−6154429 | 5613 | 48 | 2.61 | 0.07 | −1.34 | 0.20 | 2.20 | 0.20 |
| 2MASS J23513299+3937545 | 4474 | 56 | 1.04 | 0.08 | −1.75 | 0.20 | 1.97 | 0.20 |

**Note.** Table B2 is published in its entirety in the electronic edition of the *Astrophysical Journal*. Note that the machine-readable version also contains the contents of Tables B3–B6.

(This table is available in its entirety in machine-readable form in the online article.)

**Table B3**
Atomic Parameters of Transitions Neighboring the Th II Transition at 4019.13 Å

| Wavelength (Å) | Species[a] | $\chi$ (eV) | log *gf* | Major Blend (?) |
|---|---|---|---|---|
| 4018.820 | 60.1 | 0.064 | −0.850 | Yes |
| 4018.836 | 106.00112 | 1.399 | −5.954 | ... |
| 4018.887 | 26.0 | 4.253 | −1.940 | ... |
| 4018.915 | 106.00113 | 1.204 | −4.569 | ... |
| 4018.924 | 106.00113 | 1.204 | −2.737 | ... |
| 4018.929 | 23.0 | 2.579 | −0.650 | ... |
| 4018.952 | 106.00112 | 1.509 | −2.349 | ... |
| 4018.956 | 606.01212 | 0.451 | −8.789 | ... |
| 4018.978 | 607.01314 | 3.186 | −1.306 | ... |
| 4018.983 | 106.00113 | 0.460 | −1.371 | Yes |
| 4018.986 | 92.1 | 0.036 | −1.390 | ... |
| 4018.990 | 106.00112 | 1.509 | −4.755 | ... |
| 4018.990 | 607.01314 | 3.186 | −2.738 | ... |
| 4018.999 | 25.0 | 4.350 | −1.500 | ... |
| 4019.000 | 106.00113 | 1.204 | −2.660 | ... |
| 4019.003 | 26.0 | 4.317 | −3.920 | ... |
| 4019.006 | 607.01314 | 3.186 | −1.193 | ... |
| 4019.009 | 106.00113 | 1.204 | −4.989 | ... |
| 4019.036 | 23.1 | 3.750 | −1.730 | ... |
| 4019.042 | 25.0 | 4.662 | −0.560 | ... |
| 4019.042 | 26.0 | 2.609 | −2.720 | Yes |
| 4019.052 | 106.00112 | 1.509 | −5.330 | ... |
| 4019.057 | 58.1 | 1.013 | −0.530 | Yes |
| 4019.067 | 28.0 | 1.934 | −3.400 | Yes |
| 4019.090 | 106.00112 | 1.509 | −2.437 | ... |
| 4019.114 | 27.0059 | 2.278 | −2.272 | Yes |
| 4019.114 | 27.0059 | 2.278 | −2.448 | Yes |
| 4019.119 | 27.0059 | 2.278 | −2.147 | Yes |
| 4019.119 | 27.0059 | 2.278 | −2.272 | Yes |
| 4019.126 | 27.0059 | 2.278 | −2.147 | Yes |
| 4019.126 | 27.0059 | 2.278 | −2.261 | Yes |
| 4019.126 | 27.0059 | 2.278 | −2.466 | Yes |
| 4019.129 | 26.0 | 4.317 | −4.450 | ... |
| 4019.129 | 90.1 | 0.000 | −0.230 | Yes |
| 4019.136 | 27.0059 | 2.278 | −1.850 | ... |
| 4019.136 | 27.0059 | 2.278 | −2.261 | ... |
| 4019.138 | 23.0 | 1.802 | −2.150 | ... |
| 4019.143 | 42.0 | 3.396 | −1.390 | ... |
| 4019.146 | 106.00113 | 0.461 | −1.363 | Yes |
| 4019.163 | 27.0 | 2.868 | −3.140 | ... |
| 4019.228 | 74.0 | 0.412 | −2.200 | ... |
| 4019.229 | 106.00112 | 1.490 | −4.444 | ... |
| 4019.245 | 606.01212 | 0.252 | −9.130 | ... |
| 4019.295 | 30.0 | 4.078 | −1.120 | Yes (fake/unidentified) |
| 4019.255 | 27.0059 | 0.581 | −4.436 | Yes |
| 4019.261 | 27.0059 | 0.581 | −4.436 | Yes |
| 4019.261 | 27.0059 | 0.581 | −4.612 | Yes |
| 4019.264 | 27.0059 | 0.629 | −4.336 | Yes |
| 4019.270 | 27.0059 | 0.581 | −4.272 | Yes |
| 4019.270 | 27.0059 | 0.581 | −4.737 | Yes |
| 4019.270 | 27.0059 | 0.581 | −4.862 | Yes |
| 4019.283 | 27.0059 | 0.581 | −4.264 | Yes |
| 4019.283 | 27.0059 | 0.581 | −4.298 | Yes |
| 4019.283 | 27.0059 | 0.581 | −5.290 | Yes |
| 4019.288 | 27.0059 | 0.629 | −4.552 | Yes |
| 4019.289 | 24.1 | 5.326 | −5.600 | ... |
| 4019.289 | 27.0059 | 0.629 | −4.962 | Yes |
| 4019.298 | 27.0059 | 0.581 | −4.015 | Yes |
| 4019.298 | 27.0059 | 0.581 | −4.425 | Yes |
| 4019.308 | 27.0059 | 0.629 | −4.801 | Yes |
| 4019.308 | 27.0059 | 0.629 | −4.835 | Yes |
| 4019.309 | 27.0059 | 0.629 | −5.827 | Yes |
| 4019.316 | 27.0059 | 0.581 | −3.799 | Yes |
| 4019.324 | 27.0059 | 0.629 | −4.809 | Yes |
| 4019.324 | 27.0059 | 0.629 | −5.274 | Yes |
| 4019.324 | 27.0059 | 0.629 | −5.399 | Yes |
| 4019.336 | 27.0059 | 0.629 | −4.973 | Yes |
| 4019.336 | 27.0059 | 0.629 | −5.149 | Yes |
| 4019.344 | 27.0059 | 0.629 | −4.973 | Yes |
| 4019.400 | 607.01314 | 3.184 | −1.461 | ... |
| 4019.407 | 607.01314 | 3.184 | −2.607 | ... |
| 4019.420 | 607.01314 | 3.184 | −1.306 | ... |
| 4019.422 | 28.1 | 6.324 | −4.920 | ... |
| 4019.448 | 23.0 | 2.581 | −1.220 | ... |
| 4019.464 | 23.0 | 3.111 | −2.890 | ... |
| 4019.471 | 58.1 | 0.875 | −1.460 | Modified log *gf* |

**Note.**
[a] The species are coded so that for atoms the number to the left of the decimal point denotes the atomic number and the number to the right denotes neutral (=0) or ionized species (=1). For molecules, the first digit denotes the atomic number of the first atom of the molecule (e.g., 1 denotes H), the following two digits denote the atomic number of the second atom of the molecule; the first three digits after the decimal point denote the mass number of the first atom, and the following two digits denote the mass number of the second atom.

**Table B4**
Fe Abundances and Uncertainties of Stars in This Work

| Star Name | [Fe II/H] | $\sigma_{T_{\rm eff}}$ | $\sigma_{\log\ g}$ | $\sigma_{[\rm M/H]}$ | $\sigma_{\xi}$ | $\sigma_{\rm stat}$ | $\sigma_{\rm stddev}$ | $\sigma_{\rm tot}$ |
|---|---|---|---|---|---|---|---|---|
| 2MASS J00101758−1735387 | −2.26 | 0.01 | 0.02 | 0.00 | −0.03 | 0.01 | 0.04 | 0.06 |
| 2MASS J00193176+3141441 | −1.88 | −0.01 | 0.03 | 0.03 | −0.07 | 0.02 | 0.09 | 0.13 |
| 2MASS J00524174−0902235 | −1.41 | −0.00 | 0.02 | 0.03 | −0.09 | 0.01 | 0.09 | 0.13 |
| 2MASS J01320993+3300431 | −1.70 | −0.01 | 0.04 | 0.02 | −0.08 | 0.02 | 0.08 | 0.12 |
| 2MASS J01425445−0904162 | −2.00 | −0.01 | 0.04 | 0.04 | −0.06 | 0.01 | 0.10 | 0.13 |
| 2MASS J01530024−3417360 | −1.65 | 0.00 | 0.02 | 0.02 | −0.07 | 0.01 | 0.08 | 0.11 |
| 2MASS J01542216+0341454 | −2.16 | −0.01 | 0.05 | 0.05 | −0.07 | 0.01 | 0.08 | 0.13 |
| 2MASS J03422816−6500355 | −2.15 | 0.01 | 0.03 | 0.02 | −0.05 | 0.01 | 0.04 | 0.08 |
| 2MASS J04141624−6009048 | −2.56 | 0.00 | 0.03 | 0.01 | −0.03 | 0.04 | 0.11 | 0.13 |
| 2MASS J06320130−2026538 | −1.66 | −0.00 | 0.03 | 0.02 | −0.07 | 0.01 | 0.09 | 0.12 |
| 2MASS J07103110−7121522 | −1.56 | −0.01 | 0.03 | 0.04 | −0.09 | 0.01 | 0.11 | 0.15 |
| 2MASS J07114252−3432368 | −2.02 | 0.00 | 0.03 | 0.02 | −0.06 | 0.02 | 0.10 | 0.12 |
| 2MASS J07352232−4425010 | −1.74 | −0.01 | 0.04 | 0.02 | −0.06 | 0.00 | 0.05 | 0.09 |
| 2MASS J08474871−2204315 | −1.69 | −0.01 | 0.04 | 0.04 | −0.08 | 0.01 | 0.09 | 0.13 |
| 2MASS J09185890−2311511 | −2.01 | −0.01 | 0.03 | 0.04 | −0.07 | 0.01 | 0.08 | 0.12 |
| 2MASS J09574607−3923072 | −1.64 | −0.00 | 0.06 | 0.04 | −0.08 | 0.04 | 0.14 | 0.18 |
| 2MASS J10251539−3554026 | −1.78 | −0.02 | 0.03 | 0.02 | −0.08 | 0.01 | 0.10 | 0.14 |
| 2MASS J10401894−4106124 | −1.60 | −0.00 | 0.03 | 0.03 | −0.09 | 0.01 | 0.10 | 0.14 |
| 2MASS J11093699−2005278 | −1.74 | 0.00 | 0.02 | 0.04 | −0.08 | 0.01 | 0.10 | 0.14 |
| 2MASS J11404944−1615396 | −1.92 | −0.02 | 0.04 | 0.04 | −0.07 | 0.01 | 0.06 | 0.11 |
| 2MASS J12044314−2911051 | −2.43 | −0.01 | 0.05 | 0.04 | −0.06 | 0.02 | 0.10 | 0.13 |
| 2MASS J12091322−1415313 | −2.02 | −0.01 | 0.04 | 0.04 | −0.07 | 0.01 | 0.09 | 0.13 |
| 2MASS J12170829+0415146 | −2.69 | −0.01 | 0.08 | 0.01 | −0.03 | 0.02 | 0.05 | 0.10 |
| 2MASS J12355013−3131112 | −2.38 | 0.01 | 0.03 | 0.01 | −0.04 | 0.01 | 0.04 | 0.06 |
| 2MASS J13494713−7423395 | −2.57 | 0.00 | 0.03 | 0.01 | −0.04 | 0.01 | 0.04 | 0.07 |
| 2MASS J14100568−0701443 | −2.14 | −0.04 | 0.08 | 0.03 | −0.06 | 0.04 | 0.11 | 0.16 |
| 2MASS J14534137+0040467 | −2.41 | −0.02 | 0.04 | 0.05 | −0.05 | 0.01 | 0.11 | 0.14 |
| 2MASS J15383085−1804242 | −2.02 | 0.00 | 0.03 | 0.02 | −0.06 | 0.01 | 0.07 | 0.10 |
| 2MASS J16110508−1107125 | −1.76 | 0.00 | 0.04 | 0.03 | −0.07 | 0.02 | 0.09 | 0.12 |
| 2MASS J17405736−5339473 | −2.15 | −0.02 | 0.05 | 0.04 | −0.07 | 0.01 | 0.09 | 0.13 |
| 2MASS J17541561−5148268 | −1.85 | −0.01 | 0.04 | 0.03 | −0.07 | 0.03 | 0.07 | 0.11 |
| 2MASS J18050641−4907579 | −2.68 | 0.00 | 0.06 | 0.02 | −0.05 | 0.01 | 0.07 | 0.11 |
| 2MASS J18294122−4504000 | −2.53 | 0.01 | 0.11 | 0.02 | −0.03 | 0.01 | 0.05 | 0.13 |
| 2MASS J19192768−5959140 | −2.72 | −0.02 | 0.04 | 0.02 | −0.04 | 0.01 | 0.07 | 0.09 |
| 2MASS J19215077−4452545 | −2.79 | 0.00 | 0.05 | 0.03 | −0.04 | 0.01 | 0.09 | 0.11 |
| 2MASS J19291910−5528181 | −2.17 | −0.05 | 0.03 | 0.03 | −0.06 | 0.02 | 0.10 | 0.14 |
| 2MASS J20050670−3057445 | −2.96 | 0.01 | 0.04 | 0.02 | −0.03 | 0.00 | 0.05 | 0.07 |
| 2MASS J20554594−3155159 | −2.73 | −0.00 | 0.02 | −0.02 | −0.03 | 0.01 | 0.06 | 0.07 |
| 2MASS J21051293−0439557 | −1.58 | 0.00 | 0.04 | 0.04 | −0.09 | 0.02 | 0.14 | 0.18 |
| 2MASS J22041814−0232101 | −1.75 | −0.01 | 0.05 | 0.05 | −0.07 | 0.01 | 0.09 | 0.13 |
| 2MASS J22132050−5137385 | −2.19 | 0.00 | 0.03 | 0.00 | −0.04 | 0.00 | 0.08 | 0.09 |
| 2MASS J22190836−2333467 | −2.48 | 0.01 | 0.03 | 0.01 | −0.04 | 0.01 | 0.05 | 0.07 |
| 2MASS J22242502+0111262 | −1.38 | −0.03 | 0.04 | 0.05 | −0.09 | 0.01 | 0.09 | 0.15 |
| 2MASS J22372037−4741375 | −2.71 | 0.00 | 0.04 | 0.01 | −0.03 | 0.01 | 0.05 | 0.07 |
| 2MASS J22562536−0719562 | −2.14 | −0.03 | 0.03 | 0.03 | −0.07 | 0.01 | 0.09 | 0.12 |
| 2MASS J23451760−6154429 | −1.34 | 0.01 | 0.04 | 0.01 | −0.07 | 0.00 | 0.07 | 0.11 |
| 2MASS J23513299+3937545 | −1.76 | −0.02 | 0.03 | 0.05 | −0.08 | 0.01 | 0.09 | 0.14 |

**Note.** The data from this table are included in the machine-readable version of Table B2.

**Table B5**
Eu Abundances and Uncertainties of Stars from This Work

| Star Name | log $\epsilon$(Eu) | [Eu/Fe] | $\sigma_{T_{eff}}$ | $\sigma_{\log g}$ | $\sigma_{[M/H]}$ | $\sigma_{\xi}$ | $\sigma_{stat}$ | $\sigma_{stddev}$ | $\sigma_{tot}$ |
|---|---|---|---|---|---|---|---|---|---|
| 2MASS J00101758−1735387 | −0.33 | +1.41 | 0.06 | 0.02 | 0.01 | −0.01 | 0.01 | 0.00 | 0.07 |
| 2MASS J00193176+3141441 | −0.87 | +0.49 | 0.04 | 0.04 | 0.03 | −0.02 | 0.02 | 0.04 | 0.08 |
| 2MASS J00524174−0902235 | −0.26 | +0.63 | 0.03 | 0.03 | 0.05 | −0.00 | 0.01 | 0.01 | 0.07 |
| 2MASS J01320993+3300431 | −0.47 | +0.71 | 0.04 | 0.04 | 0.03 | −0.02 | 0.02 | 0.01 | 0.07 |
| 2MASS J01425445−0904162 | −0.80 | +0.68 | 0.01 | 0.02 | 0.03 | −0.04 | 0.01 | 0.01 | 0.05 |
| 2MASS J01530024−3417360 | −0.48 | +0.65 | 0.05 | 0.01 | 0.03 | −0.01 | 0.01 | 0.02 | 0.06 |
| 2MASS J01542216+0341454 | −1.30 | +0.34 | 0.02 | 0.02 | 0.03 | −0.04 | 0.01 | 0.03 | 0.06 |
| 2MASS J03422816−6500355 | −0.66 | +0.97 | 0.04 | 0.03 | 0.02 | −0.02 | 0.00 | 0.00 | 0.06 |
| 2MASS J04141624−6009048 | −0.31 | +1.73 | 0.05 | 0.04 | 0.02 | −0.01 | 0.02 | 0.01 | 0.07 |
| 2MASS J06320130−2026538 | −0.35 | +0.79 | 0.03 | 0.02 | 0.02 | −0.02 | 0.00 | 0.00 | 0.05 |
| 2MASS J07103110−7121522 | −0.37 | +0.67 | 0.03 | 0.03 | 0.05 | −0.01 | 0.00 | 0.01 | 0.07 |
| 2MASS J07114252−3432368 | −0.11 | +1.39 | 0.04 | 0.03 | 0.03 | −0.04 | 0.00 | 0.01 | 0.07 |
| 2MASS J07352232−4425010 | −0.45 | +0.77 | 0.01 | 0.04 | 0.02 | −0.01 | 0.01 | 0.01 | 0.05 |
| 2MASS J08474871−2204315 | −0.23 | +0.94 | 0.02 | 0.02 | 0.04 | −0.03 | 0.01 | 0.01 | 0.06 |
| 2MASS J09185890−2311511 | −0.93 | +0.56 | 0.04 | 0.03 | 0.05 | −0.01 | 0.00 | 0.01 | 0.07 |
| 2MASS J09574607−3923072 | −0.69 | +0.43 | 0.02 | 0.03 | 0.04 | −0.02 | 0.04 | 0.01 | 0.07 |
| 2MASS J10251539−3554026 | −0.49 | +0.77 | 0.05 | 0.04 | 0.04 | −0.02 | 0.00 | 0.01 | 0.07 |
| 2MASS J10401894−4106124 | −0.35 | +0.73 | 0.01 | 0.01 | 0.03 | −0.03 | 0.00 | 0.11 | 0.12 |
| 2MASS J11093699−2005278 | −0.66 | +0.56 | 0.03 | 0.01 | 0.05 | −0.02 | 0.01 | 0.02 | 0.07 |
| 2MASS J11404944−1615396 | −0.68 | +0.72 | 0.01 | 0.01 | 0.02 | −0.05 | 0.03 | 0.01 | 0.06 |
| 2MASS J12044314−2911051 | −1.30 | +0.61 | 0.01 | 0.03 | 0.03 | −0.03 | 0.01 | 0.01 | 0.05 |
| 2MASS J12091322−1415313 | −0.95 | +0.55 | 0.03 | 0.03 | 0.04 | −0.02 | 0.01 | 0.02 | 0.07 |
| 2MASS J12170829+0415146 | −1.37 | +0.80 | 0.02 | 0.07 | 0.01 | −0.01 | 0.02 | 0.01 | 0.08 |
| 2MASS J12355013−3131112 | −1.35 | +0.51 | 0.04 | 0.01 | 0.00 | −0.02 | 0.02 | 0.00 | 0.05 |
| 2MASS J13494713−7423395 | −1.27 | +0.78 | 0.04 | 0.02 | 0.01 | −0.02 | 0.01 | 0.00 | 0.05 |
| 2MASS J14100568−0701443 | −0.97 | +0.65 | −0.00 | 0.07 | 0.03 | −0.01 | 0.02 | 0.01 | 0.08 |
| 2MASS J14534137+0040467 | −0.22 | +1.67 | −0.02 | 0.01 | −0.02 | −0.06 | 0.00 | 0.02 | 0.07 |
| 2MASS J15383085−1804242 | −0.06 | +1.44 | 0.03 | 0.03 | 0.02 | −0.03 | 0.02 | 0.01 | 0.06 |
| 2MASS J16110508−1107125 | −0.67 | +0.57 | 0.05 | 0.05 | 0.05 | −0.02 | 0.01 | 0.04 | 0.09 |
| 2MASS J17405736−5339473 | −1.19 | +0.44 | 0.03 | 0.03 | 0.03 | −0.02 | 0.01 | 0.02 | 0.06 |
| 2MASS J17541561−5148268 | −0.88 | +0.45 | 0.04 | 0.03 | 0.04 | −0.02 | 0.01 | 0.05 | 0.08 |
| 2MASS J18050641−4907579 | −1.59 | +0.57 | 0.05 | 0.06 | 0.03 | −0.01 | 0.00 | 0.03 | 0.09 |
| 2MASS J18294122−4504000 | −1.31 | +0.70 | 0.04 | 0.18 | 0.02 | −0.01 | 0.01 | 0.00 | 0.19 |
| 2MASS J19192768−5959140 | −1.95 | +0.25 | 0.01 | 0.03 | 0.02 | −0.01 | 0.01 | 0.02 | 0.05 |
| 2MASS J19215077−4452545 | −1.75 | +0.52 | 0.05 | 0.05 | 0.04 | −0.01 | 0.00 | 0.00 | 0.08 |
| 2MASS J19291910−5528181 | −1.33 | +0.32 | 0.00 | 0.02 | 0.03 | −0.02 | 0.01 | 0.00 | 0.04 |
| 2MASS J20050670−3057445 | −1.42 | +1.02 | 0.04 | 0.04 | 0.02 | −0.01 | 0.00 | 0.01 | 0.06 |
| 2MASS J20554594−3155159 | −1.44 | +0.77 | 0.04 | 0.02 | 0.08 | −0.01 | 0.00 | 0.00 | 0.09 |
| 2MASS J21051293−0439557 | −0.39 | +0.67 | 0.05 | 0.04 | 0.05 | −0.02 | 0.01 | 0.00 | 0.09 |
| 2MASS J22041814−0232101 | −0.84 | +0.39 | 0.03 | 0.03 | 0.05 | −0.01 | 0.02 | 0.07 | 0.10 |
| 2MASS J22132050–5137385 | 0.80 | +2.47 | −0.04 | 0.00 | −0.04 | −0.10 | 0.00 | 0.01 | 0.12 |
| 2MASS J22190836−2333467 | −0.99 | +0.97 | 0.04 | 0.02 | 0.01 | −0.01 | 0.01 | 0.00 | 0.05 |
| 2MASS J22242502+0111262 | −0.31 | +0.55 | 0.05 | 0.03 | 0.04 | −0.02 | 0.01 | 0.02 | 0.08 |
| 2MASS J22372037−4741375 | −1.46 | +0.73 | 0.05 | 0.04 | 0.01 | −0.01 | 0.01 | 0.00 | 0.07 |
| 2MASS J22562536−0719562 | −0.73 | +0.89 | 0.02 | 0.03 | 0.03 | −0.02 | 0.00 | 0.04 | 0.07 |
| 2MASS J23451760−6154429 | 0.35 | +1.17 | 0.01 | 0.01 | 0.00 | −0.04 | 0.01 | 0.04 | 0.06 |
| 2MASS J23513299+3937545 | −0.63 | +0.61 | −0.01 | −0.00 | 0.02 | −0.06 | 0.02 | 0.05 | 0.08 |

**Note.** The data from this table are included in the machine-readable version of Table B2.

**Table B6**
Dy Abundances and Uncertainties of Stars from This Work

| Star Name | log $\epsilon$(Dy) | [Dy/Fe] | $\sigma_{T_{\rm eff}}$ | $\sigma_{\log g}$ | $\sigma_{[M/H]}$ | $\sigma_{\xi}$ | $\sigma_{\rm stat}$ | $\sigma_{\rm stddev}$ | $\sigma_{\rm tot}$ |
|---|---|---|---|---|---|---|---|---|---|
| 2MASS J00101758−1735387 | 0.41 | +1.57 | 0.07 | 0.02 | 0.01 | −0.03 | 0.00 | 0.01 | 0.08 |
| 2MASS J00193176+3141441 | −0.16 | +0.62 | 0.05 | 0.05 | 0.03 | −0.04 | 0.01 | 0.03 | 0.09 |
| 2MASS J00524174−0902235 | 0.47 | +0.78 | 0.02 | 0.01 | 0.03 | −0.05 | 0.00 | 0.11 | 0.13 |
| 2MASS J01320993+3300431 | 0.23 | +0.83 | 0.03 | 0.02 | 0.03 | −0.05 | 0.02 | 0.01 | 0.07 |
| 2MASS J01425445−0904162 | −0.21 | +0.69 | 0.02 | 0.03 | 0.05 | −0.05 | 0.00 | 0.03 | 0.08 |
| 2MASS J01530024−3417360 | 0.22 | +0.77 | 0.08 | 0.01 | 0.04 | −0.04 | 0.00 | 0.01 | 0.10 |
| 2MASS J01542216+0341454 | −0.58 | +0.48 | 0.04 | 0.05 | 0.05 | −0.05 | 0.02 | 0.03 | 0.10 |
| 2MASS J03422816−6500355 | 0.12 | +1.17 | 0.05 | 0.02 | 0.02 | −0.04 | 0.00 | 0.01 | 0.08 |
| 2MASS J04141624−6009048 | 0.48 | +1.94 | 0.02 | 0.01 | 0.00 | −0.06 | 0.03 | 0.08 | 0.10 |
| 2MASS J06320130−2026538 | 0.37 | +0.93 | 0.06 | 0.03 | 0.04 | −0.04 | 0.03 | 0.01 | 0.09 |
| 2MASS J07103110−7121522 | 0.33 | +0.79 | 0.05 | 0.03 | 0.07 | −0.05 | 0.02 | 0.03 | 0.11 |
| 2MASS J07114252−3432368 | 0.63 | +1.55 | 0.04 | 0.03 | 0.03 | −0.06 | 0.01 | 0.01 | 0.09 |
| 2MASS J07352232−4425010 | 0.26 | +0.90 | 0.01 | 0.02 | 0.02 | −0.05 | 0.00 | 0.04 | 0.07 |
| 2MASS J08474871−2204315 | 0.52 | +1.11 | 0.02 | 0.02 | 0.04 | −0.07 | 0.00 | 0.02 | 0.09 |
| 2MASS J09185890−2311511 | −0.24 | +0.67 | 0.03 | 0.02 | 0.04 | −0.06 | 0.00 | 0.03 | 0.08 |
| 2MASS J09574607−3923072 | 0.23 | +0.77 | 0.00 | −0.02 | 0.04 | −0.10 | 0.06 | 0.08 | 0.14 |
| 2MASS J10251539−3554026 | 0.26 | +0.94 | 0.04 | 0.02 | 0.03 | −0.06 | 0.00 | 0.02 | 0.09 |
| 2MASS J10401894−4106124 | 0.48 | +0.98 | 0.02 | 0.01 | 0.04 | −0.07 | 0.00 | 0.02 | 0.08 |
| 2MASS J11093699−2005278 | 0.01 | +0.65 | 0.05 | 0.04 | 0.08 | −0.00 | 0.08 | 0.03 | 0.13 |
| 2MASS J11404944−1615396 | −0.01 | +0.81 | 0.04 | 0.04 | 0.05 | −0.06 | 0.01 | 0.01 | 0.09 |
| 2MASS J12044314−2911051 | −0.57 | +0.76 | 0.00 | 0.02 | 0.02 | −0.06 | 0.00 | 0.00 | 0.07 |
| 2MASS J12091322−1415313 | −0.20 | +0.72 | 0.03 | 0.03 | 0.05 | −0.06 | 0.01 | 0.01 | 0.09 |
| 2MASS J12170829+0415146 | −0.68 | +0.91 | 0.00 | 0.05 | −0.00 | −0.04 | 0.03 | 0.05 | 0.09 |
| 2MASS J12355013−3131112 | −0.64 | +0.64 | 0.06 | 0.02 | 0.02 | −0.01 | 0.02 | 0.03 | 0.08 |
| 2MASS J13494713−7423395 | −0.53 | +0.94 | 0.05 | 0.02 | 0.02 | −0.03 | 0.01 | 0.02 | 0.07 |
| 2MASS J14100568−0701443 | −0.19 | +0.85 | −0.00 | 0.07 | 0.03 | −0.05 | 0.02 | 0.03 | 0.10 |
| 2MASS J14534137+0040467 | 0.28 | +1.59 | −0.03 | −0.04 | −0.03 | −0.12 | 0.00 | 0.02 | 0.14 |
| 2MASS J15383085−1804242 | 0.68 | +1.60 | 0.04 | 0.02 | 0.03 | −0.07 | 0.00 | 0.00 | 0.08 |
| 2MASS J16110508−1107125 | 0.11 | +0.77 | 0.05 | 0.04 | 0.04 | −0.05 | 0.06 | 0.08 | 0.14 |
| 2MASS J17405736−5339473 | −0.49 | +0.56 | 0.06 | 0.07 | 0.07 | −0.00 | 0.08 | 0.04 | 0.15 |
| 2MASS J17541561−5148268 | −0.21 | +0.54 | 0.02 | 0.01 | 0.01 | −0.08 | 0.02 | 0.00 | 0.08 |
| 2MASS J18050641−4907579 | −1.02 | +0.56 | 0.05 | 0.05 | 0.02 | −0.04 | 0.00 | 0.00 | 0.08 |
| 2MASS J18294122−4504000 | −0.63 | +0.80 | 0.05 | 0.20 | 0.02 | −0.01 | 0.03 | 0.01 | 0.20 |
| 2MASS J19192768−5959140 | −1.33 | +0.29 | 0.04 | 0.06 | 0.04 | −0.01 | 0.02 | 0.05 | 0.10 |
| 2MASS J19215077−4452545 | −1.01 | +0.68 | 0.07 | 0.07 | 0.06 | −0.00 | 0.00 | 0.00 | 0.11 |
| 2MASS J19291910−5528181 | −0.60 | +0.47 | −0.03 | −0.02 | 0.00 | −0.08 | 0.00 | 0.00 | 0.09 |
| 2MASS J20050670−3057445 | −0.74 | +1.12 | 0.05 | 0.04 | 0.02 | −0.03 | 0.00 | 0.02 | 0.07 |
| 2MASS J20554594−3155159 | −0.86 | +0.77 | 0.08 | 0.06 | 0.08 | 0.02 | 0.24 | 0.04 | 0.27 |
| 2MASS J21051293−0439557 | 0.35 | +0.83 | 0.05 | 0.03 | 0.05 | −0.06 | 0.02 | 0.07 | 0.12 |
| 2MASS J22041814−0232101 | −0.22 | +0.43 | 0.06 | 0.07 | 0.07 | −0.03 | 0.03 | 0.02 | 0.13 |
| 2MASS J22132050–5137385 | 1.29 | +2.38 | 0.06 | 0.06 | 0.04 | −0.00 | 0.00 | 0.03 | 0.09 |
| 2MASS J22190836−2333467 | −0.38 | +1.00 | 0.03 | 0.00 | −0.00 | −0.03 | 0.00 | 0.00 | 0.05 |
| 2MASS J22242502+0111262 | 0.37 | +0.65 | 0.06 | 0.04 | 0.06 | −0.04 | 0.02 | 0.02 | 0.11 |
| 2MASS J22372037−4741375 | −0.68 | +0.93 | 0.06 | 0.04 | 0.02 | −0.01 | 0.02 | 0.03 | 0.08 |
| 2MASS J22562536−0719562 | −0.03 | +1.01 | 0.00 | 0.02 | 0.03 | −0.05 | 0.00 | 0.02 | 0.07 |
| 2MASS J23451760−6154429 | 1.06 | +1.30 | 0.04 | 0.03 | 0.03 | −0.03 | 0.00 | 0.02 | 0.06 |
| 2MASS J23513299+3937545 | −0.04 | +0.62 | 0.02 | 0.04 | 0.06 | −0.05 | 0.01 | 0.01 | 0.09 |

**Note.** The data from this table are included in the machine-readable version of Table B2.

**Table B7**
Th Abundances and Uncertainties of Stars from This Work

| Star Name | log ϵ (Th) | [Th/Fe] | $\sigma_{T_{\rm eff}}$ | $\sigma_{\log\ g}$ | $\sigma_{[M/H]}$ | $\sigma_{\xi}$ | $\sigma_{\rm stat}$ | $\sigma_{\rm blend}$ | $\sigma_{\rm tot}$ |
|---|---|---|---|---|---|---|---|---|---|
| 2MASS J00101758−1735387 | −0.76 | +1.48 | 0.06 | 0.02 | 0.01 | 0.00 | 0.01 | 0.03 | 0.07 |
| 2MASS J00193176+3141441 | −1.32 | +0.54 | 0.06 | 0.07 | 0.06 | 0.02 | 0.05 | 0.04 | 0.13 |
| 2MASS J00524174−0902235 | −0.75 | +0.64 | 0.01 | 0.02 | 0.05 | 0.05 | 0.01 | 0.08 | 0.11 |
| 2MASS J01320993+3300431 | −0.89 | +0.79 | 0.05 | 0.06 | 0.07 | 0.01 | 0.04 | 0.03 | 0.12 |
| 2MASS J01425445−0904162 | −1.36 | +0.62 | 0.02 | 0.03 | 0.05 | 0.00 | 0.06 | 0.05 | 0.10 |
| 2MASS J01530024−3417360 | −1.18 | +0.45 | 0.04 | 0.02 | 0.02 | 0.07 | 0.05 | 0.13 | 0.16 |
| 2MASS J01542216+0341454 | −2.01 | +0.13 | 0.06 | 0.09 | 0.11 | 0.10 | 0.08 | 0.13 | 0.23 |
| 2MASS J03422816−6500355 | −1.36 | +0.77 | 0.04 | 0.03 | 0.04 | 0.03 | 0.04 | 0.04 | 0.09 |
| 2MASS J04141624−6009048 | −0.57 | +1.97 | 0.04 | 0.02 | 0.01 | −0.04 | 0.06 | 0.02 | 0.09 |
| 2MASS J06320130−2026538 | −0.91 | +0.73 | 0.02 | 0.01 | 0.03 | 0.00 | 0.01 | 0.11 | 0.12 |
| 2MASS J07103110−7121522 | −0.95 | +0.59 | 0.03 | 0.02 | 0.05 | 0.03 | 0.02 | 0.04 | 0.08 |
| 2MASS J07114252−3432368 | −0.68 | +1.32 | 0.02 | 0.01 | 0.02 | −0.01 | 0.01 | 0.03 | 0.05 |
| 2MASS J07352232−4425010 | −1.08 | +0.64 | 0.00 | 0.03 | 0.02 | 0.03 | 0.03 | 0.09 | 0.11 |
| 2MASS J08474871−2204315 | −0.83 | +0.84 | 0.04 | 0.05 | 0.07 | 0.06 | 0.04 | 0.03 | 0.13 |
| 2MASS J09185890−2311511 | −1.47 | +0.52 | 0.03 | 0.02 | 0.05 | 0.04 | 0.01 | 0.09 | 0.12 |
| 2MASS J09574607−3923072 | −1.03 | +0.59 | 0.03 | 0.02 | 0.05 | 0.04 | 0.18 | 0.12 | 0.23 |
| 2MASS J10251539−3554026 | −1.15 | +0.61 | 0.08 | 0.08 | 0.09 | 0.05 | 0.02 | 0.08 | 0.17 |
| 2MASS J10401894−4106124 | −0.79 | +0.79 | 0.01 | 0.01 | 0.04 | 0.02 | 0.02 | 0.00 | 0.05 |
| 2MASS J11093699−2005278 | −1.10 | +0.62 | 0.01 | −0.01 | 0.05 | 0.00 | 0.03 | 0.07 | 0.09 |
| 2MASS J11404944−1615396 | −1.21 | +0.69 | −0.00 | −0.01 | 0.03 | 0.00 | 0.12 | 0.12 | 0.17 |
| 2MASS J12044314−2911051 | −2.09 | +0.32 | 0.05 | 0.08 | 0.08 | 0.06 | 0.03 | 0.05 | 0.15 |
| 2MASS J12091322−1415313 | −1.62 | +0.38 | −0.00 | 0.01 | 0.05 | 0.03 | 0.02 | 0.08 | 0.10 |
| 2MASS J12170829+0415146 | −1.97 | +0.70 | 0.03 | 0.08 | 0.03 | 0.02 | 0.08 | 0.03 | 0.13 |
| 2MASS J12355013−3131112 | −1.86 | +0.50 | 0.10 | 0.06 | 0.06 | 0.05 | 0.12 | 0.07 | 0.20 |
| 2MASS J13494713−7423395 | −1.54 | +1.01 | 0.06 | 0.04 | 0.03 | 0.01 | 0.03 | 0.06 | 0.10 |
| 2MASS J14100568−0701443 | −1.48 | +0.64 | 0.01 | 0.09 | 0.04 | 0.03 | 0.06 | 0.07 | 0.14 |
| 2MASS J14534137+0040467 | −1.20 | +1.19 | 0.06 | 0.05 | 0.05 | 0.01 | 0.09 | 0.01 | 0.13 |
| 2MASS J15383085−1804242 | −0.60 | +1.40 | 0.04 | 0.04 | 0.04 | −0.02 | 0.01 | 0.00 | 0.07 |
| 2MASS J16110508−1107125 | −1.26 | +0.48 | 0.08 | 0.08 | 0.04 | 0.02 | 0.10 | 0.04 | 0.16 |
| 2MASS J17405736−5339473 | −1.73 | +0.40 | 0.03 | 0.04 | 0.05 | 0.05 | 0.02 | 0.09 | 0.13 |
| 2MASS J17541561−5148268 | −1.30 | +0.53 | 0.08 | 0.08 | 0.10 | 0.03 | 0.06 | 0.03 | 0.17 |
| 2MASS J18050641−4907579 | −2.20 | +0.46 | 0.08 | 0.10 | 0.07 | 0.04 | 0.05 | 0.05 | 0.17 |
| 2MASS J18294122−4504000 | −1.68 | +0.83 | 0.03 | 0.23 | 0.00 | −0.03 | 0.02 | 0.07 | 0.25 |
| 2MASS J19192768−5959140 | −2.62 | +0.08 | 0.04 | 0.07 | 0.08 | 0.06 | 0.08 | 0.11 | 0.18 |
| 2MASS J19215077−4452545 | −2.23 | +0.54 | 0.07 | 0.08 | 0.07 | 0.02 | 0.03 | 0.05 | 0.14 |
| 2MASS J19291910−5528181 | −2.21 | −0.06 | 0.10 | 0.12 | 0.17 | 0.16 | 0.08 | 0.18 | 0.34 |
| 2MASS J20050670−3057445 | −1.99 | +0.95 | 0.06 | 0.05 | 0.03 | 0.01 | 0.00 | 0.11 | 0.14 |
| 2MASS J20554594−3155159 | −1.94 | +0.77 | 0.06 | 0.03 | 0.12 | 0.02 | 0.07 | 0.04 | 0.16 |
| 2MASS J21051293−0439557 | −0.83 | +0.73 | 0.01 | −0.00 | 0.04 | 0.04 | 0.05 | 0.09 | 0.12 |
| 2MASS J22041814−0232101 | −1.38 | +0.35 | 0.01 | 0.02 | 0.06 | 0.04 | 0.07 | 0.15 | 0.18 |
| 2MASS J22132050–5137385 | 0.20 | +2.37 | 0.00 | 0.00 | 0.00 | −0.03 | 0.01 | 0.01 | 0.03 |
| 2MASS J22190836−2333467 | −1.37 | +1.09 | 0.06 | 0.03 | 0.02 | 0.02 | 0.03 | 0.02 | 0.08 |
| 2MASS J22242502+0111262 | −0.85 | +0.51 | 0.05 | 0.06 | 0.07 | 0.09 | 0.06 | 0.07 | 0.16 |
| 2MASS J22372037−4741375 | −1.80 | +0.89 | 0.06 | 0.04 | 0.02 | 0.00 | 0.03 | 0.05 | 0.10 |
| 2MASS J22562536−0719562 | −1.31 | +0.81 | 0.00 | 0.04 | 0.05 | 0.01 | 0.00 | 0.05 | 0.08 |
| 2MASS J23451760−6154429 | −0.19 | +1.13 | 0.04 | 0.02 | 0.02 | −0.01 | 0.00 | 0.02 | 0.06 |
| 2MASS J23513299+3937545 | −1.35 | +0.39 | 0.02 | 0.02 | 0.08 | 0.01 | 0.11 | 0.06 | 0.16 |

**Note.** The data from this table are included in the machine-readable version of Table B2.

## ORCID iDs

Shivani P. Shah https://orcid.org/0000-0002-3367-2394
Rana Ezzeddine https://orcid.org/0000-0002-8504-8470
Erika M. Holmbeck https://orcid.org/0000-0002-5463-6800
Alexander P. Ji https://orcid.org/0000-0002-4863-8842
Vinicius M. Placco https://orcid.org/0000-0003-4479-1265
Ian U. Roederer https://orcid.org/0000-0001-5107-8930
Mohammad K. Mardini https://orcid.org/0000-0001-9178-3992
Sam A. Usman https://orcid.org/0000-0003-0918-7185
Avrajit Bandyopadhyay https://orcid.org/0000-0002-8304-5444
Timothy C. Beers https://orcid.org/0000-0003-4573-6233
Anna Frebel https://orcid.org/0000-0002-2139-7145
Terese T. Hansen https://orcid.org/0000-0001-6154-8983
Charli M. Sakari https://orcid.org/0000-0002-5095-4000
Chris Sneden https://orcid.org/0000-0002-3456-5929

## References

Abohalima, A., & Frebel, A. 2018, ApJS, 238, 36
Alencastro Puls, A., Kuske, J., Hansen, C. J., et al. 2025, A&A, 693, A294
Alonso, A., Arribas, S., & Martínez-Roger, C. 1999, A&AS, 140, 261
Andrews, J. J., & Mandel, I. 2019, ApJL, 880, L8
Aoki, M., Ishimaru, Y., Aoki, W., & Wanajo, S. 2017, ApJ, 837, 8
Aoki, W., Honda, S., Sadakane, K., & Arimoto, N. 2007, PASJ, 59, L15

Argast, D., Samland, M., Thielemann, F.-K., & Qian, Y.-Z. 2004, A&A, 416, 997
Asplund, M., Grevesse, N., Sauval, A. J., & Scott, P. 2009, ARA&A, 47, 481
Astropy Collaboration, Robitaille, T. P., Tollerud, E. J., et al. 2013, A&A, 558, A33
Azhari, A., Matsuno, T., Aoki, W., Ishigaki, M. N., & Tolstoy, E. 2025, A&A, 699, A276
Bandyopadhyay, A., Beers, T. C., Ezzeddine, R., et al. 2024a, MNRAS, 529, 2191
Bandyopadhyay, A., Ezzeddine, R., Allende Prieto, C., et al. 2024b, ApJS, 274, 39
Bandyopadhyay, A., Sivarani, T., Beers, T. C., et al. 2022, ApJ, 937, 52
Barklem, P. S., Christlieb, N., Beers, T. C., et al. 2005, A&A, 439, 129
Battistini, C., & Bensby, T. 2016, A&A, 586, A49
Bauswein, A., Goriely, S., & Janka, H.-T. 2013, ApJ, 773, 78
Beers, T. C., & Christlieb, N. 2005, ARA&A, 43, 531
Belokurov, V., & Kravtsov, A. 2024, MNRAS, 528, 3198
Beniamini, P., Hotokezaka, K., & Piran, T. 2016, ApJL, 829, L13
Beniamini, P., & Piran, T. 2024, ApJ, 966, 17
Bernstein, R., Shectman, S. A., Gunnels, S. M., Mochnacki, S., & Athey, A. E. 2003, SPIE, 4841, 1694
Beun, J., McLaughlin, G. C., Surman, R., & Hix, W. R. 2008, PhRvC, 77, 035804
Bonaca, A., Conroy, C., Cargile, P. A., et al. 2020, ApJL, 897, L18
Botelho, R. B., Milone, A. d. C., Meléndez, J., et al. 2019, MNRAS, 482, 1690
Bovy, J., Hogg, D. W., & Roweis, S. T. 2011, AnApS, 5, 1657
Brauer, K., Ji, A. P., Drout, M. R., & Frebel, A. 2021, ApJ, 915, 81
Broekgaarden, F. S., Berger, E., Stevenson, S., et al. 2022, MNRAS, 516, 5737
Burbidge, E. M., Burbidge, G. R., Fowler, W. A., & Hoyle, F. 1957, RvMP, 29, 547
Busso, M., Gallino, R., Lambert, D. L., Travaglio, C., & Smith, V. V. 2001, ApJ, 557, 802
Cain, M., Frebel, A., Ji, A. P., et al. 2020, ApJ, 898, 40
Cameron, A. G. W. 1957, PASP, 69, 201
Casagrande, L., Ramírez, I., Meléndez, J., Bessell, M., & Asplund, M. 2010, A&A, 512, A54
Casey, A. R. 2014, PhD thesis, Australian National University, Canberra
Castelli, F., & Kurucz, R. L. 2003, IAUS, 210, A20
Cavallo, L., Cescutti, G., & Matteucci, F. 2023, A&A, 674, A130
Cayrel, R. 1988, IAUS, 132, 345
Cayrel, R., Depagne, E., Spite, M., et al. 2004, A&A, 416, 1117
Cayrel, R., Hill, V., Beers, T. C., et al. 2001, Natur, 409, 691
Cescutti, G., Bonifacio, P., Caffau, E., et al. 2022, A&A, 668, A168
Chen, H.-Y., Landry, P., Read, J. S., & Siegel, D. M. 2025, ApJ, 985, 154
Choplin, A., Goriely, S., & Siess, L. 2022, A&A, 667, L13
Choplin, A., Goriely, S., Siess, L., & Martinet, S. 2025, EPJA, 61, 68
Clayton, D. D., & Rassbach, M. E. 1967, ApJ, 148, 69
Collette, A. 2013, Python and HDF, 5 (O'Reilly)
Côté, B., Eichler, M., Arcones, A., et al. 2019, ApJ, 875, 106
Cowan, J. J., & Rose, W. K. 1977, ApJ, 212, 149
Cowan, J. J., Sneden, C., Burles, S., et al. 2002, ApJ, 572, 861
Cowan, J. J., Sneden, C., Lawler, J. E., et al. 2021, RvMP, 93, 015002
Cutri, R. M., Skrutskie, M. F., van Dyk, S., et al. 2003, yCat, II/246,
Dalton, G., Trager, S. C., Abrams, D. C., et al. 2012, SPIE, 8446, 84460P
de Jong, R. S., Agertz, O., Berbel, A. A., et al. 2019, Msngr, 175, 3
del Peloso, E. F., da Silva, L., & Arany-Prado, L. I. 2005, A&A, 434, 301
Di Matteo, P., Spite, M., Haywood, M., et al. 2020, A&A, 636, A115
Dvorkin, I., Daigne, F., Goriely, S., Vangioni, E., & Silk, J. 2021, MNRAS, 506, 4374
Eichler, M., Arcones, A., Käppeli, R., et al. 2016, JPhCS, 665, 012054
Eichler, M., Sayar, W., Arcones, A., & Rauscher, T. 2019, ApJ, 879, 47
Ezzeddine, R., Rasmussen, K., Frebel, A., et al. 2020, ApJ, 898, 150
Farouqi, K., Kratz, K. L., Pfeiffer, B., et al. 2010, ApJ, 712, 1359
Fitzpatrick, M., Placco, V., Bolton, A., et al. 2025, ASPC, 541, 461
Foreman-Mackey, D. 2016, JOSS, 1, 24
Foreman-Mackey, D., Hogg, D. W., Lang, D., & Goodman, J. 2013, PASP, 125, 306
François, P., Depagne, E., Hill, V., et al. 2007, A&A, 476, 935
Frebel, A. 2018, ARNPS, 68, 237
Frebel, A., Christlieb, N., Norris, J. E., et al. 2007, ApJL, 660, L117
Frebel, A., & Ji, A. P. 2023, arXiv:2302.09188
Freiburghaus, C., Rosswog, S., & Thielemann, F.-K. 1999, ApJL, 525, L121
Fujibayashi, S., Kiuchi, K., Wanajo, S., et al. 2023, ApJ, 942, 39
Gaia Collaboration, Vallenari, A., Brown, A. G. A., et al. 2023, A&A, 674, A1
Goriely, S. 2015, EPJA, 51, 22
Gottlieb, O., Metzger, B. D., Issa, D., et al. 2025, ApJL, 993, L54
Gratton, R. G., & Sneden, C. 1994, A&A, 287, 927
Gratton, R. G., Sneden, C., Carretta, E., & Bragaglia, A. 2000, A&A, 354, 169
Green, G. M. 2018, JOSS, 3, 695
Halevi, G., & Mösta, P. 2018, MNRAS, 477, 2366
Hampel, M., Stancliffe, R. J., Lugaro, M., & Meyer, B. S. 2016, ApJ, 831, 171
Hansen, C. J., Primas, F., Hartman, H., et al. 2012, A&A, 545, A31
Hansen, T. T., Holmbeck, E. M., Beers, T. C., et al. 2018, ApJ, 858, 92
Hansen, T. T., Ji, A. P., Da Costa, G. S., et al. 2021, ApJ, 915, 103
Hansen, T. T., Roederer, I. U., Shah, S. P., et al. 2025, A&A, 697, A127
Harris, C. R., Millman, K. J., van der Walt, S. J., et al. 2020, Natur, 585, 357
Hayek, W., Wiesendahl, U., Christlieb, N., et al. 2009, A&A, 504, 511
Haynes, C. J., & Kobayashi, C. 2019, MNRAS, 483, 5123
Henkel, A., Foucart, F., Raaijmakers, G., & Nissanke, S. 2023, PhRvD, 107, 063028
Hill, V., Christlieb, N., Beers, T. C., et al. 2017, A&A, 607, A91
Hill, V., Plez, B., Cayrel, R., et al. 2002, A&A, 387, 560
Hirai, Y., Beers, T. C., Lee, Y. S., et al. 2025, ApJ, 990, 125
Holmbeck, E. M., & Andrews, J. J. 2024, ApJ, 963, 110
Holmbeck, E. M., Beers, T. C., Roederer, I. U., et al. 2018, ApJL, 859, L24
Holmbeck, E. M., Frebel, A., McLaughlin, G. C., et al. 2019a, ApJ, 881, 5
Holmbeck, E. M., Hansen, T. T., Beers, T. C., et al. 2020, ApJS, 249, 30
Holmbeck, E. M., Sprouse, T. M., Mumpower, M. R., et al. 2019b, ApJ, 870, 23
Honda, S., Aoki, W., Ishimaru, Y., & Wanajo, S. 2007, ApJ, 666, 1189
Honda, S., Aoki, W., Ishimaru, Y., Wanajo, S., & Ryan, S. G. 2006, ApJ, 643, 1180
Honda, S., Aoki, W., Kajino, T., et al. 2004, ApJ, 607, 474
Hotokezaka, K., Beniamini, P., & Piran, T. 2018, IJMPD, 27, 1842005
Huber, D., Slumstrup, D., Hon, M., et al. 2024, ApJ, 975, 19
Hunter, J. D. 2007, CSE, 9, 90
Ishimaru, Y., Wanajo, S., & Prantzos, N. 2015, ApJL, 804, L35
Issa, D., Gottlieb, O., Metzger, B. D., et al. 2025, ApJL, 985, L26
Ivans, I. I., Simmerer, J., Sneden, C., et al. 2006, ApJ, 645, 613
Jaupart, C., Labrosse, S., Lucazeau, F., & Mareschal, J. 2007, Treatise on geophysics, 7 (Elsevier), 223
Jean-Baptiste, I., Di Matteo, P., Haywood, M., et al. 2017, A&A, 604, A106
Ji, A. P., Casey, A. R., Ting, Y.-S., et al. 2025, LESSPayne: Labeling Echelle Spectra with SMHR and Payne, Astrophysics Source Code Library, ascl:2503.025
Ji, A. P., & Frebel, A. 2018, ApJ, 856, 138
Ji, A. P., Li, T. S., Hansen, T. T., et al. 2020a, AJ, 160, 181
Ji, A. P., Li, T. S., Simon, J. D., et al. 2020b, ApJ, 889, 27
Ji, A. P., Simon, J. D., Roederer, I. U., et al. 2023, AJ, 165, 100
Johnson, J. A. 2002, ApJS, 139, 219
Johnson, J. A., & Bolte, M. 2001, ApJ, 554, 888
Just, O., Aloy, M. A., Obergaulinger, M., & Nagataki, S. 2022, ApJL, 934, L30
Kelson, D. D. 1998, PhD thesis, University of California, Santa Cruz
Kelson, D. D. 2003, PASP, 115, 688
Kelson, D. D., Illingworth, G. D., van Dokkum, P. G., & Franx, M. 2000, ApJ, 531, 159
Kirby, E. N., Lanfranchi, G. A., Simon, J. D., Cohen, J. G., & Guhathakurta, P. 2011, ApJ, 727, 78
Kobayashi, C., Karakas, A. I., & Lugaro, M. 2020, ApJ, 900, 179
Kobayashi, C., Mandel, I., Belczynski, K., et al. 2023, ApJL, 943, L12
Komiya, Y., & Shigeyama, T. 2016, ApJ, 830, 76
Koppelman, H. H., Helmi, A., Massari, D., Price-Whelan, A. M., & Starkenburg, T. K. 2019, A&A, 631, L9
Korobkin, O., Rosswog, S., Arcones, A., & Winteler, C. 2012, MNRAS, 426, 1940
Kramida, A., Ralchenko, Y., Reader, J. & NIST ASD Team 2022, NIST Atomic Spectra Database 5.12, National Institute of Standards and Technology National Institute of Standards and Technology
Krüger, C. J., & Foucart, F. 2020, PhRvD, 101, 103002
Kurucz, R. L. 2011, CaJPh, 89, 417
Lai, D. K., Bolte, M., Johnson, J. A., et al. 2008, ApJ, 681, 1524
Larsen, J. R., Rørsted, J. L., Aguirre Børsen-Koch, V., et al. 2025, A&A, 697, A153
Lin, Y., Li, H., Jiang, R., et al. 2025, ApJL, 984, L43
Lombardo, L., Hansen, C. J., Rizzuti, F., et al. 2025, A&A, 693, A293
Luna, A. M., Ji, A. P., Chiti, A., et al. 2025, OJAp, 8, 47696
Lund, K. A., Engel, J., McLaughlin, G. C., et al. 2023, ApJ, 944, 144
Lund, K. A., McLaughlin, G. C., Miller, J. M., & Mumpower, M. R. 2024, ApJ, 964, 111
Luo, H., O'Rourke, J. G., & Deng, J. 2024, SciA, 10, eado7603

Mandel, I., & Broekgaarden, F. S. 2022, LRR, 25, 1
Maoz, D., & Nakar, E. 2025, ApJ, 982, 179
Mashonkina, L., Christlieb, N., Barklem, P. S., et al. 2010, A&A, 516, A46
Mashonkina, L., Christlieb, N., & Eriksson, K. 2014, A&A, 569, A43
Mashonkina, L., Ryabtsev, A., & Frebel, A. 2012, A&A, 540, A98
Matas Pinto, A. M., Spite, M., Caffau, E., et al. 2021, A&A, 654, A170
McWilliam, A. 1998, AJ, 115, 1640
McWilliam, A., Preston, G. W., Sneden, C., & Searle, L. 1995, AJ, 109, 2757
Mendoza-Temis, J. d. J., Wu, M.-R., Langanke, K., et al. 2015, PhRvC, 92, 055805
Miller, J. M., Sprouse, T. M., Fryer, C. L., et al. 2020, ApJ, 902, 66
Mishenina, T., Pignatari, M., Gorbaneva, T., et al. 2022, MNRAS, 516, 3786
Montes, F., Beers, T. C., Cowan, J., et al. 2007, ApJ, 671, 1685
Monty, S., Belokurov, V., Sanders, J. L., et al. 2024, MNRAS, 533, 2420
Mösta, P., Roberts, L. F., Halevi, G., et al. 2018, ApJ, 864, 171
Mucciarelli, A., Bellazzini, M., & Massari, D. 2021, A&A, 653, A90
Naidu, R. P., Conroy, C., Bonaca, A., et al. 2020, ApJ, 901, 48
Nilsson, H., Ivarsson, S., Johansson, S., & Lundberg, H. 2002, A&A, 381, 1090
Nimmo, F., Primack, J., Faber, S. M., Ramirez-Ruiz, E., & Safarzadeh, M. 2020, ApJL, 903, L37
Nishimura, N., Sawai, H., Takiwaki, T., Yamada, S., & Thielemann, F. 2017, ApJL, 836, L21
Nishimura, N., Takiwaki, T., & Thielemann, F.-K. 2015, ApJ, 810, 109
Ojima, T., Ishimaru, Y., Wanajo, S., Prantzos, N., & François, P. 2018, ApJ, 865, 87
Okada, H., Aoki, W., Tominaga, N., & Honda, S. 2026, ApJ, 997, 119
Ou, X., Ji, A. P., Frebel, A., Naidu, R. P., & Limberg, G. 2024, ApJ, 974, 232
Pedregosa, F., Varoquaux, G., Gramfort, A., et al. 2011, JMLR, 12, 2825
Placco, V. M., Almeida-Fernandes, F., Holmbeck, E. M., et al. 2023, ApJ, 959, 60
Placco, V. M., Holmbeck, E. M., Frebel, A., et al. 2017, ApJ, 844, 18
Placco, V. M., Sneden, C., Roederer, I. U., et al. 2021a, linemake: Line list generator, Astrophysics Source Code Library, ascl:2104.027
Placco, V. M., Sneden, C., Roederer, I. U., et al. 2021b, RNAAS, 5, 92
Prantzos, N., Abia, C., Cristallo, S., Limongi, M., & Chieffi, A. 2020, MNRAS, 491, 1832
Qiu, Y., Radice, D., Richers, S., et al. 2025, PhRvD, 112, 123039
Racca, M., Hansen, T. T., Roederer, I. U., et al. 2025, A&A, 704, A282
Ramírez, I., & Meléndez, J. 2005, ApJ, 626, 465
Reichert, M., Bugli, M., Guilet, J., et al. 2024, MNRAS, 529, 3197
Reichert, M., Obergaulinger, M., Aloy, M. Á., et al. 2023, MNRAS, 518, 1557
Reichert, M., Obergaulinger, M., Eichler, M., Aloy, M. Á., & Arcones, A. 2021, MNRAS, 501, 5733
Ren, J., Christlieb, N., & Zhao, G. 2012, A&A, 537, A118
Roederer, I. U. 2013, AJ, 145, 26
Roederer, I. U., Beers, T. C., Hattori, K., et al. 2024, ApJ, 971, 158
Roederer, I. U., Cowan, J. J., Pignatari, M., et al. 2022a, ApJ, 936, 84
Roederer, I. U., Karakas, A. I., Pignatari, M., & Herwig, F. 2016, ApJ, 821, 37
Roederer, I. U., Kratz, K.-L., Frebel, A., et al. 2009, ApJ, 698, 1963
Roederer, I. U., Lawler, J. E., Den Hartog, E. A., et al. 2022b, ApJS, 260, 27
Roederer, I. U., Preston, G. W., Thompson, I. B., et al. 2014, AJ, 147, 136
Roederer, I. U., Sakari, C. M., Placco, V. M., et al. 2018, ApJ, 865, 129
Roederer, I. U., Sneden, C., Lawler, J. E., & Cowan, J. J. 2010, ApJL, 714, L123
Roederer, I. U., Vassh, N., Holmbeck, E. M., et al. 2023, Sci, 382, 1177
Safarzadeh, M., Ramirez-Ruiz, E., Andrews, J. J., et al. 2019a, ApJ, 872, 105
Safarzadeh, M., Sarmento, R., & Scannapieco, E. 2019b, ApJ, 876, 28
Sakari, C. M., Placco, V. M., Farrell, E. M., et al. 2018, ApJ, 868, 110
Saleem, M., Chen, H.-Y., Siegel, D. M., et al. 2025, arXiv:2508.06020
Saraf, P., Allende Prieto, C., Sivarani, T., et al. 2023, MNRAS, 524, 5607
Schatz, H., Toenjes, R., Pfeiffer, B., et al. 2002, ApJ, 579, 626
Schlafly, E. F., & Finkbeiner, D. P. 2011, ApJ, 737, 103
Seeger, P. A., Fowler, W. A., & Clayton, D. D. 1965, ApJS, 11, 121
Shah, S. P., Ezzeddine, R., Ji, A. P., et al. 2023, ApJ, 948, 122
Shah, S. P., Ezzeddine, R., Roederer, I. U., et al. 2024, MNRAS, 529, 1917
Shibata, M., Fujibayashi, S., Wanajo, S., et al. 2025, PhRvD, 111, 123017
Siegel, D. M., Barnes, J., & Metzger, B. D. 2019, Natur, 569, 241
Simmerer, J., Sneden, C., Cowan, J. J., et al. 2004, ApJ, 617, 1091
Simonetti, P., Matteucci, F., Greggio, L., & Cescutti, G. 2019, MNRAS, 486, 2896
Siqueira Mello, C., Hill, V., Barbuy, B., et al. 2014, A&A, 565, A93
Skúladóttir, Á., Hansen, C. J., Salvadori, S., & Choplin, A. 2019, A&A, 631, A171
Sneden, C., Bean, J., Ivans, I., Lucatello, S., & Sobeck, J. 2012, MOOG: LTE Line Analysis and Spectrum Synthesis, Astrophysics Source Code Library, ascl:1202.009
Sneden, C., Cowan, J. J., & Gallino, R. 2008, ARA&A, 46, 241
Sneden, C., Cowan, J. J., Ivans, I. I., et al. 2000a, ApJL, 533, L139
Sneden, C., Cowan, J. J., Lawler, J. E., et al. 2003, ApJ, 591, 936
Sneden, C., Johnson, J., Kraft, R. P., et al. 2000b, ApJL, 536, L85
Sneden, C., Lawler, J. E., Cowan, J. J., Ivans, I. I., & Den Hartog, E. A. 2009, ApJS, 182, 80
Sneden, C. A. 1973, PhD thesis, University of Texas, Austin
Sobeck, J. S., Kraft, R. P., Sneden, C., et al. 2011, AJ, 141, 175
Soker, N. 2026, IJAA, 16, 94
Sprouse, T. M., Lund, K. A., Miller, J. M., McLaughlin, G. C., & Mumpower, M. R. 2024, ApJ, 962, 79
Storm, N., Bergemann, M., Eitner, P., et al. 2025, MNRAS, 538, 3284
Ting, Y.-S., Conroy, C., Rix, H.-W., & Cargile, P. 2019, ApJ, 879, 69
Tody, D. 1986, SPIE, 627, 733
Tody, D. 1993, ASPC, 52, 173
Travaglio, C., Galli, D., Gallino, R., et al. 1999, ApJ, 521, 691
Travaglio, C., Gallino, R., Arnone, E., et al. 2004, ApJ, 601, 864
Tsujimoto, T., & Shigeyama, T. 2014, ApJL, 795, L18
Tull, R. G., MacQueen, P. J., Sneden, C., & Lambert, D. L. 1995, PASP, 107, 251
Unterborn, C. T., Johnson, J. A., & Panero, W. R. 2015, ApJ, 806, 139
Valentini, M., Chiappini, C., Bossini, D., et al. 2019, A&A, 627, A173
Vanderplas, J. T., Connolly, A. J., Ivezić, Ž., & Gray, A. 2012, in Introduction to astroML: Machine learning for astrophysics (CIDU), 47
van de Voort, F., Pakmor, R., Grand, R. J. J., et al. 2020, MNRAS, 494, 4867
Vassh, N., Mumpower, M. R., McLaughlin, G. C., Sprouse, T. M., & Surman, R. 2020, ApJ, 896, 28
Venn, K. A., Irwin, M., Shetrone, M. D., et al. 2004, AJ, 128, 1177
Virtanen, P., Gommers, R., Oliphant, T. E., et al. 2020, NatMe, 17, 261
Wanajo, S. 2007, ApJL, 666, L77
Wanajo, S., Fujibayashi, S., Hayashi, K., et al. 2024, PhRvL, 133, 241201
Wanajo, S., Hirai, Y., & Prantzos, N. 2021, MNRAS, 505, 5862
Wanajo, S., Itoh, N., Ishimaru, Y., Nozawa, S., & Beers, T. C. 2002, ApJ, 577, 853
Wanajo, S., Kajino, T., Mathews, G. J., & Otsuki, K. 2001, ApJ, 554, 578
Wehmeyer, B., Pignatari, M., & Thielemann, F.-K. 2015, MNRAS, 452, 1970
Westin, J., Sneden, C., Gustafsson, B., & Cowan, J. J. 2000, ApJ, 530, 783
Winteler, C., Käppeli, R., Perego, A., et al. 2012, ApJL, 750, L22
Wu, M.-R., Tamborra, I., Just, O., & Janka, H.-T. 2017, PhRvD, 96, 123015
Xiang, M., & Rix, H.-W. 2022, Natur, 603, 599
Xing, Q., Zhao, G., Aoki, W., et al. 2024, ApJ, 965, 79
Xylakis-Dornbusch, T., Hansen, T. T., Beers, T. C., et al. 2024, A&A, 688, A123
Yong, D., Karakas, A. I., Lambert, D. L., Chieffi, A., & Limongi, M. 2008a, ApJ, 689, 1031
Yong, D., Kobayashi, C., Da Costa, G. S., et al. 2021, Natur, 595, 223
Yong, D., Lambert, D. L., Paulson, D. B., & Carney, B. W. 2008b, ApJ, 673, 854
Zha, S., Müller, B., & Powell, J. 2024, ApJ, 969, 141